\documentclass[a4paper,11pt]{article}
\pdfoutput=1 
\usepackage{jheppub} 
\usepackage[T1]{fontenc} 
\usepackage{hyperref}

         \let\g = \gamma     \let\e = \epsilon

\newcommand{\nn}{\nonumber}

\newcommand{\beq}{\begin{equation}}
\newcommand{\eeq}{\end{equation}}
\newcommand{\bea}{\begin{eqnarray}}
\newcommand{\eea}{\end{eqnarray}}
\newcommand{\beqa}{\begin{eqnarray}}
\newcommand{\eeqa}{\end{eqnarray}}

\newcommand{\AL}[1]{\langle#1|}
\newcommand{\AR}[1]{|#1\rangle}
\newcommand{\SL}[1]{[#1|}
\newcommand{\SR}[1]{|#1]}
\renewcommand{\AA}[1]{\langle#1\rangle}
\newcommand{\SSS}[1]{[#1]}
\newcommand{\AS}[1]{\langle#1]}

\newcommand{\lid}[2]{#1\!\cdot\!#2}
\newcommand{\slashk}{k \! \! \!  /}
\newcommand{\slashK}{K \! \! \!  /}
\newcommand{\slashp}{p \! \! \!  /}

\newcommand{\slashr}{r \! \! \!  /}
\newcommand{\slasheps}{\epsilon \! \! \!  /}
\newcommand{\kapp}{\kappa}
\newcommand{\kapphat}{\hat{\kappa}}
\newcommand{\kstr}{\kappa^*}
\newcommand{\kstrhat}{\hat{\kappa}^*}
\newcommand{\qb}{\bar{q}}
\newcommand{\Amp}{\mathcal{A}}

\newcommand{\imag}{\mathrm{i}}

\newcommand{\vep}{\varepsilon}
\newcommand{\graph}[3]{\raisebox{-#3ex}{\epsfig{file=figures/#1.pdf,width=#2ex}}}

\title{\boldmath BCFW recursion for TMD parton scattering}

\author{A. van Hameren}
\author{and M. Serino}
\affiliation{The H. Niewodnicza\'nski Institute of Nuclear Physics, Polish Academy of Sciences, \\
ul. Radzikowskiego 152, 31-342, Cracow, Poland}

\emailAdd{hameren@ifj.edu.pl}
\emailAdd{mirko.serino@ifj.edu.pl.edu}

\abstract{
We investigate the application of the BCFW recursion relation to scattering amplitudes with one off-shell particle in a Yang-Mills theory with fermions.
We provide a set of conditions of applicability of the BCFW recursion, stressing some important differences with respect to the pure on-shell case.
We show how the formulas for Maximally-Helicity-Violating (MHV) configurations with any number of partons, which are well known in the fully on-shell case,
are generalised to this kinematic regime.
We also derive analytic expressions for all the helicity configurations of the 5-point color-stripped tree-level amplitudes 
for any of the partons being off the mass shell. We release with the arXiv submission a Mathematica notebook containing all the results derived in the paper.}

\begin{document} 

\begin{flushright}
IFJPAN-IV-2015-5
\end{flushright}

\maketitle
\flushbottom

\section{Introduction}

One of the remarkable features of quantum chromodynamics (QCD) is that analytic expressions 
for tree-level scattering amplitudes of this theory turn out to be much simpler than one would expect,
starting their construction from the Feynman rules.
Originally, the inventive use of the gauge freedom inherent to the theory~\cite{%
 Kleiss:1985yh%
,Gunion:1985vca%
,Xu:1986xb%
}
in combination with off-shell recursion allowed for reaching compact expressions~\cite{%
 Berends:1987me%
,Mangano:1987xk%
,Mangano:1990by%
,Berends:1989hf%
,Kosower:1989xy%
}.
With the introduction of the on-shell recursion by Britto, Cachazo, Feng and Witten (BCFW)~\cite{%
 Britto:2004ap%
,Britto:2005fq%
},
the occurrence of the compact expression for multi-gluon amplitudes became more natural, 
since it is a recursion of the gauge invariant compact expressions themselves, rather than of gauge non-invariant vertices or off-shell currents.
This type of recursion found its use in the calculation of other amplitudes, 
in particular amplitudes involving massless quark-antiquark pairs~\cite{%
Luo:2005rx%
}.

Another, recent, application of BCFW recursion is the calculation of multi-gluon amplitudes with off-shell gluons~\cite{%
 vanHameren:2014iua%
}.
These are relevant for the calculation of cross sections with factorization prescriptions that demand off-shell partonic initial states, like High Energy Factorization (HEF)~\cite{%
 Catani:1990eg%
,Collins:1991ty%
} (also known as TMD Factorization, i.e. Transverse Momentum Dependent Factorization),
and for example the parton Reggeization approach~\cite{%
 Fadin:1993wh%
,Fadin:1996nw%
}.
Such amplitudes can be defined in a gauge invariant manner~\cite{%
 Lipatov:1995pn%
,Antonov:2004hh%
,vanHameren:2012if%
,Kotko:2014aba%
},
in the sense that on-shell gluons satisfy the usual Ward identities, and any gauge can be chosen for the internal gluon propagators.

Besides for off-shell initial-state gluons, the amplitudes can also be defined for off-shell initial-state quarks~\cite{%
 Lipatov:2000se%
,vanHameren:2013csa%
}.
Recent applications of such amplitudes can be found in~\cite{%
 Nefedov:2013ywa%
,Kniehl:2014qva%
,Nefedov:2014qea%
,Karpishkov:2014epa%
}.
In this paper, we present how multi-gluon amplitudes with one quark-antiquark pair and one off-shell parton can be calculated via BCFW recursion.
The off-shell parton may be a gluon, a quark or an antiquark.

The paper is organized as follows.
In section \ref{definitions} we recall the conventions and parametrizations introduced in \cite{vanHameren:2014iua}, to explain our notation.
In section \ref{conditions} we briefly recall the BCFW recursion in the on-shell case, discuss in detail its extension to color-ordered tree-level amplitudes with off-shell partons and provide conditions on the complex shifts which can be used.
All the possible kinds of contributions to the recursive algorithm are classified in section \ref{residuesec}.
In section \ref{MHVsec} we show how the structure of Maximally Helicity Violating (MHV) amplitudes with one off-shell parton can be recovered by induction, as in the on-shell case.
Then, in section \ref{five}, the recursion procedure is extended to the 5-point amplitudes which, in contrast with the on-shell case, are not MHV.
Section \ref{summary} is a short summary of the results presented, while a few technical details are left to the appendices.

\section{Definitions}\label{definitions}%

We will always consider scattering amplitudes with all particles outgoing.
The details of the construction briefly outlined here are given in \cite{vanHameren:2014iua}
and we will not repeat them in detail.

It suffices to say that for every particle with momentum $k^\mu_i$ 
we have to specify a direction $p^\mu_i$ which is orthogonal to its momentum,
\begin{align}
k_1^\mu + k_2^\mu + \cdots + k_n^\mu = 0
&\qquad\textrm{momentum conservation}\label{Eq:momcons}\\
p_1^2 = p_2^2 = \cdots = p_n^2 = 0
&\qquad\textrm{light-likeness}\label{momcon1}\\
\lid{p_1}{k_1} = \lid{p_2}{k_2}=\cdots=\lid{p_n}{k_n}=0
&\qquad\textrm{eikonal condition}
\label{Eq:eikcon}
\end{align}
With the help of an auxiliary light-like four-vector $q^\mu$, the momentum $k^\mu$ 
can be decomposed in terms of its light-like direction $p^\mu$, satisfying $\lid{p}{k}=0$, and a transversal part, following
\begin{equation}
k^\mu = x(q)p^\mu 
- \frac{\kapp}{2}\,\frac{\AS{p|\gamma^\mu|q}}{\SSS{pq}} 
- \frac{\kstr}{2}\,\frac{\AS{q|\gamma^\mu|p}}{\AA{qp}} \, ,
\end{equation}
with
\begin{equation}
x(q)=\frac{\lid{q}{k}}{\lid{q}{p}}
\quad,\quad
\kapp = \frac{\AS{q|\slashk|p}}{\AA{qp}}
\quad,\quad
\kstr = \frac{\AS{p|\slashk|q}}{\SSS{pq}} \, .
\end{equation}
If $k$ is on-shell, then $p=k$.
The coefficients $\kapp$ and $\kstr$ are independent of the auxiliary momentum $q^\mu$ (see appendix \ref{AppSchouten}) and
\begin{equation}
k^2 = -\kapp\kstr \ .
\end{equation}
We will consider the calculation of {\em color-ordered\/} or {\em dual\/} amplitudes,
so that, despite our jargon being mostly the QCD one, the results hold for any Yang-Mills theories with fermions
which are parity-invariant.
By construction, these amplitudes contain only planar Feynman graphs 
and are constructed with color-stripped Feynman rules.
Gauge-invariant amplitudes with off-shell particles also contain auxiliary eikonal particles~\cite{vanHameren:2012if,vanHameren:2013csa}.
Eikonal quarks will be denoted by dashed oriented lines.
An off-shell fermion is represented by a double oriented solid line;
an off shell gluon by a double solid line.
In order to connect off-shell particles with their diagrammatic representation in terms of the auxiliary quarks or photons,
one can imagine that the double lines representing them can be bent open.
To put it another way, when one sees an external particles represented by a double line in our conventions, one has to imagine that,
in the case of an off-shell gluon, they must be intended as eikonal quarks pair $(q_A,\qb_A)$;
in the case of an off-shell fermion, the double line splits into an eikonal quark line and an auxiliary null momentum photon ($\gamma_A$).
Once this is done, these particles can be combined in all the possible ways using the Feynman rules reported below to obtain gauge-invariant amplitudes.
For further details on this construction, we refer the reader to the original papers~\cite{vanHameren:2012if,vanHameren:2013csa},
where  the Feynman rules were worked out in detail. \\
We stress that, in order to apply the BCFW recursion, {\em Feynman rules are not at all necessary}: it is enough to know $3$-point amplitudes,
which can be built diagrammatically but also derived from first principles (in the on-shell case) or derived via the recursion itself,
as we will briefly point out in Appendix \ref{App3Point}, particularly in \ref{App3PointSpecial}. \\
We report the Feynman rules for color-ordered amplitudes here in the Feynman gauge :
\begin{eqnarray}
&&
\graph{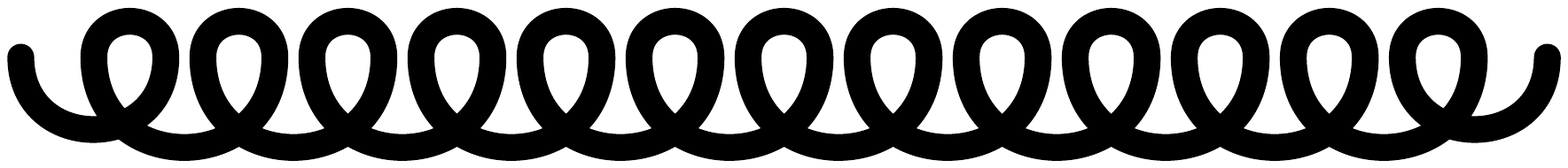}{15}{10} = \frac{-\eta^{\mu\nu}}{K^2} \, ,
\quad\quad\quad
\graph{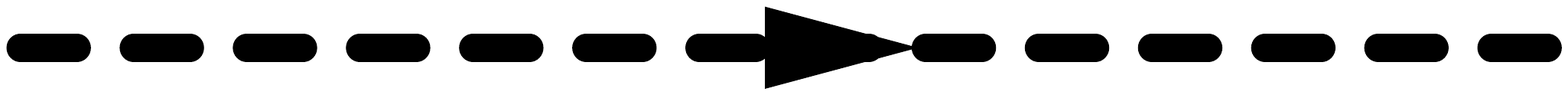}{15}{10} = \frac{\slashp}{2\lid{p}{K}} \, ,
\quad\quad\quad
\graph{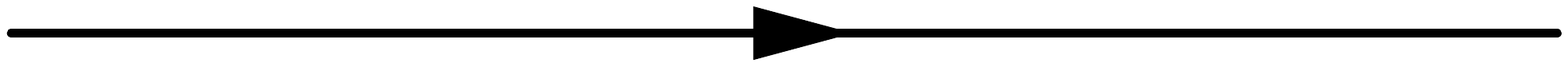}{15}{10} = \frac{\slashk}{k^2} \, ,
\nn \\
&&
\graph{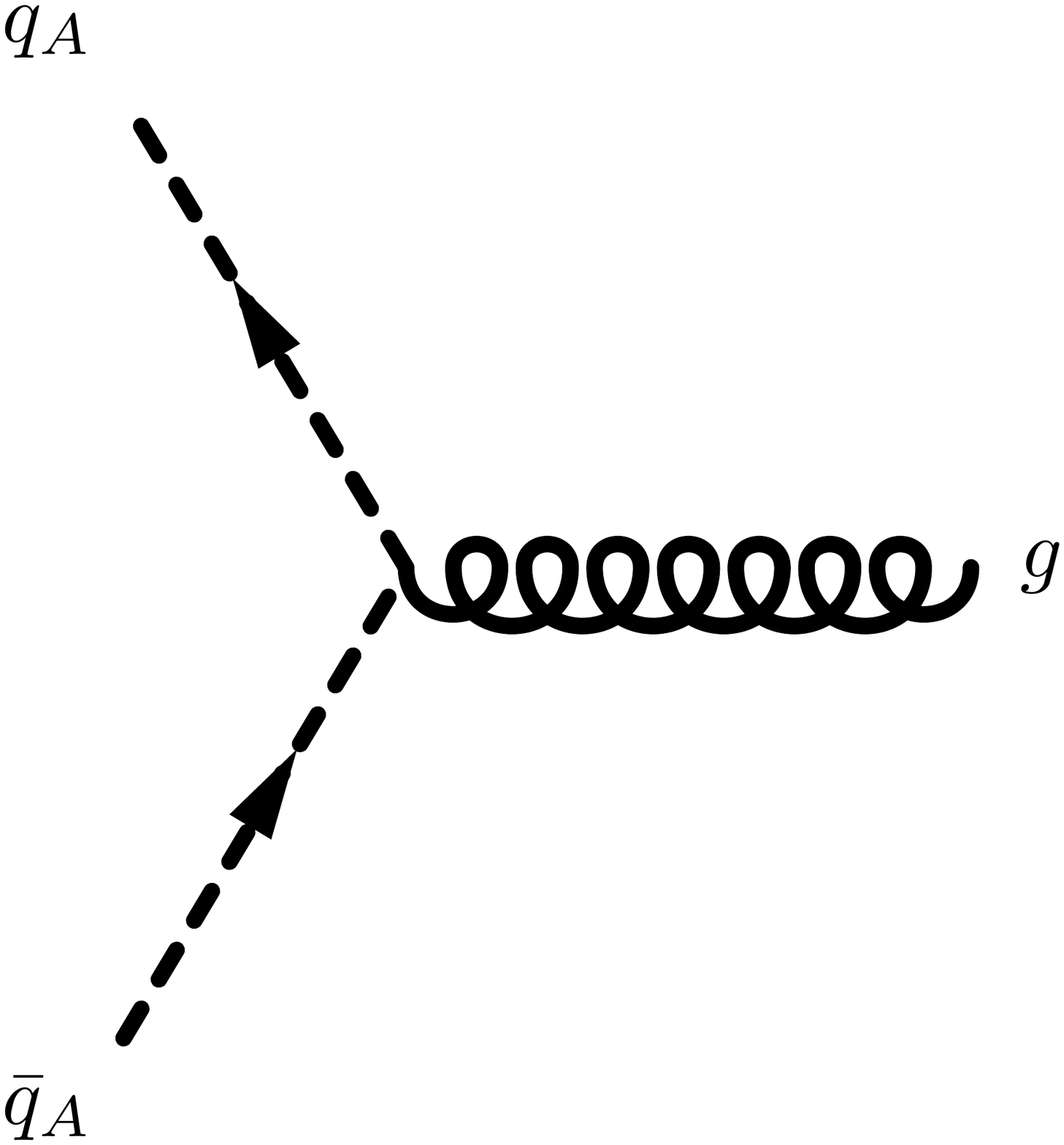}{20}{13} = \frac{\gamma^\mu}{\sqrt{2}}
\quad\quad\quad
\graph{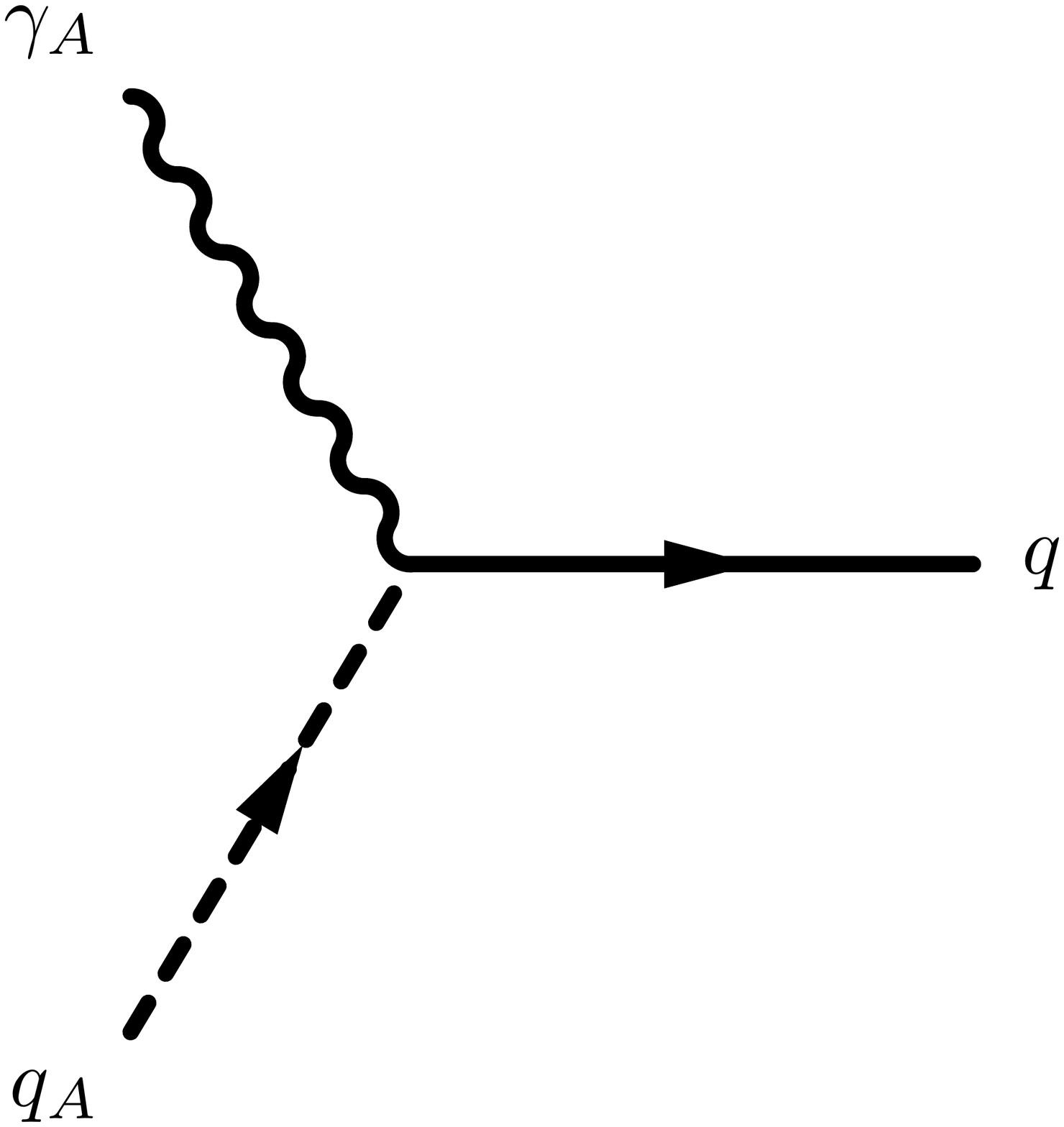}{20}{13}= \frac{\gamma^\mu}{\sqrt{2}} \, ,
\nn
\end{eqnarray}
\begin{eqnarray}
&& 
\graph{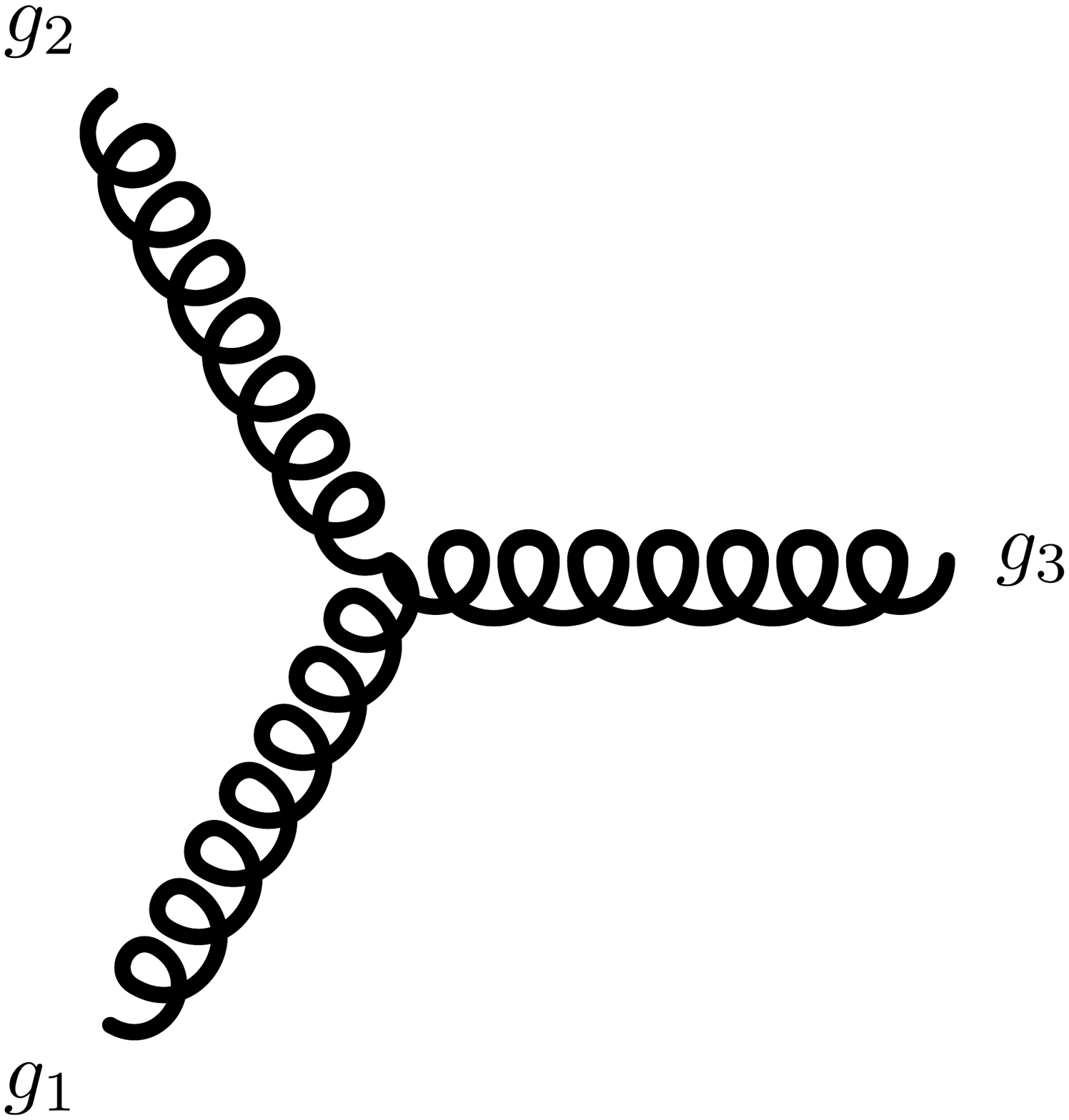}{20}{13} = \frac{1}{\sqrt{2}} \big[(K_1-K_2)^{\mu_3}\eta^{\mu_1\mu_2} + (K_2-K_3)^{\mu_1}\eta^{\mu_2\mu_3} + (K_3-K_1)^{\mu_2}\eta^{\mu_3\mu_1}\big]
\nn \\
&&
\graph{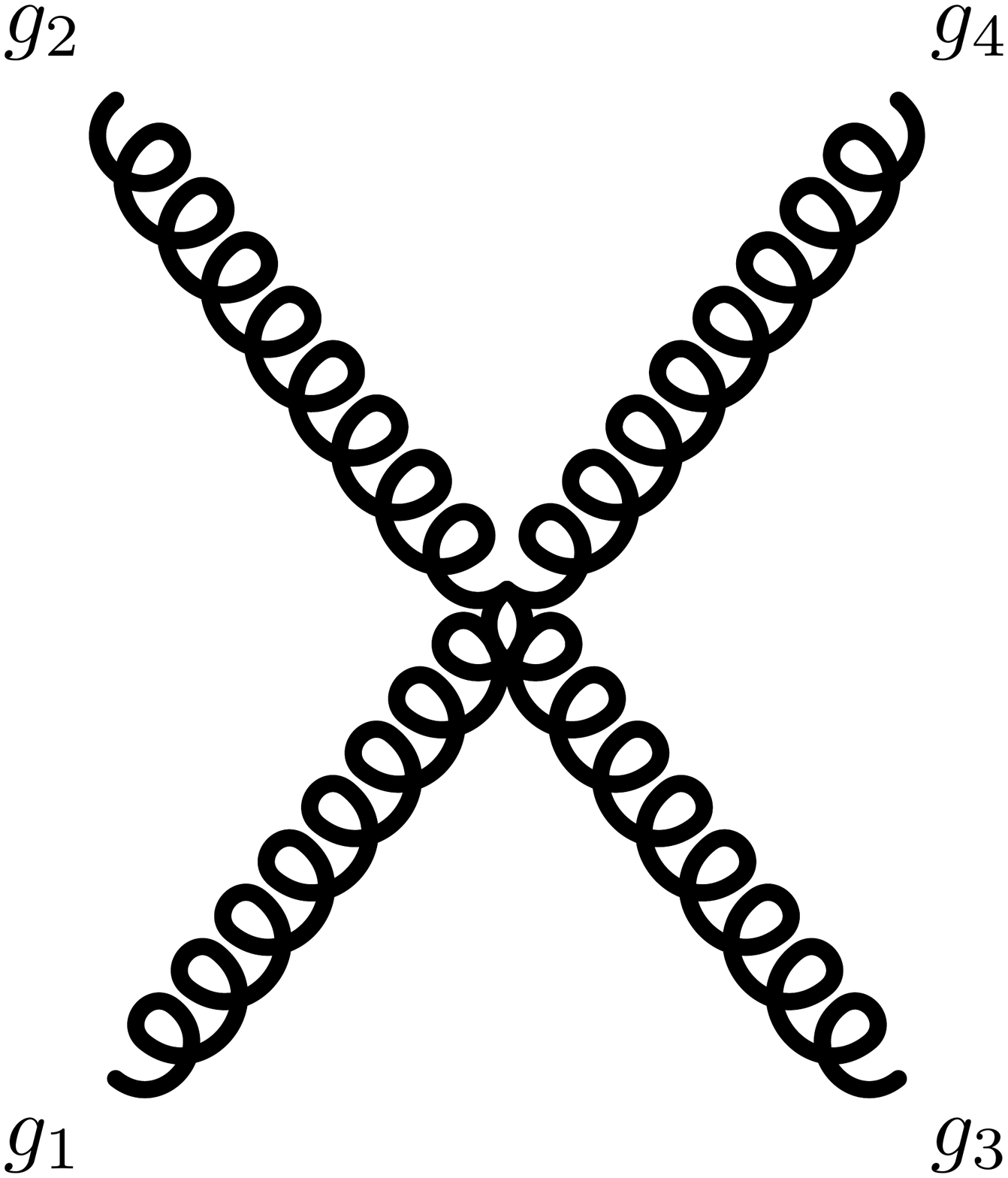}{20}{13}= -\frac{1}{2}\,\big[ 2\,\eta^{\mu_1\mu_3}\eta^{\mu_2\mu_4} - \eta^{\mu_1\mu_2}\eta^{\mu_3\mu_4} - \eta^{\mu_1\mu_4}\eta^{\mu_2\mu_3} \big]
\nn
\end{eqnarray}
The symbol $p$ refers to the direction associated with the eikonal quark line,
$K$ stands for the momentum flowing through a propagator or into a vertex, and $\eta^{\mu\nu}$ is the Minkowski metric tensor.
We remind that the polarization vectors for gluons can be expressed as
\begin{equation} \label{polarization}
\e^\mu_{+ } = \frac{\AS{q | \g^\mu | g}}{\sqrt{2}\AA{q g}} \, , 
\quad
\e^\mu_{-} = \frac{\AS{g |\g^\mu | q}}{\sqrt{2}\SSS{g q}} \, ,
\end{equation}
where $q$ is the auxiliary light-like vector and $g$ stands for the gluon momentum.
In the expressions of the amplitudes, gluons will be denoted by the number of the corresponding particle, 
whereas we will always explicitly distinguish quarks and antiquarks with $q$ and $\qb$ respectively.

\section{The BCFW construction}\label{conditions}

This is the most important part of the paper, because we identify the conditions allowing to perform the BCFW recursion
and clarify the differences between the off-shell case and the on-shell case, which are particularly significant for fermions.
For the sake of completeness, we will report the results discussed in \cite{vanHameren:2014iua} and then move to the fermion case,
which was discussed in the on-shell kinematics in \cite{Luo:2005rx}.

We start by recalling the general idea of the BCFW relation \cite{Britto:2004ap,Britto:2005fq}.
The starting point is Cauchy's residue theorem, stating that
\begin{equation}
\lim_{z \to \infty} f(z) = 0 \Rightarrow \oint \frac{dz}{2\pi\imag} \, \frac{f(z)}{z} = 0 \, ,
\end{equation}
where the integration countour encloses all the poles of the rational function $f(z)$ and extends to infinity, implying that 
the function at the origin $f(0)$ can be represented as a sum over the residues at the single poles in the complex plane,
\begin{equation}
f(0) = - \sum_{i} \frac{\lim_{z\rightarrow z_i} f(z)\, (z-z_i) }{z_i} \, .
\label{residues}
\end{equation}
The idea underlying the BCFW recursion relation is to apply (\ref{residues}) to the evaluation of a tree-level
scattering amplitude in a massless theory. This can be accomplished very naturally in the spinor helicity formalism,
preserving both momentum conservation and on-shellness of all the momenta in the scattering amplitude (\ref{Eq:eikcon}),
by adding a complex shift to the momenta of two of the particles, so that the amplitude is turned into a rational function
of the complex variable z.

The crucial step, for our purposes, is to cast this into a form that suits 
the application to off-shell processes and reduces to the more known formulas in the literature about on-shell recursion relations
if all the particles are on the mass shell.
Let us pick up two particles, which we label $i$ and $j$ for convenience, and let us choose the direction of each to
be the reference vector for the other, so that their momenta with transverse (off-shell) component are
\begin{eqnarray}
k_i^\mu = x_i(p_j)\, p_i^\mu - \frac{\kappa_i}{2}\, \frac{\AL{i} \g^\mu \SR{j}}{\SSS{ij}} - \frac{\kstr_i}{2}\, \frac{\AL{j} \g^\mu \SR{i}}{\AA{ji}}
\nn \\
k_j^\mu = x_j(p_i)\, p_j^\mu - \frac{\kappa_j}{2}\, \frac{\AL{j} \g^\mu \SR{i}}{\SSS{ji}} - \frac{\kstr_j}{2}\, \frac{\AL{i} \g^\mu \SR{j}}{\AA{ij}} \, .
\end{eqnarray}
We adopt the notation $\SR{p_a}(\AR{p_a}) \equiv \SR{a} (\AR{a})$, where $g_a$ is any gluon, 
whereas for quarks $\SR{p_q} (\AR{p_q}) \equiv \SR{q} (\AR{q})$ and $\SR{p_{\qb}} (\AR{p_{\qb}}) \equiv \SR{\qb} (\AR{\qb})$.
Like in \cite{vanHameren:2014iua}, we choose the shift vector to be
\begin{equation}
e^\mu = \frac{1}{2}\, \AL{i} \g^\mu \SR{j}
~,
\end{equation}
which satisfies
\beq
p_i \cdot e = p_j \cdot e = e \cdot e = 0 \, .
\eeq
Shifted quantities will be denoted by a hat symbol.
Shifted momenta are thus
\bea
\hat{k}_i^\mu = k_i + z e^\mu 
&=&
x_i(p_j)\, p_i^\mu - \frac{\kappa_i - z\SSS{ij}}{2}\, \frac{\AL{i} \g^\mu \SR{j}}{\SSS{ij}} - \frac{\kstr_i}{2}\, \frac{\AL{j} \g^\mu \SR{i}}{\AA{ji}}
\nn \\
&=& 
x_i(p_j)\, p_i^\mu - \frac{\hat{\kappa}_i}{2}\, \frac{\AL{i} \g^\mu \SR{j}}{\SSS{ij}} - \frac{\kstr_i}{2}\, \frac{\AL{j} \g^\mu \SR{i}}{\AA{ji}} \, ,
\nn \\
\hat{k}_j^\mu = k_j - z e^\mu 
&=&
x_j(p_i)\, p_j^\mu - \frac{\kappa_j}{2}\, \frac{\AL{j} \g^\mu \SR{i}}{\SSS{ji}} - \frac{\kstr_j + z \AA{ij}}{2}\, \frac{\AL{i} \g^\mu \SR{j}}{\AA{ij}}
\nn \\
&=&
x_j(p_i)\, p_j^\mu - \frac{\kappa_j}{2}\, \frac{\AL{j} \g^\mu \SR{i}}{\SSS{ji}} - \frac{\kstrhat_j}{2}\, \frac{\AL{i} \g^\mu \SR{j}}{\AA{ij}} \, ,
\label{shifts}
\eea
a construction which manifestly preserves momentum conservation and the eikonal conditions $p_i \cdot \hat{k}_i = 0$ and $p_j \cdot \hat{k}_j = 0$.
Of course, if we had chosen $e^\mu = 1/2\, \AL{j}\g^\mu\SR{i}$ the shifting coefficients would have been $\kstrhat_i$ and $\kappa_j$.
Below we discus in detail the constraints imposed on the choice of shift vectors by the hypothesis of the BCFW recursion relation. \\
Notice that, if the particles are on-shell, then applying the relation $\g_\mu \AL{i}\g^\mu\SR{j} = 2\, \left( \SL{j}\AR{i} + \AL{i}\SR{j}  \right)$ 
to (\ref{shifts}) immediately leads to
\begin{equation}
\SR{\hat{i}} = \SR{i} + z\, \SR{j} \, , \quad \AL{\hat{j}} = \AL{j} - z\, \SL{i} \, ,
\end{equation}
which is the form in which the shifts are usually presented in the literature and to which the reader might be accustomed.
We may summarize the analysis so far by saying that 
{\em for an off-shell momentum only one of the scalar coefficients ($\kappa$ or $\kstr$) in the transverse momentum shifts, whereas the directions do not shift}. 
Let us schematically collect the changes induced either in the momenta or in the directions by the two possible shift vectors, for definitness.
First of all, for any shift vector $e^\mu$ and for any couple of particles $i$ and $j$ that we may pick up, we always convey that
\begin{eqnarray}
\hat{k}_i &=& k_i + z\, e^\mu \, ,\nn \\
\hat{k}_j &=& k_j  -  z\, e^\mu \, .
\end{eqnarray}
Now we summarise separately the results of the choice of either shift vector:
\begin{eqnarray}
e^\mu &=& \frac{1}{2}\AL{i}\g^\mu\SR{j} 
\Leftrightarrow 
\left\{ \begin{array}{c}
\textrm{$i$ off-shell:} \quad \kapphat_i = \kappa_i - z \SSS{ij} 
\\
\textrm{$i$ on-shell:} \quad \SR{\hat{i}} = \SR{i} + z \SR{j} 
\\
\textrm{$j$ off-shell:} \quad \kstrhat_j = \kstr_j + z \AA{ij} 
\\
\textrm{$j$ on-shell:} \quad \AR{\hat{j}} = \AR{j} - z \AR{i}
\end{array} \right.
\label{shift1}
\\
e^\mu &=& \frac{1}{2}\AL{j}\g^\mu\SR{i}
\Leftrightarrow 
\left\{ \begin{array}{c}
\textrm{$i$ off-shell:} \quad \kstrhat_i = \kstr_i - z \AA{ji} 
\\
\textrm{$i$ on-shell:} \quad \AR{\hat{i}} = \AR{i} + z \AR{j} 
\\
\textrm{$j$ off-shell:} \quad \kapphat_j = \kappa_j + z \SSS{ji} 
\\
\textrm{$j$ on-shell:} \quad \SR{\hat{j}} = \SR{j} - z \SR{i}
\end{array} \right.
\label{shift2}
\end{eqnarray}
It is important to notice that (\ref{shift1}) and (\ref{shift2}), taking (\ref{polarization}) into account,
imply the following large $z$ behaviors for the polarization vectors of the on-shell shifted gluons:
\begin{eqnarray}
e^\mu &=& \frac{1}{2}\AL{i}\g^\mu\SR{j} \Rightarrow \epsilon^\mu_{i-} \sim \frac{1}{z} \quad \textrm{and} \quad \epsilon^\mu_{j+} \sim \frac{1}{z} \, ,
\nn \\
e^\mu &=& \frac{1}{2}\AL{j}\g^\mu\SR{i} \Rightarrow \epsilon^\mu_{i+} \sim \frac{1}{z} \quad \textrm{and} \quad \epsilon^\mu_{j-} \sim \frac{1}{z} \, ,
\label{shift3}
\end{eqnarray}
whereas the polarization vectors of shifted gluons with opposite helicity stay constant.

Now, the important point is that \emph{not all of these choices are suitable to apply the BCFW recursion}.
This is because some of them, depending also on the nature of the shifted particles, imply violation of the asymptotic condition
\begin{equation}
\lim_{z\to\infty} \Amp(z) = 0 \, ,
\end{equation}
which is essential in order to apply the residue theorem.
In the following, we recall the results of the analysis of the asymptotic behaviour of the complexified
amplitudes for the on-shell case and then present the general analysis in our kinematics.

\subsection{Purely gluonic amplitudes}%

A clear analysis of the on-shell purely gluonic case is contained in \cite{Britto:2005fq}
and it is found that, if the shift vector is $e^\mu = \frac{1}{2} \AL{i}\g^\mu\SR{j}$, 
then $\Amp(z) \stackrel{z\to\infty}{\longrightarrow} 0$  for three possible helicity choices of the shifted particles, namely $(h_i,h_j) = (-,+),(-,-),(+,+)$.
The argument proving the validity of the two equal helicity cases requires the knowledge of the construction of tree amplitudes through MHV
diagrams presented in \cite{Cachazo:2004kj} and we will not go through it; it is easy to obtain a diagrammatic proof for the first case, that we will call the
\emph{original BCF prescription} from now on, as it was the only one discussed in \cite{Britto:2004ap}.
In fact, any Yang-Mills tree-level amplitude is a product of vertices and propagators, and thus a shifted amplitude is a rational function of $z$.
Flowing along the diagram from the $i$-th through the $j$-th external gluon, every propagator will contribute a factor $1/z$, whereas vertices
can contribute a factor $z$ only if they are cubic; in the case in which all vertices are cubic, there can be at most one more vertex than there are propagators,
so the shifted amplitude behaves $\sim z$. But then, due to the shifts in (\ref{shift1}), the polarization vectors $\epsilon^\mu_{i\,-}$ and $\epsilon^\mu_{j\,+}$ 
both behave like $\sim 1/z$ (see( \ref{polarization})), so in the end the amplitude is $\sim 1/z$ asymptotically, completing the proof.
This proof seems to be gauge-dependent, since it requires the auxiliary 
momenta in the definition of $\epsilon^\mu_{i\,-}$ and $\epsilon^\mu_{j\,+}$ 
not to be $k_j^\mu$ and $k_i^\mu$ respectively, because for that choice (\ref{shift3}) does not hold.
However, we have already proved the correct large-$z$ behavior graph-by-graph, and gauge cancellations 
in the sum of all graphs guarantee the correct large-$z$ behavior for any choice of auxiliary momenta.
Of course, in the case $e^\mu = \frac{1}{2} \AL{j} \g^\mu \SR{i}$, the allowed helicity pairs are $(h_i,h_j) = (+,-), (-,-), (+,+) $.

In the general discussion of scattering amplitudes with off-shell particles, we will avoid to choose shift vectors which reduce,
in the on-shell limit, to the equal helicity cases; this will allow us to always keep our arguments at the level of the intuitive and familiar Feynman diagrams picture.
In the derivations in the main text, however, we will use such a choice once, for a 5-point amplitude, to show explicitly its effectiveness.\\
It should be clear that the proved validity in the on-shell case transfers directly to the off-shell kinematics, because
eikonal quark vertices only depend on the direction and eikonal quark propagators can only contribute powers of $z$ to the denominator.
So, if some of the gluons are off-shell and two on-shell gluons are shifted, everything stays exactly the same. \\

If one of the shifted gluons is off-shell and the other one is on-shell, we assume that the helicity of the latter agrees with the original BCF prescription.
It is then easy to realize that the BCFW argument holds in this case as well, provided we include the off-shell gluons' propagators in the amplitude,
so that both external lines contribute a factor $1/z$. 
In particular, the fact that the shifts preserve the eikonal conditions implies that 
the eikonal propagators associated to shifted off-shell gluons do not contribute powers of $z$ to the denominator. \\

Finally, if both shifted gluons are off-shell, they both will contribute a factor $1/z$ with either shift vector choice.
Actually, this is quite general: \emph{if both shifted particles are off-shell, the BCFW recursion works with both shift vectors}.

\subsection{Amplitudes with one fermion pair: shifting gluons}

%
\begin{figure}[h]
\begin{center}
\includegraphics[scale=0.7]{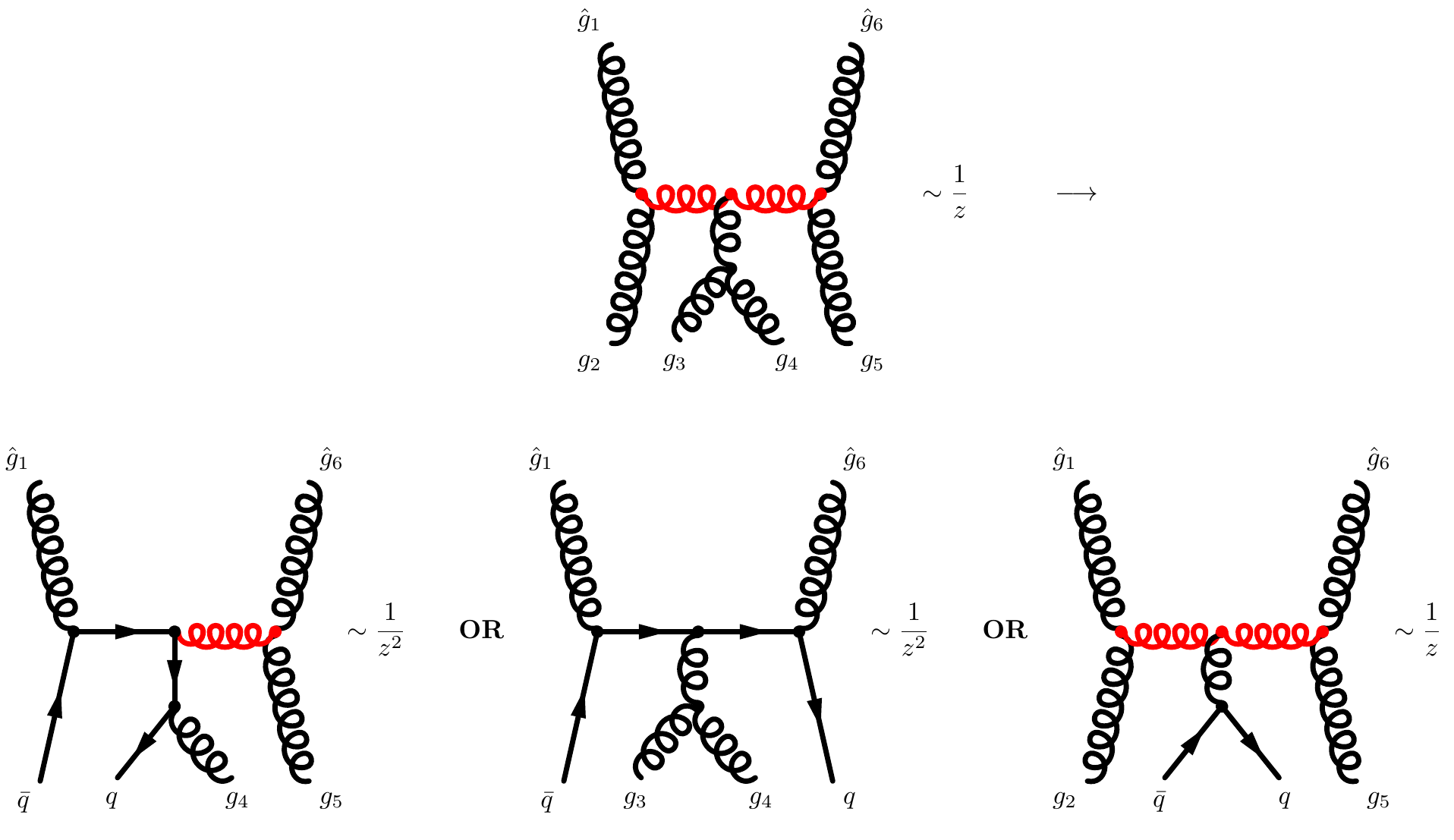}
\caption{These are the only three possible modifications (modulo specular ones) leading from the $6$-gluon diagram
to a diagram with 4-gluons and a fermion pair. We depict in red the propagators and vertices which are not asymptotically constant as functions of z.
We observe that the large $z$ behaviour stays the same only if no fermion propagator is introduced by the switch; otherwise, it always improves by one power.}
\label{z_flow}
\end{center}
\end{figure}

In this case, it is apparent that, if we shift only gluons in such a way as to recover the original BCF prescription in the on-shell limit,
the proof is straightforward, because fermion-gluon vertices do not depend on momenta, whereas fermion propagators with shifted momenta
are constant for $z \to \infty$. This implies that, in the worst case, the behaviour of the product of vertices and propagators 
will be the same as in the case with only gluons, whereas off-shell gluon propagators or the on-shell gluon polarization vectors 
make the all amplitude behave like $\sim 1/z$. \\
There is a clear-cut way to visualise this, which is illustrated in fig. \ref{z_flow}.
Consider a Feynman diagram for $n$-gluon scattering which contains only triple gluon vertices, i.e.\ one of the possibly worst behaved
contributions to the $n$-gluon scattering amplitude. In the example depicted in the figure, we have $6$ external gluons, 
$4$ internal vertices and $3$ propagators. 
We shift two of the external gluons according to the original BCF prescription, so as to guarantee the overall $1/z$ behavior.
Now we turn two unshifted gluons into a fermion and an antifermion, 
in order to generate a diagram contributing to the $n$-point amplitude with $n-2$ gluons and one fermion pair:  
it is seen  in fig. \ref{z_flow} for a $6$-point diagram (and it is easy to understand it in general) that, 
in order to connect the fermions, the number of triple gluon vertices which have to turn into gluon-fermion vertices 
is always higher by one unit than the number of gluon propagators becoming fermion propagators.
Notice that every diagram contributing to an amplitude with a fermion pair can be thought as the product
of the contributions generated in this way, through which fermion number flows, 
times a purely gluonic sub-diagram, which can contain also $4$-gluon vertices.
So we are not disregarding any diagram.  \\

\subsection{Amplitudes with one fermion pair: shifting one gluon and one fermion}\label{fancy_shift}

Now we discuss what happens if we also shift  fermion lines.
It is best to briefly review the discussion of the on-shell case presented in \cite{Luo:2005rx}; 
this reference also treats the cases when multiple fermion pairs and species are present, that we disregard.
The qualification stressed there and relevant to this work is that
{\em for on-shell amplitudes it is not allowed to shift one fermion and one gluon with the same helicity sign}.
The reason for this proviso is the following: an amplitude with one fermion pair is given by
\begin{equation}
\Amp(\qb^+,q^-,g_1,\dots, g_n) =  \AL{q} \dots \SR{\qb} \, ,
\end{equation}
where the dots stand for the product of propagators, vertices and gluon polarization vectors
(inverting the fermion pair helicities would simply amount to $\qb \leftrightarrow q$).
If one shifts the fermion $q^-$ and one gluon with negative helicity $g^-$, 
then the shift vector is either $e^\mu = \AL{q} \g^\mu \SR{g}/2 $ or $e^\mu = \AL{g} \g^\mu \SR{q}/2 $.
In the first case neither $\AR{q}$ nor $\epsilon^\mu_{g\,-}$ are shifted, 
while in the second their large $z$ behaviours compensate, 
so there are not enough negative powers of $z$ to tame $\Amp(z)$.
On the other hand, $e^\mu = \AL{q} \g^\mu \SR{g}/2$  for $h_g= +1$
or $\AL{g} \g^\mu \SR{\qb}/2 $ for $h_g= -1$ are manifestly good shift vectors, because 
the gluon contributes a factor $1/z$ whereas the fermion spinor in the amplitude stays inert;
beside, there will always be at least one fermion-gluon vertex on the path along which
$z$ flows from one shifted particle to the other, so that the product of propagators and vertices
is either asymptotically constant or vanishes. This can be checked graphically exactly like in fig. \ref{z_flow}.

We can then choose to shift an off-shell (anti) fermion and an on-shell gluon
provided that, in the on-shell limit, the (anti) fermion has opposite helicity sign w.r.t.\ the gluon.
We dub this the {\em original LW prescription} because, to the extent of our knowledge,
it was first formulated in \cite{Luo:2005rx}.

However, there is another possibility, which is typical of the fermion case: 
we can choose the shift vector in such a way that the shifted particles will have the same helicity sign in the on-shell limit, 
provided the gluon polarization behaves like $1/z$: this is because, as long as the fermion is off-shell, its direction
does not shift, whereas its transverse momentum does, and this ensures that the amplitude will have the correct
asymptotic behaviour; an interesting discontinuity appears in this case, taking the on-shell limit, because the recursion
keeps working if the fermion has an arbitrary small transverse momentum, but stops working as soon as this vanishes. 

Let us clarify this last point with an example: 
suppose we want to evaluate $\Amp(g_1^+,g_2^+,\qb^*,q^-,g_3^-)$ recursively.
Fermion helicity conservation implies that, in the on-shell limit, $\qb$ must have negative helicity.
We have two kinds of choices at hand:
\begin{enumerate}
\item we choose $e^\mu = \frac{1}{2} \AL{3} \g^\mu \SR{\qb}$, which would work for $\Amp(g_1^+,g_2^+,\qb^+,q^-,g_3^-)$ as well;
\item we choose $e^\mu = \frac{1}{2} \AL{\qb} \g^\mu \SR{2}$, which \emph{would not work for} $\Amp(g_1^+,g_2^+,\qb^+,q^-,g_3^-)$, 
         but it does yield the correct result as long as $k_{\qb}^2 \neq 0$.
\end{enumerate}
We will explicitly perform the computation of this specific amplitude in three ways in section \ref{five}.

\section{Classification of all the possible residues}\label{residuesec}

In this section, we classify the possible residues which appear due to single poles in $z$.
These poles appear because of denominators of the gluon or fermion propagators vanishing.
We convey that the momentum $K^\mu$ will always denote the momentum flowing to the 
propagator exhibiting a pole.

Our scattering amplitude $\Amp(0)$, according to (\ref{residues}), is given by
\begin{equation}
\Amp(0) = \sum_{s=g,f} \left(  \sum_{p} \sum_{h=+,-} \mathrm{A}^s_{p,h} + \sum_{i} \mathrm{B}^s_i  + \mathrm{C}^s + \mathrm{D}^s \right) \, ,
\end{equation}
where the index $s$ refers to the particle species, which can be gluon or fermion, and $h$ to the helicity.
Let us discuss in some detail these contributions.

\subsection{Gluon residues}

\begin{align}
\mathrm{A}^g_{p,h} &= \graph{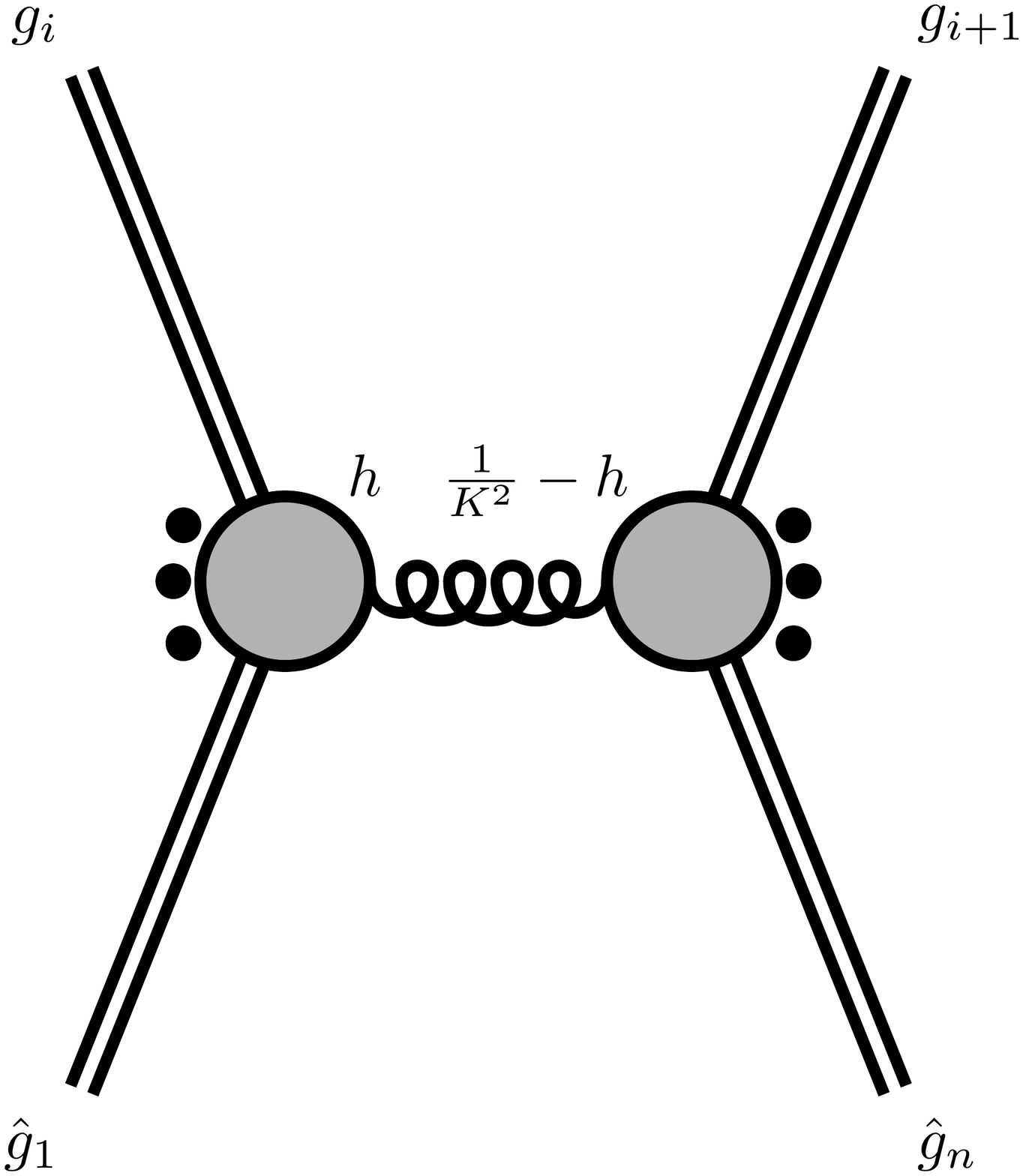}{25}{18}
\quad&
\mathrm{B}^g_{i} &= \graph{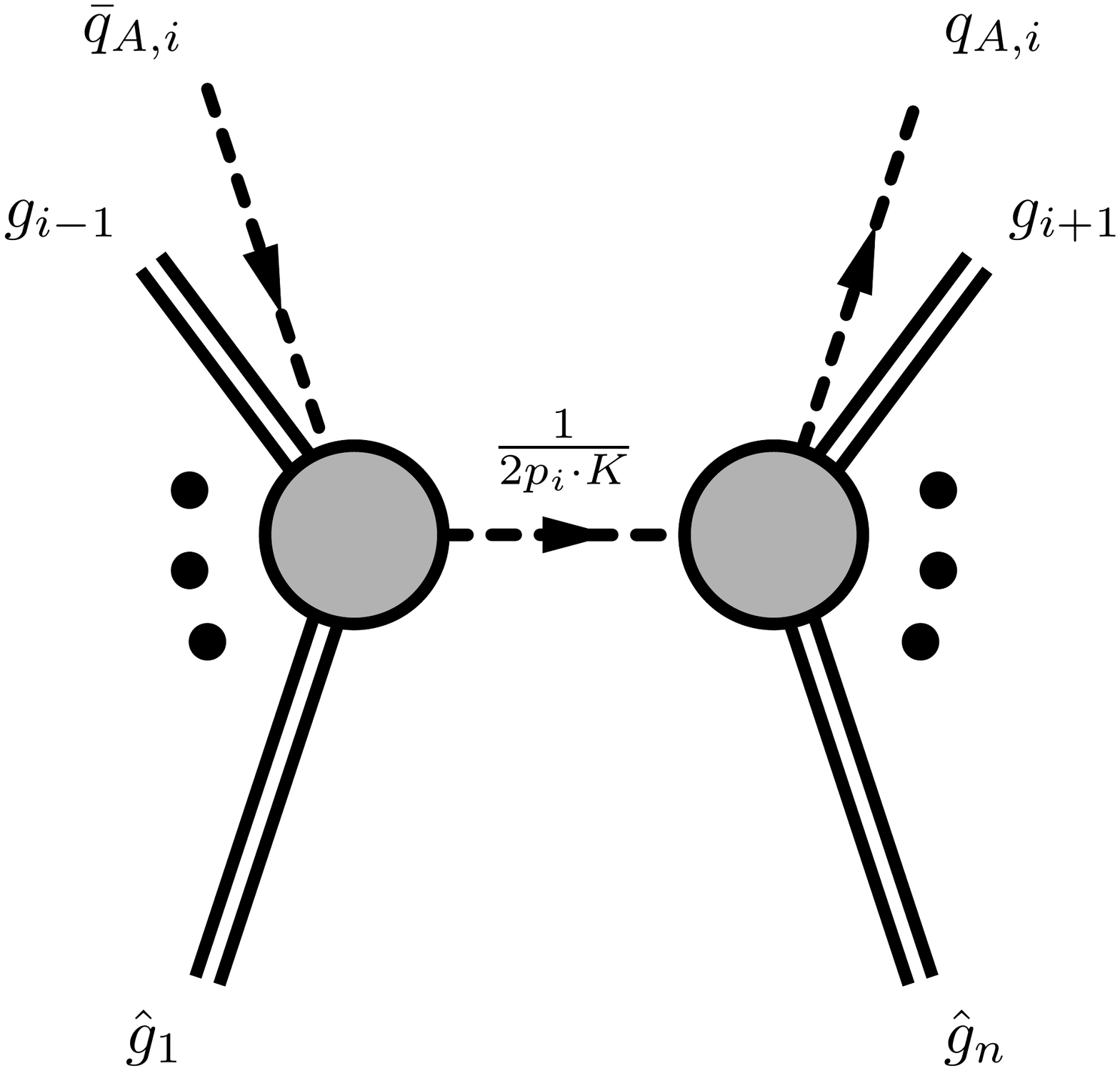}{27}{20}
\notag\\
\mathrm{C}^g &= \graph{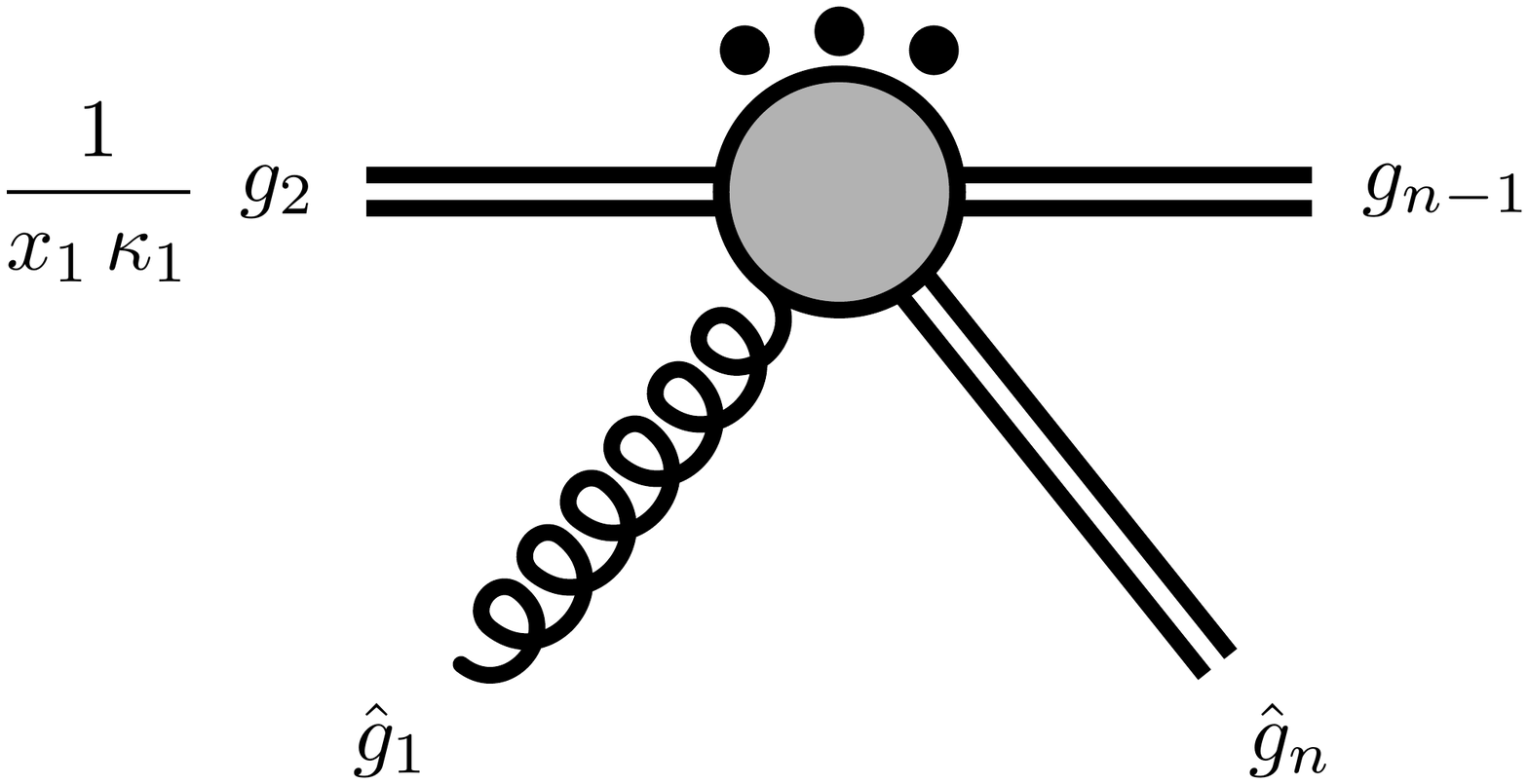}{28}{23}
\quad&
\mathrm{D}^g &= \graph{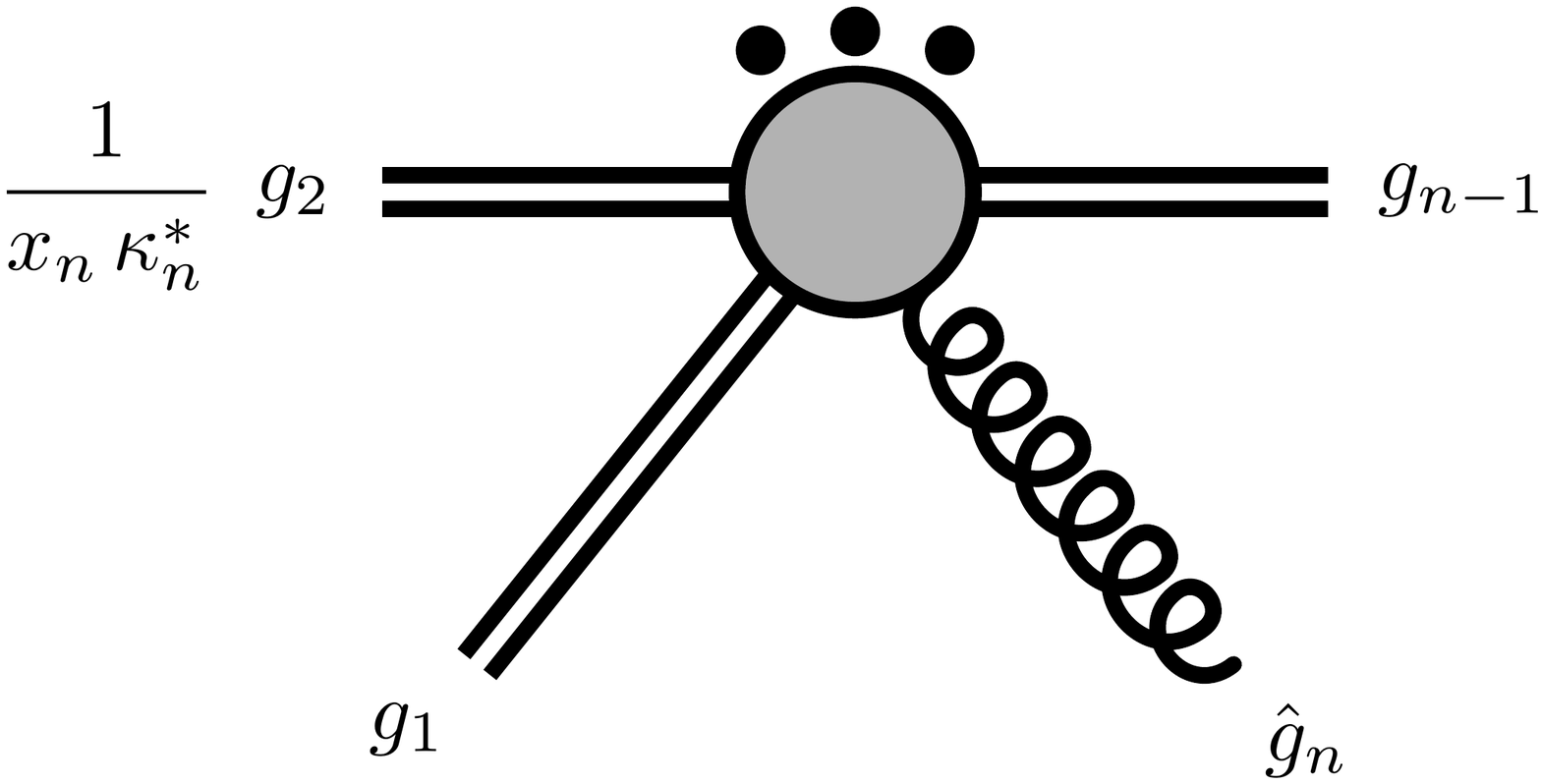}{28}{23}
\nn
\end{align}

\begin{itemize}
\item  $\mathrm{A}^g_{p,h}$ are due to the poles which appear in the original BCFW recursion for on-shell amplitudes. 
         The index $p$ runs over all possible cyclically ordered distributions of the particles into two ordered subsets with shifted particles
         being never on the same side of the propagator. 
         The pole appears because the shifted momentum squared of one of the intermediate virtual gluons, $K^2(z)$, 
         goes on-shell for $$ z = - \frac{K^2}{2 \, e \cdot K} \, . $$
\item $\mathrm{B}^g_i$ are due to the poles appearing in the propagator of auxiliary eikonal quarks, which account for the off-shell gluon.
         This means that $p_i \cdot \hat{K}(z) = 0$, where $\hat{K}$ is the momentum flowing through the eikonal propagator. 
         The value of $z$ for which this happens is $$  z = - \frac{2\, p_i \cdot K}{2\,p_i \cdot e}\, . $$
         If the $i$-th particle is on-shell, this term is not present.
\item $\mathrm{C}^g$ and $\mathrm{D}^g$ denote the same kind of residues: they appear respectively when the shifted $i$-th or $j$-th gluons are off-shell.
         They are due to the vanishing of the shifted momentum in the propagator of the off-shell particle: $k_i^2(z) = 0 \quad \text{or} \quad k_j^2(z)=0$.
         In the figure the case $(i,j)=(1,n)$ is depicted, explicitly showing the prefactors times on-shell amplitudes.
\end{itemize}
Table \ref{CD} summarizes the results for $\mathrm{C}^g$ and $\mathrm{D}^g$ terms which are needed to carry on computations and we recall
the derivation presented in \cite{vanHameren:2014iua} for the $\mathrm{C}^g$ residue with $e^\mu =\AL{i}\g^\mu\SR{j}/2$.
It appears due to the pole when the square $\hat{k}_i^2(z)$ of the external momentum vanishes, so that the gluon becomes on-shell.
The square $\hat{k}_i^2(z)$ vanishes because $\hat{\kapp}_i(z)$ vanishes, so we have
\begin{equation}
\hat{k}_i^\mu \equiv \hat{k}_i^\mu\big(z=\kapp_i/\SSS{ij} \big) = x_i(p_j)\, p_i^\mu - \frac{\kstr_i}{2}\,\frac{\AS{j|\gamma^\mu|i}}{\AA{ji}} \, .
\label{Eq:kTermC}
\end{equation}
All graphs contributing to C have the propagator with momentum $\hat{k}_i^\mu$, which is connected to the only eikonal vertex with $p_i^\mu$ each of these graphs contains.
This means that this vertex can be considered to be contracted with a current satisfying current conservation times
the intermediate propagator with unshifted momentum, as dictated by BCFW,
\begin{equation}
\mathrm{C} = \sqrt{2}\,p_i^\mu\,\frac{1}{\kapp_i\kstr_i}\,J_\mu \qquad\textrm{where}\qquad \hat{k}_i^\mu\,J_\mu = 0 \, ,
\end{equation}
where we have explicitly taken the propagator outside of $J_\mu$.
Using (\ref{Eq:kTermC}), we find
\begin{equation}
\mathrm{C} = \frac{1}{x_i(p_j)\kapp_i}\,\vep_i^\mu\,J_\mu
\quad,\quad
\vep_i^\mu = \frac{\AS{j|\gamma^\mu|i}}{\sqrt{2}\,\AA{ji}} \, .
\end{equation}
So C is given by $1/\left(x_i\,\kapp_i\right)$ times the amplitude for which gluon $i$ is on-shell with $h_i = +$.
Remember that its momentum is $\hat{k}_i^\mu$ and not $p_i^\mu$ and that $\lid{\hat{k}_i}{\vep_i}=0$.
Using the explicit expression of the longitudinal momentum fraction $x_i(p_j) = \lid{p_j}{k_i}/\lid{p_j}{p_i}$ 
and strategic choices for the auxiliary vectors for $\kapp_i,\kstr_i,\kstr_j$, we get the shifted quantities
listed in the corresponding column of Table \ref{CD}.

With respect to the figures in \cite{vanHameren:2014iua}, we have explicitly included the
factor $1/x_g$ in the figures describing the $C^g$ and $D^g$ poles, but the conventions are the same.
\begin{table}
$$
\begin{array}{|c|c|}\hline
&    e^\mu = \AL{i}\g^\mu \SR{j} / 2
\\ 
\hline
&  (h_i, h_j) = (+,* \vee +) 
\\
\mathrm{C}^g 
& \SR{\hat{k}_i} = \sqrt{x_i}\, \SR{i} \, , \,\,\, \AR{\hat{k}_i} = \frac{\slashk_i\SR{j}}{\sqrt{x_i}\SSS{ij}} 
\\ 
& \kstrhat_j =   \frac{\AL{j} \slashk_i + \slashk_j \SR{i}}{\SSS{ji}}  \,\,\, \vee \,\,\, \AR{\hat{j}} = \frac{\left( \slashk_i + \slashp_j \right)\SR{i}}{\SSS{ji}} 
\\
\hline
&  (h_i, h_j) = (*\vee -,-) 
\\
\mathrm{D}^g 
& \AR{\hat{k}_j} = \sqrt{x_j}\, \AR{j} \, , \,\,\, \SR{\hat{k}_j} = \frac{\slashk_j\AR{i}}{\sqrt{x_j}\AA{ji}}
\\ 
& \kapphat_i =  \frac{\AL{j} \slashk_j + \slashk_i \SR{i}}{\AA{ij}}  \,\,\, \vee \,\,\, \SR{\hat{i}} = \frac{\left( \slashk_j + \slashp_i \right)\AR{i}}{\AA{ij}}
\\
\hline
\end{array}
$$
\caption{
Shifted quantities needed for the evaluation of $\mathrm{C}^g$ and $\mathrm{D}^g$ residues.
An obviously similar table holds for $ e^\mu = \AL{j}\g^\mu \SR{i} / 2  $ }
\label{CD}
\end{table} 
%

\subsection{Fermion residues}

\begin{align}
\mathrm{A}^f_{p,h} &= \graph{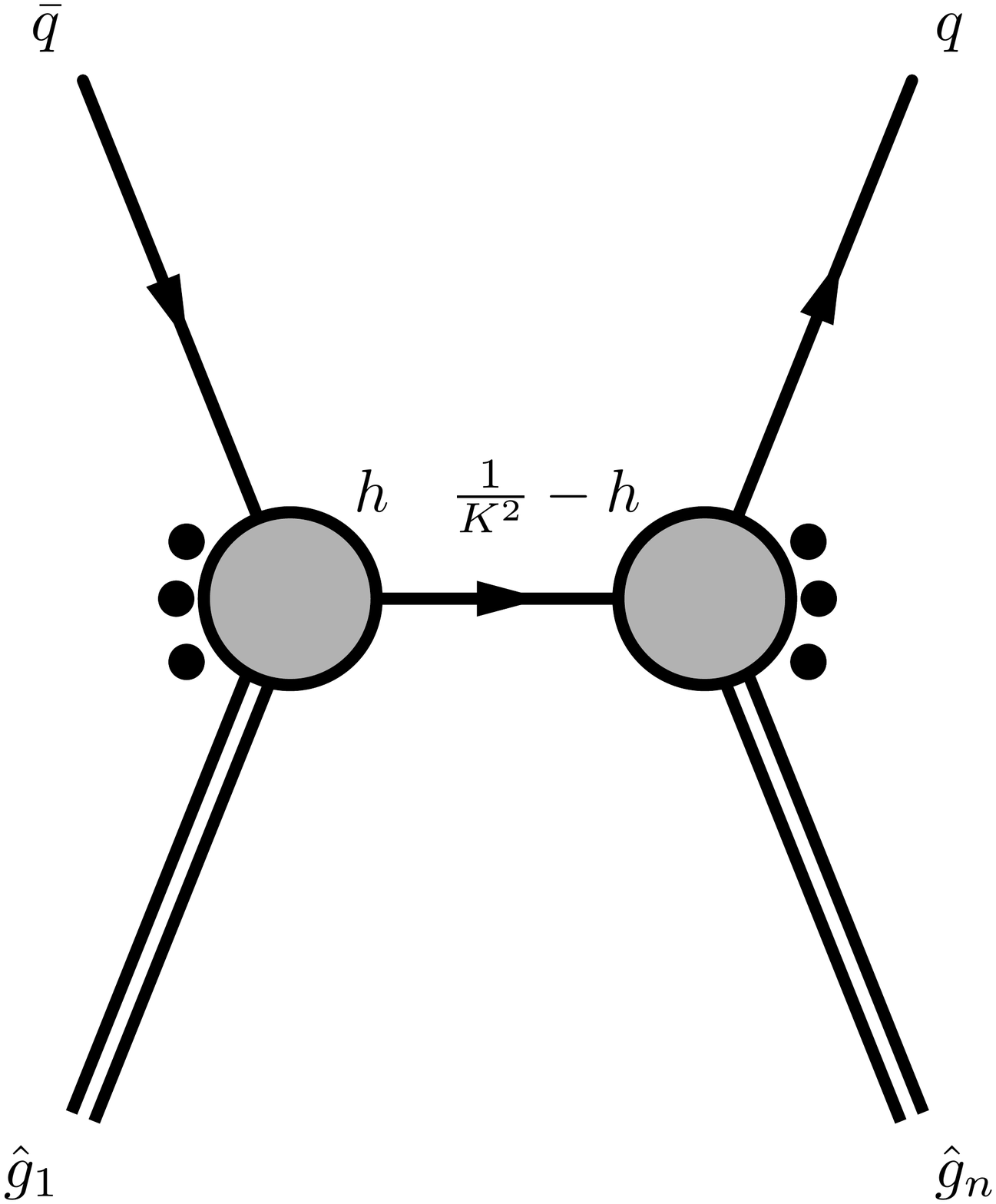}{25}{17}
\quad&
\mathrm{B}^f_{i} &= \graph{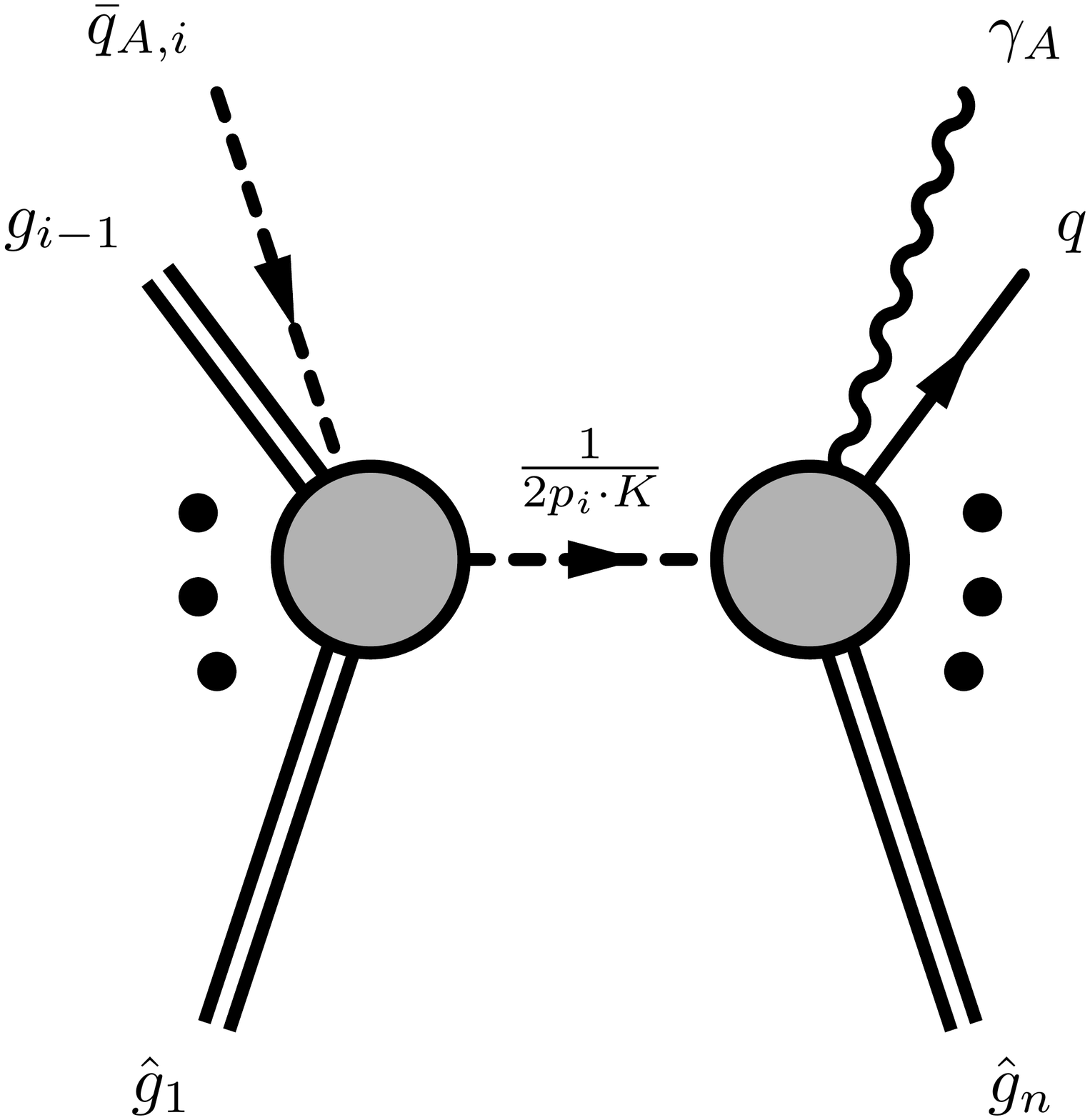}{29}{20}
\notag\\
\mathrm{C}^f &= \graph{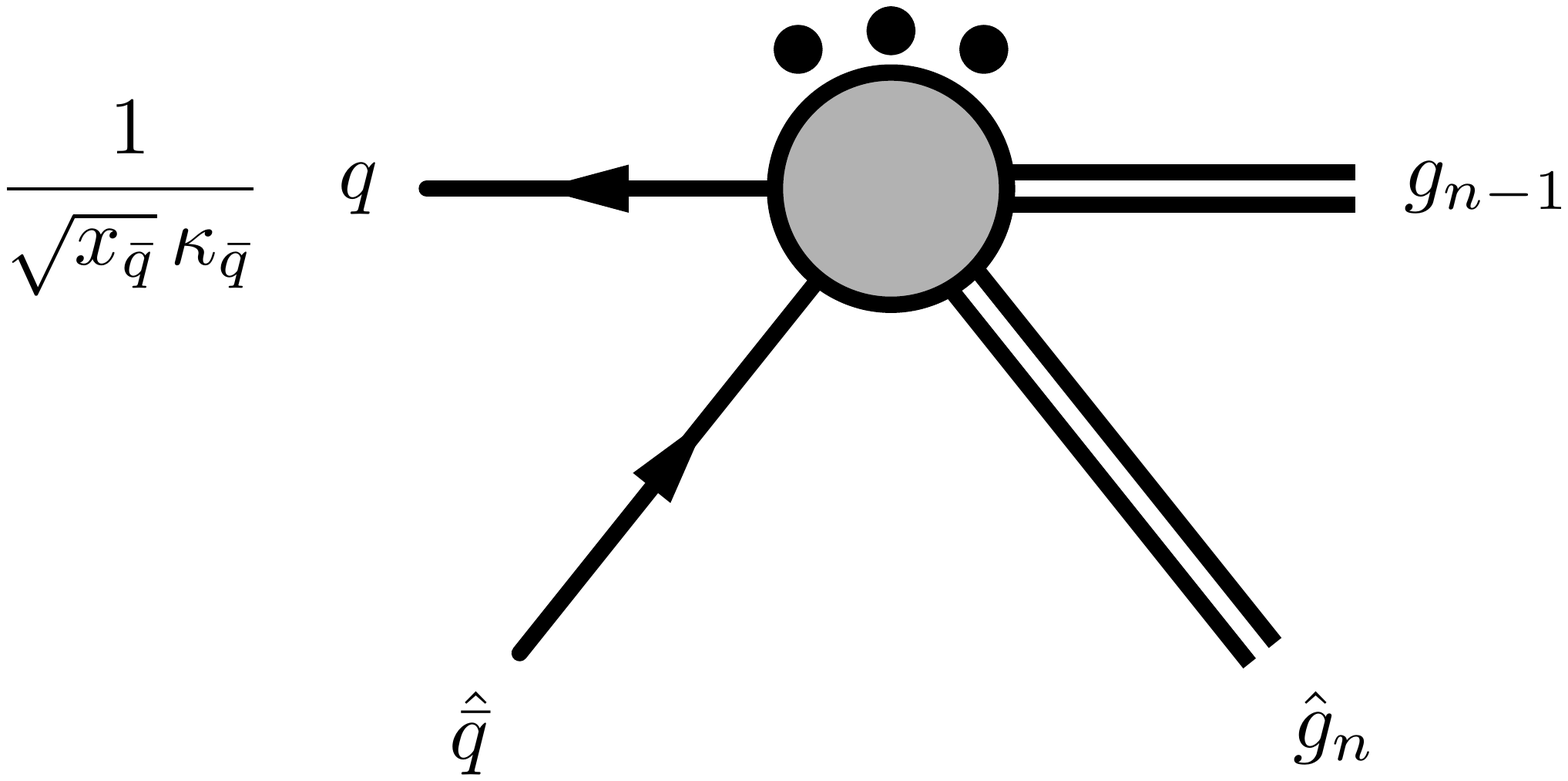}{33}{27}
\quad&
\mathrm{D}^f &= \graph{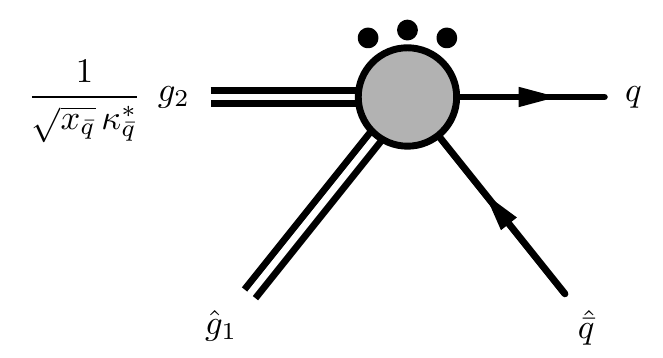}{37}{14}
\nn
\end{align}

\begin{itemize}
\item  The terms $\mathrm{A}^f_{p,h}$ and $\mathrm{B}^f_i$ are exactly the same as in the gluon case.
          The denominators of propagators behave in the same way and the numerator in the case of the $\mathrm{A}^f_{p,h}$ terms ensures 
          that there are opposite helicities at the ends of the propagator.
          In the $\mathrm{B}^f_i$ terms we have an eikonal (anti)quark and a zero-momentum photon needed 
          to account for the off-shell (anti)quark. If the $i$-th particle is an on-shell fermion, they are not present.
\item $\mathrm{C}^f$ and $\mathrm{D}^f$ denote exactly the same kind of residues as in the case of gluons. 
         We summarize the results for $\mathrm{C}^f$ and $\mathrm{D}^f$ terms in Table \ref{CDf} assuming we shift the antiquark.
         Shifting the quark works exactly the same way, with obvious changes of labels.
\end{itemize}
As mentioned in the previous section, as far as we are in the off-shell regime, 
we can use shift vectors which either stay legitimate or not in the on-shell limit.
But now this has implications on the $\mathrm{C}^f$ and $\mathrm{D}^f$ terms. 
In particular, these terms identically vanish if the shift vectors are chosen 
in such a way that in the on-shell limit they respect the original LW prescription.

We illustrate this explicitly. Suppose we want to compute $\Amp(\qb^*,q^-,g_1,\dots,g_n)$.
We choose $e^\mu = \AL{g} \g^\mu \SR{\qb}/2$ with $g$ being any gluon with $h_g = -1$.
Then the $\mathrm{D}^f$ residue will be given by
\begin{equation}
\AL{q} \dots \frac{\hat{\slashk}_{\qb}}{k^2_{\qb}} \times \sum_{h=+,-} \frac{\slasheps_{\gamma_A,h}}{\sqrt{2}} \SR{\qb} \, ,
\label{DtermAmp}
\end{equation}
where the dots stand for the rest of vertices and propagators, 
$\epsilon^\mu_{\gamma_A}$ is the polarization vector of the auxiliary photon and $\hat{\slashk}_{\qb}$ is the
on-shell shifted momentum of the antifermion, evaluated for 
$$
z = \frac{\kstr_{\qb}}{\AA{\qb g}} \, ,
$$
which is thus equal to
\begin{equation}
\hat{\slashk}_{\qb} = \AR{\hat{k}_{\qb}}\SL{\hat{k}_{\qb}} + \SR{\hat{k}_{\qb}}\AL{\hat{k}_{\qb}} \, ,
\label{slashed}
\end{equation}
where the hatted spinors can be found in Table \ref{CDf}, whereas the polarization vectors for the 
auxiliary photon are given by, following \cite{vanHameren:2013csa},
\begin{eqnarray}
\epsilon^\mu_{\gamma_A,+} 
&=&
\frac{\AL{a_{\qb}} \g^\mu \SR{\qb} }{\sqrt{2}\, \AA{a_{\qb} \qb}} \, ,
\nn \\ 
\epsilon^\mu_{\gamma_A,-} 
&=&
\frac{ \AL{\qb} \g^\mu \SR{a_{\qb}} }{\sqrt{2}\,\SSS{\qb a_{\qb}}} \, ,
\label{AuxPol}
\end{eqnarray}
with $a_{\qb}$ the auxiliary vector, the dependence from which must eventually disappear.
Inserting (\ref{slashed}) and (\ref{AuxPol}) into (\ref{DtermAmp}), we get
\begin{equation}
\frac{1}{\kappa_{\qb}\, \kstr_{\qb}} \, \AL{q} \dots \SR{\hat{k}_{\qb} } \AA{\hat{k}_{\qb} \qb} \frac{\SSS{ a_{\qb} \qb} }{\SSS{\qb a_{\qb}}}  = 
\frac{1}{\kappa_{\qb}\, \kstr_{\qb}} \, \AL{q} \dots \SR{\hat{k}_{\qb} } \AA{ \qb \hat{k}_{\qb}} 
= 0
\end{equation}
because of the relation in Table \ref{CDf}: $ \AR{\hat{k}_{\qb}} = \sqrt{x_{\qb}} \, \AR{\qb} $.

Exactly the same kind of analysis gives, for the $\textrm{C}^f$ term with shift vector $e^\mu = \frac{1}{2} \AL{\qb} \g^\mu \SR{g}$,
$h_g=+1$ and $h_{\qb} = +1$ in the on-shell limit,
\begin{eqnarray}
\AL{q} \dots \sum_{h=+,-} \frac{ \slashk_{\qb} }{k^2_{\qb}} \times \frac{\slasheps_{\gamma_A,h}}{\sqrt{2}} \SR{\qb} 
&=& 
- \frac{1}{\kappa_{\qb} \,\kstr_{\qb}}\, \AL{q}  \dots \SR{\hat{k}_{\qb}} \AA{ \hat{k}_{\qb} \, \qb} \nn \\
&=&
\frac{\AL{q} \dots \SR{\hat{k}_{\qb}} }{\kappa_{\qb} \,\kstr_{\qb}} \, \frac{ \AL{\qb} \slashk_{\qb} \SR{g} }{\sqrt{x_{\qb}} \, \SSS{\qb g} } =
\frac{1}{ \sqrt{x_{\qb}}\, \kappa_{\qb} } \, \AL{q} \dots \SR{\hat{k}_{\qb}}
\end{eqnarray}
where we have exploited strategically the relations in Table \ref{CDf} and we can recognize on the r.h.s.
the on-shell amplitude evaluated for shifted momenta of the antifermion and the gluon.
The analysis can be extended in the same way to the two remaining cases.
\begin{table}
$$
\begin{array}{|c|c|c|}\hline
&   & \text{Admitted $(!)$ or not ($ \times $) on-shell }
\\ 
\hline
\mathrm{C}^f \Leftrightarrow e^\mu = \AL{\qb}\g^\mu \SR{g}/2
& \SR{\hat{k}_{\qb}} = \sqrt{x_{\qb}}\, \SR{\qb} \, , \,\,\, \AR{\hat{k}_{\qb}} = \frac{\slashk_{\qb} \SR{g}}{\sqrt{x_{\qb}} \SSS{\qb g}}  
&  (h_{\qb}, h_g) = (-,* \vee + ) \quad !
\\ 
& \kstrhat_g =   \frac{\AL{g} \slashk_{\qb} + \slashk_g \SR{\qb}}{\SSS{g \qb}}  \,\,\, \vee \,\,\, \AR{\hat{g}} = \frac{\left( \slashk_{\qb} + \slashp_g \right)\SR{{\qb}}}{\SSS{g \qb}} 
& (h_{\qb}, h_g) = (+,* \vee +) \quad \times
\\
\hline
\mathrm{D}^f \Leftrightarrow e^\mu = \AL{g}\g^\mu \SR{\qb}/2
& \AR{\hat{k}_{\qb}} = \sqrt{x_{\qb}}\, \AR{\qb} \, , \,\,\, \SR{\hat{k}_{\qb}} = \frac{\slashk_{\qb} \AR{g}}{\sqrt{x_{\bar{q}}}\AA{\qb g}}
& (h_g, h_{\qb}) = (* \vee -,+) \quad !
\\ 
& \kapphat_g =  \frac{\AL{\qb} \slashk_{\qb} + \slashk_g \SR{g}}{\AA{g \qb}}  \,\,\, \vee \,\,\, \SR{\hat{g}} = \frac{\left( \slashk_{\qb} + \slashp_g \right)\AR{g}}{\AA{g \qb}}
&  (h_g, h_{\qb}) = (* \vee -,-) \quad \times
\\
\hline
\end{array}
$$
\caption{
Shifted quantities needed for the evaluation of $\mathrm{C}^f$ and $\mathrm{D}^f$ residues in the case in which the antiquark is shifted.
A completely similar table holds for the case in which $q$ is shifted $( \bar{q} \rightarrow q)$.
}
\label{CDf}
\end{table} 
%

\section{MHV amplitudes with one off-shell parton}\label{MHVsec}

When all momenta are taken either as outgoing or incoming, 
MHV amplitudes are defined as amplitudes with all but two particles having the same helicity sign.
The structure of MHV on-shell gluon amplitudes was first guessed by Parke and Taylor in \cite{Parke:1986gb}.
Once one knows the basic 3-point amplitudes, here reported in Appendix \ref{App3Point},
it is easy to prove by induction and the BCFW recursion relation that all the MHV amplitudes have
a very simple structure.
For instance, in the mostly-plus case
\bea
\Amp(g_1^+,\dots,g_i^-,\dots,g_j^-,\dots,g_n^+) &=& \frac{\AA{ij}^4}{\AA{12}\AA{23}\dots \AA{n1}} \, , \nn \\
\Amp(\qb^+,\dots, g_i^-,\dots,g_n^+,q^-) &=& \frac{\AA{i q}^3 \AA{i \qb}}{\AA{\qb 1}\AA{12}\dots\AA{nq}\AA{q\qb}} \, , \nn \\
\Amp(\qb^-,g_1^+,g_2^+\dots,\dots,g_n^+,q^-) &=& 0 \, .
\label{MHVs}
\eea
The mostly-minus MHV amplitudes  are just obtained by $\AA{ab} \leftrightarrow \SSS{ba}$ in the results above.
In this section we show how one can derive general formulas for MHV amplitudes with one off-shell parton.
We will illustrate in detail a few cases and obtain that their structures are natural and intuitive generalizations of (\ref{MHVs}), 
including the case in which all gluons have the same helicity, which does not vanish if one of the fermions is off-shell and the other has opposite 
helicity sign w.r.t. the gluons.
We dub them {\em the subleading amplitudes} because, compared to all other amplitudes, they carry an extra factor whose
absolute value is $\propto \sqrt{|k^2|}$~\cite{vanHameren:2013csa}. \\
In order to clarify this distinction, let us point out that
\begin{center}
\emph{we call "MHV" the non-vanishing amplitudes which at the same time feature the maximum difference between 
the numbers of positive and negative helicity particles and do not vanish in the on-shell case}.
\end{center}
For instance, in the purely Yang-Mills case, $\Amp(g_1^*,g_2^-,g_3^+,\dots,g_n^+)$ is an MHV amplitude~\cite{vanHameren:2014iua}.
Now $\Amp(\bar{q}^\pm,q^-,g_1^+,\dots,g_n^+) = 0$, so that $\Amp(\bar{q}^*,q^-,g_1^+,\dots,g_n^+)$ \emph{is not MHV},
according to our nomenclature, despite the fact that the difference between positive and negative helicity gluons
is higher than, for example, $\Amp(\bar{q}^*,q^-,g_1^-,g_2^+,\dots,g_n^+)$. 
This distinction might sound peculiar, but it is specifically meant to make our classification a more intuitive extension of the on-shell case.
Let us explain this point better: it was guessed in \cite{Parke:1986gb} for the first time and proved later that on-shell MHV
amplitudes are given by (\ref{MHVs}). Ever since then, it is customary for physicists working on scattering amplitudes 
to take this structure as {\em the very definition of MHV amplitudes}.
In the case with off-shell fermions this is no longer true, as we are going to see. 
In order to make the correspondence between the two cases smoother, we define MHV amplitudes in such a way that
they feature a Parke-Taylor structure in the off-shell case as well (modulo a factor).

Apart from the subleading case, all the MHV amplitudes with off-shell particles reduce to MHV amplitudes
in the on-shell case. The converse is not true, because there are 5-point amplitudes which do not feature the maximum possible difference between the numbers of positive and
negative helicity particles and which still reduce to MHV on-shell amplitudes; examples are explicitly worked out in the next section.
Let us mention that this provides a cross-check for our calculations, because 
when the contributions which are typical of the off-shell regime are dropped, the structure of the amplitude 
must reduce to the familiar Parke-Taylor formula, which always holds for the fully on-shell MHV case.
\begin{figure}[h]
\begin{center}
\includegraphics[scale=0.7]{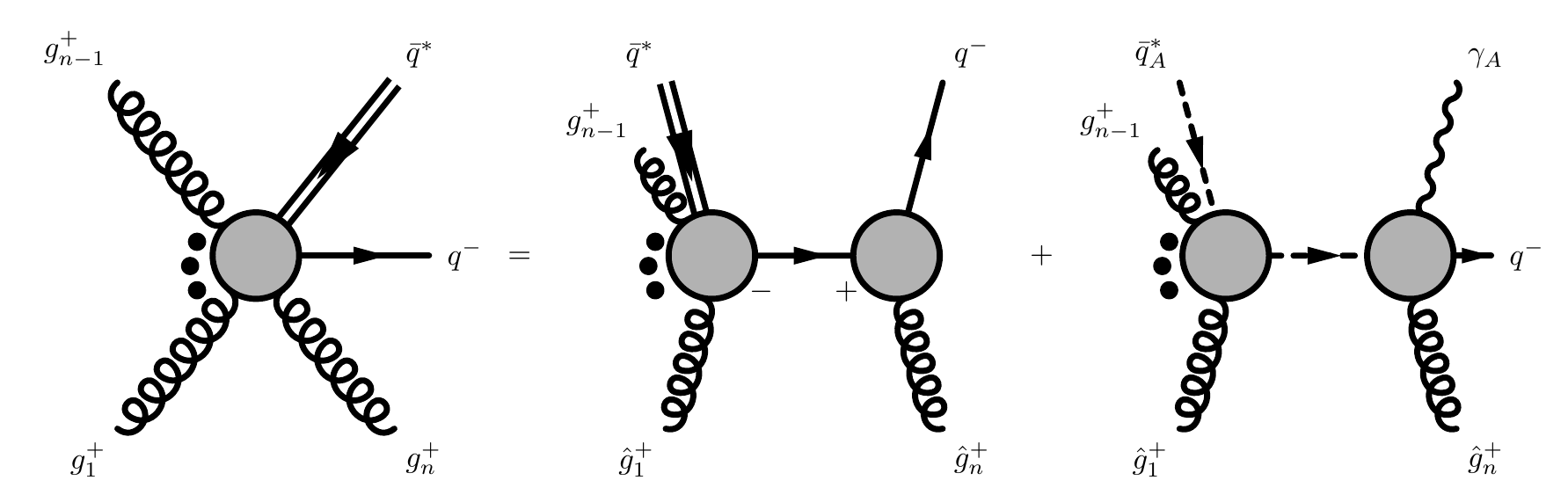}
\caption{The recursion for subleading amplitudes with an off-shell antifermion.}
\label{sublead_fig}
\end{center}
\end{figure}

Now we illustrate the first proof of this section, deriving the structure of the subleading contributions.
We assume
\beq
\Amp(g_1^+,g_2^+,\dots,g_{n-1}^+,\bar{q}^*,q^-) = \frac{ \AA{\qb q}^3}{\AA{12}\AA{23}\dots\AA{n-1|\qb}\AA{\qb q}\AA{q 1}} \, ,
\eeq
which is known to be true for $4$ particles~\cite{vanHameren:2013csa}. Notice that, with respect to the MHV amplitudes to be 
derived below, this structure misses a factor $\propto 1/\kappa \sim 1/\sqrt{|k_T^2|}$, causing a subleading behaviour. 
We choose the shift vector $e^\mu = \frac{1}{2} \AL{1}\g^\mu\SR{n} $.
It can be seen from fig. \ref{sublead_fig} that the second contribution involves, on the left edge
of the eikonal quark propagator, an amplitude with one off-shell (2 eikonal fermions) and $n-1$ on-shell gluons,
which vanishes, so that the only contribution is
\bea
\Amp(g_1^+,g_2^+,\dots,g_{n-1}^+,\qb^*,q^-,g_n^+) 
&=&
\frac{ \AA{\qb\hat{K}}^3 }{\AA{\hat{1}2}\AA{23}\dots\AA{\qb\hat{K}}\AA{\hat{K} \hat{1}} }\, \frac{1}{\AA{qn} \SSS{nq}} \, \frac{\SSS{\hat{n}\hat{K}}^2}{\SSS{\hat{K}q}} \, , 
\label{pre_amp_sublead}
\eea
where the notations were explained in sections \ref{residuesec} and \ref{conditions} and the $3$-point function is given in appendix \ref{App3Point}. 
The value of $z$ for which the complex amplitude exhibits a pole is fund by requiring that
\beq
(p_q + \hat{p}_n )^2 = 0 \Leftrightarrow \AA{nq}\SSS{qn} - z\, \AL{1}\slashp_q\SR{n} = \AA{nq}\SSS{qn} - z\,\AA{1q}\SSS{qn} = 0 \Leftrightarrow z = \frac{\AA{nq}}{\AA{1q}} \, . 
\eeq
Taking (\ref{shift1}) into account, the only shifted spinor is
\beq
\AR{\hat{n}} = \AR{n} - \frac{\AA{nq}}{\AA{1q}} \AR{1} = \frac{\AA{1n}}{\AA{1q}} \AR{q} \, ,
\label{shift_sublead}
\eeq
with the final form being obtained after using the Schouten identity (see appendix \ref{AppSchouten}).
This implies, for the intermediate propagator momentum,
\beq
\slashK \equiv \AR{\hat{K}}\SL{\hat{K}} + (\AR{} \leftrightarrow \SR{} ) = \AR{q}\, \left( \SL{q} + \frac{\AA{1n}}{\AA{1q}} \SL{n} \right) + (\AR{} \leftrightarrow \SR{} ) \, .
\label{K_sublead}
\eeq
In the following, when dealing with slashed momenta we will omit the specification $+ (\AR{} \leftrightarrow \SR)$.
Inserting (\ref{shift_sublead}) and (\ref{K_sublead}) into (\ref{pre_amp_sublead}) and working out the necessary simplifications
directly brings us to our result
\bea
\Amp(g_1^+,g_2^+,\dots,g_{n-1}^+,\qb^*,q^-,g_n^+) 
&=&
\frac{\AA{\qb q}^3}{\AA{12}\AA{23}\dots\AA{\qb q}\AA{q n}\AA{n1}} \, .
\eea
This completes the proof for the case the antiquark is off-shell, but the other case is completely analogous, of course.
\begin{figure}[h]
\begin{center}
\includegraphics[scale=0.7]{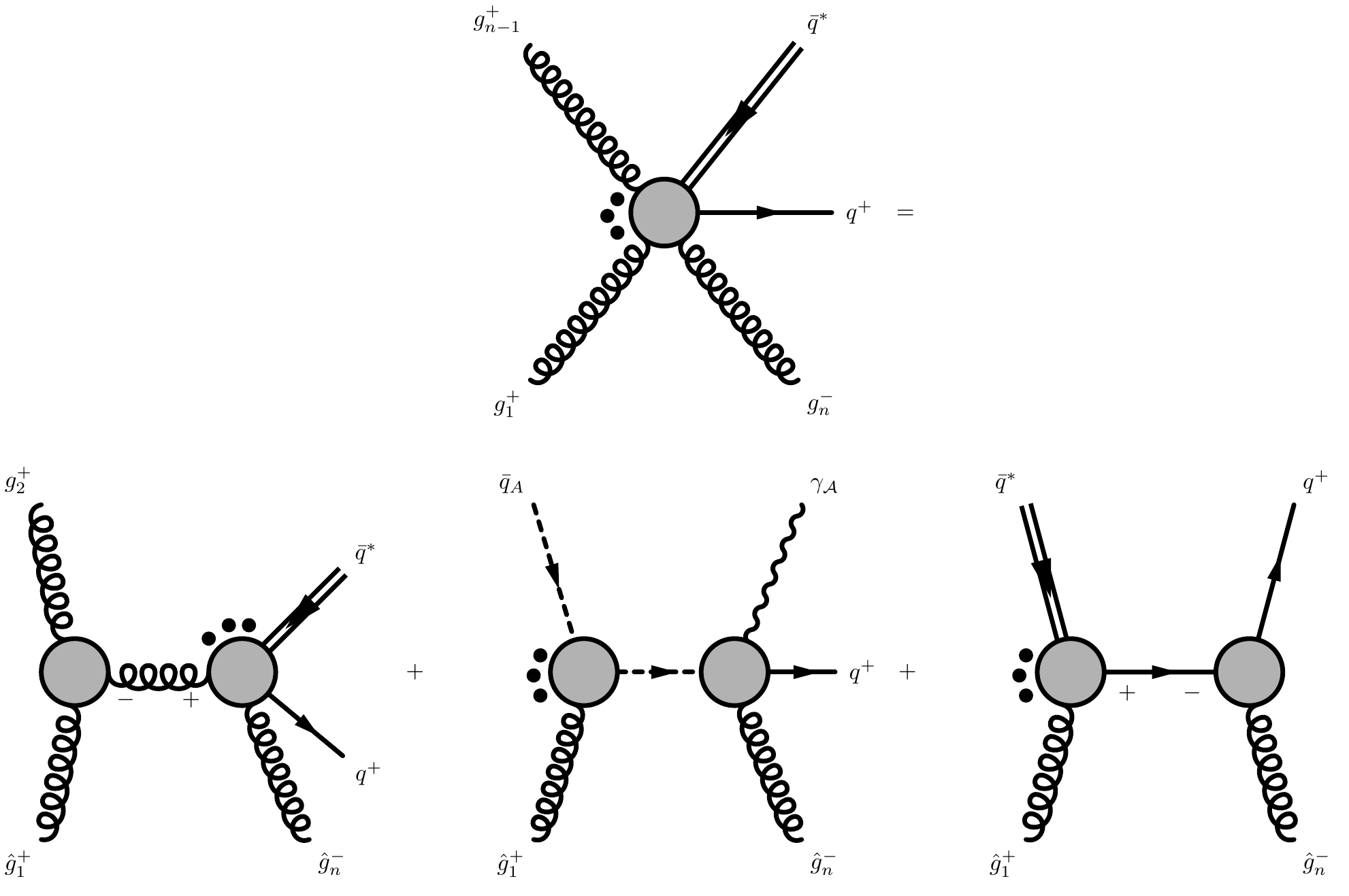}
\caption{The recursion for MHV amplitudes with an off-shell antifermion, shifting two gluons.}
\label{MHV_offs_qb_fig}
\end{center}
\end{figure}

We move on to the MHV amplitudes, starting with those with an off-shell antifermion, specifically $\Amp(g_1^+,g_2^+,\dots,g_{n-1}^+,\qb^*,q^+,g_n^-)$.
Picking up $(g_n,g_1)$ as reference particles and $e^\mu = \frac{1}{2}\AL{n}\g^\mu\SR{1}$ as shift vector, 
the BCFW recursion is illustrated by fig. \ref{MHV_offs_qb_fig}. 
Here the third contribution vanishes because the sub-amplitude on the left is less than subleading (the on-shell fermion has the same helicity as all the gluons).
Also the second is zero because all the on-shell gluons have positive helicity. 
The first contribution, instead, is given by
\bea
\Amp(g_1^+,g_2^+,\dots,g_{n-1}^+,\qb^*,q^+,g_n^-)
&=&
\frac{\SSS{ 2 \hat{1}}^3}{\SSS{\hat{1} \hat{K}} \SSS{\hat{K}\hat{2}}}\,\frac{1}{\AA{12}\SSS{21}} \,  
\frac{1}{\kstr_{\qb}} \, \frac{\AA{\qb \hat{n}}^3 \AA{q n}}{\AA{\hat{K} 3}\AA{34}\dots\AA{\qb q}\AA{q \hat{n}}\AA{\hat{n}\hat{K}}} 
\nn \\
&=&
\frac{1}{\kstr_{\qb}} \,
\frac{\SSS{ 2 \hat{1}}^3 \AA{\qb \hat{n}}^3 \AA{q n} }{\AA{34}\dots\AA{\qb q}\AA{q \hat{n}}\AA{12}\SSS{21} \AL{\hat{3}} \hat{\slashK} \SR{\hat{1}}\AL{\hat{n}}\hat{\slashK}\SR{\hat{2}}}  \,  
 \label{preamp_MHV_offs_g}
\eea
where the form of the sub-amplitude on the right on the first line is just the inductive hypothesis 
motivated by the $3$-point function in (\ref{3ptFermions}).
The location of the pole and the shifted quantities are found to be
\bea
z &=& - \frac{\AA{12}}{\AA{n2}} \, , 
\nn \\
\AR{\hat{1}} &=& \frac{\AA{n1}}{\AA{n2}} \AR{2} \, , 
\nn \\
\hat{\slashK} &=& - \AR{2} \left( \frac{\AA{n1}}{\AA{n2}} \SL{1} + \SL{2}  \right) \, .
\eea
Replacement into (\ref{preamp_MHV_offs_g}) directly leads to
\beq
\Amp(g_1^+,g_2^+,\dots,g_{n-1}^+,\qb^*,q^+,g_n^-) = 
\frac{1}{\kstr_{\qb}}\, \frac{ \AA{\qb n}^3 \AA{q n} }{ \AA{12}\AA{23}\dots\AA{\qb q}\AA{qn}\AA{n1} } \, .
\eeq
One can easily see that the proof works exactly in the same way if the fermion pair is in a different
position w.r.t. the negative helicity gluon, as the contributions of the eikonal quark pole and the fermion pole would be zero
anyway; a slight change is implied only in the structure of the non-vanishing contribution; 
another way to realise it is to pick up a different positive-helicity gluon as reference particle. 
\begin{figure}[h]
\begin{center}
\includegraphics[scale=0.7]{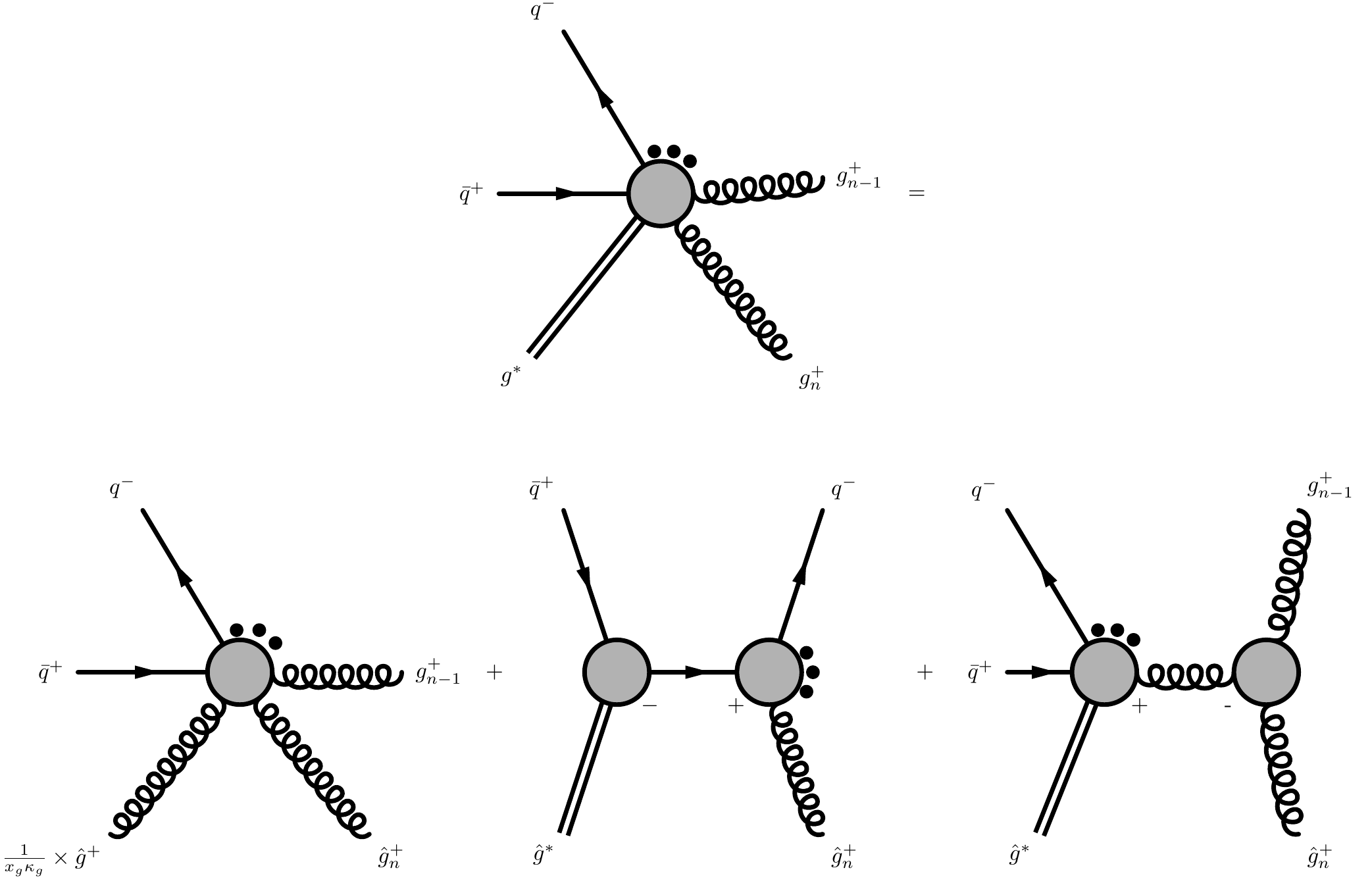}
\caption{The recursion for MHV amplitudes with an off-shell gluon, shifting the off-shell gluon itself and another gluon with positive helicity}
\label{MHV_offs_g_fig}
\end{center}
\end{figure}

The last MHV amplitude we have to investigate is the one where the off-shell particle is a gluon.
Here the situation is slightly more involved, because the amplitude depends, in principle, on the relative 
position of the fermion pair and the off-shell gluon.
We will start by showing how to derive the formula for $\Amp(g^*,\qb^+,q^-,g_1^+,g_2^+,\dots,g_n^+)$.
Let us shift $g$ and $n$, so $e^\mu = \frac{1}{2}\, \AL{g} \g^\mu \SR{n}$.
Fig. \ref{MHV_offs_g_fig} shows the three contributions which are involved:
the first is a $C^g$-residue, specifically the on-shell limit of a subleading amplitude, 
while the second contribution has one such sub-amplitudes on the right of the propagator. 
As these amplitudes vanish, we conclude that only the third contribution is non-zero and it is given by
\bea
&&
\Amp(g^*,\qb^+,q^-,g_1^+,g_2^+,\dots,g_n^+)  = 
\nn \\
&&
\frac{1}{\kstr_g}\, \frac{\AA{gq }^3 \AA{g\qb}}{\AA{g\qb}\AA{\qb q}\dots \AA{n-2|\hat{K}}\AA{\hat{K}g} } \,
\frac{1}{\AA{n-1| n} \SSS{n| n-1}} \, \frac{\SSS{n-1| n}^2}{\SSS{\hat{K}| n-1} \SSS{n| \hat{K}}} \, .
\label{preamp_offs_g}
\eea
The pole location and the shifted vectors are found to be
\bea
z &=& \frac{\AA{n-1|n}}{\AA{n-1| g}} \, , 
\nn \\
\AR{\hat{n}} &=& \AR{n} + \frac{\AA{n|n-1}}{\AA{n-1| g}}  \, \AR{g} = \frac{ \AA{n| g} }{\AA{n-1| g}} \AR{n-1} \, ,
\nn \\
\hat{\slashK} &=& \AR{n-1} \left( \SL{n-1} + \frac{\AA{n | g }}{\AA{n-1| g}}\, \SL{n} \right) 	\, ,
\eea
which give, after being used in (\ref{preamp_offs_g}), 
\beq
\Amp(g^*,\qb^+,q^-,g_1^+,g_2^+,\dots,g_n^+) = \frac{1}{\kstr_g}\, \frac{ \AA{g q}^3 \AA{g\qb} }{\AA{g \qb} \AA{\qb q} \dots \AA{n-1 |n} \AA{ n g} } \, .
\eeq
We comment on the other cases: it is very easily understood that the first two kinds of residues shown in fig. \ref{MHV_offs_g_fig} 
will always vanish as long as the fermion pair is on the left of $g_{n-1}$, because they will always incluse subleading amplitudes in the on-shell limit,
so that contributions with a gluon pole will be the only non-vanishing ones.
In the remaining case, instead, the first and third contribution are zero, whereas the fermion pole gives the expected result,
\bea
\Amp(g^*,\qb^+,q^-,g_1^+,g_2^+,\dots,g_n^+) 
&=&
\Amp(\hat{g}^*,g_1^+,\dots,g_{n-1}^+,\qb^+,\hat{K}^-)\,  \frac{1}{\AA{qn}\SSS{nq}}\, \Amp(\hat{K}^+,q^-,\hat{g}_n^+)  \
\nn \\
&=& 
\frac{1}{\kstr_g}\, \frac{ \AA{g q}^3 \AA{g\qb} }{\AA{g \qb} \AA{\qb q} \dots \AA{n-1 |n} \AA{ n g} } \, ,
\eea
after a procedure which is completely analogous to the cases worked out so far.
This completes the third proof to be presented in this section.

\section{The 5-point non-MHV amplitudes with one off-shell parton}\label{five}

In the on-shell case all amplitudes up to 5-point, both with and without fermions, exhibit the simple Parke-Taylor structure 
presented at the beginning of the previous section~\cite{Mangano:1990by}.
This does not stay true for Yang-Mills amplitudes with at least two off-shell gluons~\cite{vanHameren:2014iua}.
When we include fermions, we will see that this situation changes already if one single particle is off-shell.
In fact, all the amplitudes that we are going to study in this section are MHV
if computed in the on-shell case, i.e. for vanishing transverse component of the momentum of the off-shell particle.
All amplitudes presented in this section have been cross-checked numerically with the help of AVHLIB~\cite{Bury:2015dla},
which employs Dyson-Schwinger recursion to evaluate the amplitudes.

\subsection{Off-shell antifermion or fermion}%

Taking into account cyclic and parity invariance of color-ordered amplitudes, 
the independent contributions to be computed in the 5-point case are
\begin{eqnarray}
\textrm{subleading} \quad
&&
\left\{ \begin{array}{c}
\Amp(g_1^+,g_2^+,\bar{q}^*,q^-,g_3^+)
\end{array}\right.
\nn \\
\textrm{MHV}\quad 
&&
\left\{ \begin{array}{c}
\Amp(g_1^+,g_2^-,\bar{q}^*,q^-,g_3^-)
\\
\Amp(g_1^-,g_2^+,\bar{q}^*,q^-,g_3^-)
\\
\Amp(g_1^-,g_2^-,\bar{q}^*,q^-,g_3^+)
\end{array} \right.
\label{fivelist}
\\
\textrm{non-MHV}\quad 
&&
\left\{ \begin{array}{c}
\Amp(g_1^+,g_2^+,\bar{q}^*,q^-,g_3^-)
\\
\Amp(g_1^+,g_2^-,\bar{q}^*,q^-,g_3^+)
\\
\Amp(g_1^-,g_2^+,\bar{q}^*,q^-,g_3^+)
\end{array} \right. \, .
\nn
\end{eqnarray}
As usual, once we obtain these results, getting the parity-conjugated amplitudes 
will be just a matter of $\AA{ab}\leftrightarrow\SSS{ba}$ and $\kappa \leftrightarrow\kstr $.
To have the corresponding process with off-shell fermion and on-shell antifermion, instead, just make the exchange $\bar{q}\leftrightarrow q$.
Having worked out the MHV and subleading amplitudes in general in section \ref{MHVsec},
here we focus on the three remaining contributions.

We will quite in detail illustrate the calculation of the first non-MHV amplitude, $\Amp(g_1^+,g_2^+,\bar{q}^*,q^-,g_3^-)$,
obtaining three different representations: first by shifting two gluons; secondly by shifting the off-shell fermion and one gluon 
in a way which stays valid in the on-shell limit; finally, in order to illustrate the point made in section \ref{fancy_shift}, 
we will get the result also in a third way, which would not be correct in the on-shell kinematics.
For the other two amplitudes we just give the results, which can be worked out similarly.

\begin{figure}[h]
\begin{center}
\includegraphics[scale=0.7]{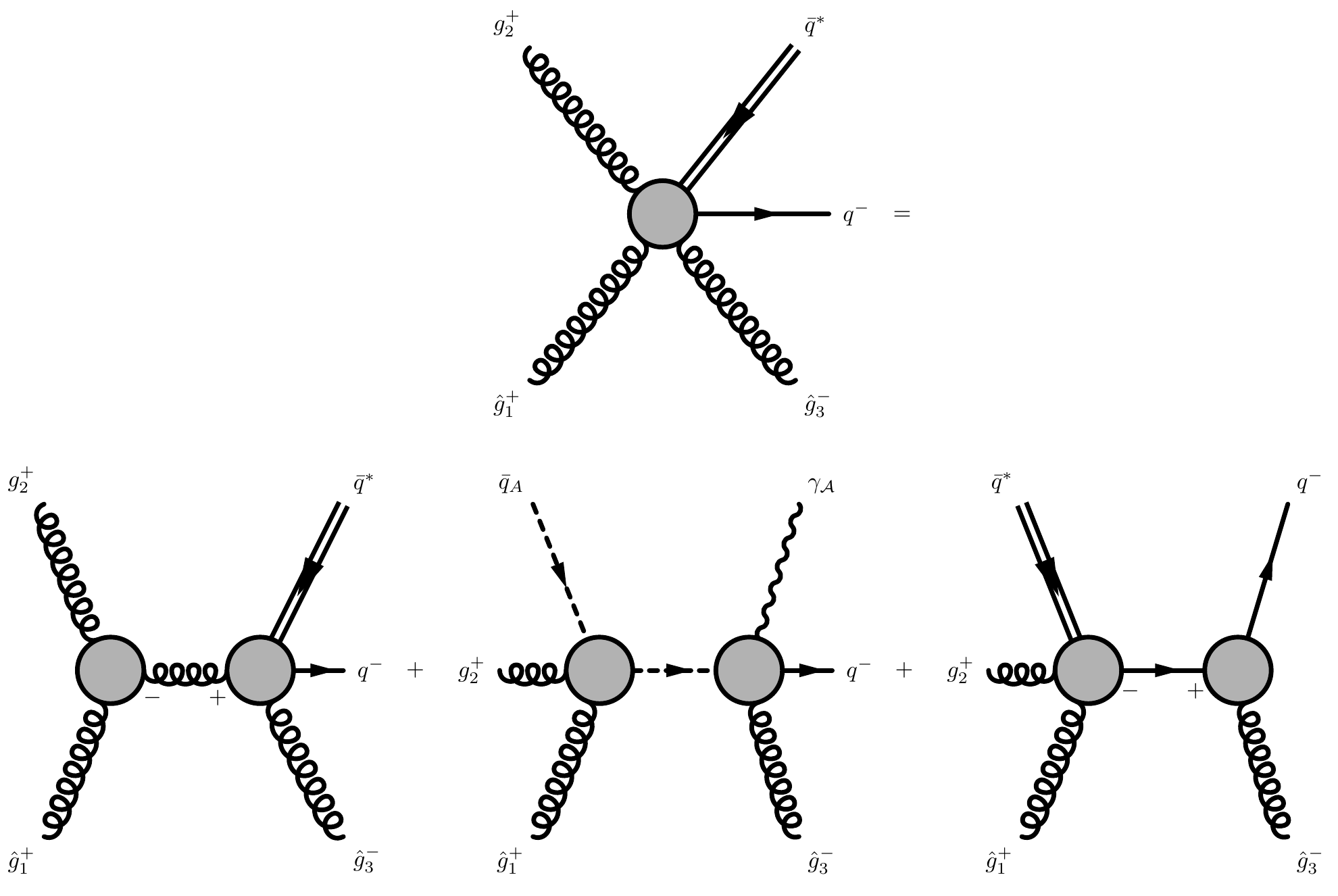}
\caption{The recursion for $\Amp(g_1^+,g_2^+,\bar{q}^*,q^-,g_3^-)$ with $e^\mu = \frac{1}{2}\AL{3}\g^\mu\SR{1}$.}
\label{5_point_recursion_1_fig}
\end{center}
\end{figure}
The first way to perform the recursion is illustrated in fig. \ref{5_point_recursion_1_fig}, where the second contribution is zero,
because the sub-amplitude on the left vanishes.
The first contribution is instead
\beq
\frac{\SSS{21}^3}{\SSS{1\hat{K}}\SSS{\hat{K} 2}}\, \frac{1}{\AA{12}\SSS{21}}\,
\frac{1}{\kappa_{\qb}}\,\frac{\SSS{\qb\hat{K}}^2\SSS{q\hat{K}}}{\SSS{\hat{K}\hat{3}}\SSS{\hat{3}q}\SSS{q\qb}} \, .
\label{pre_amp_qb_1}
\eeq
Requiring the shifted intermediate momentum to vanish, we get the location of the pole and the corresponding shifted variables
\bea
&&
(\hat{p}_1+p_2)^2 = 0 \Leftrightarrow \AA{12}\SSS{21} + z\, \AL{3} \slashp_2 \SR{1} = 0 \Leftrightarrow z= \frac{\AA{12}}{\AA{23}} \, ,
\nn \\
&&
\AR{\hat{1}} = \AR{1} + \frac{\AA{12}}{\AA{23}} \AR{3} = \frac{\AA{13}}{\AA{23}}\AR{2} \, , \quad \SR{\hat{3}} = \SR{3} - \frac{\AA{12}}{\AA{23}} \SR{1} \, ,
\nn \\
&&
\AR{\hat{K}}\SL{\hat{K}} = - \AR{2}\left(  \frac{\AA{13}}{\AA{23}}\SL{1} + \SL{2} \right)
\, , \quad
\SSS{2\hat{K}} = \frac{\SSS{12}\AA{13}}{\AA{23}} \, , 
\nn \\
&&
\SSS{\hat{K} 1} = \SSS{12} \, ,  \quad
\SSS{\hat{3}\hat{K}} = - \frac{(p_q+k_{\qb})^2}{\AA{23}} \, ,
\nn \\
&&
\SSS{\bar{q}\hat{K}} = \frac{\AL{3}\slashp_q+\slashk_{\qb}}{\AA{23}}\, , 
\quad
\SSS{q \hat{K}} = \frac{\AL{3}\slashk_{\qb}+\slashp_q\SR{\qb}}{\AA{23}}\, ,
\nn \\
&&
\SSS{q\hat{3}} = \frac{\SSS{q3}\AA{23} - \AA{12}\SSS{q1}}{\AA{23}} = \frac{\AL{2}\slashk_{\qb}\SR{q}}{\AA{23}} \, .
\eea
Collecting these in formula (\ref{pre_amp_qb_1}) we get
\beq
\frac{1}{\kappa_{\qb}}\,\frac{1}{(p_q+k_{\qb})^2}\,  
\frac{\AL{3}\slashp_1+\slashk_{\qb}\SR{\qb}^2\AL{3}\slashk_{\qb}\SR{q}}{\AA{12}\AA{13}\SSS{q\qb}\AL{2}\slashk_{\qb}\SR{q}} \, .
\eeq
The third terms is
\beq
\frac{\AA{\qb\hat{K}}^2}{\AA{\hat{1}2}\AA{2\qb}\AA{\hat{K}1}}\, \frac{1}{\AA{q3}\SSS{3q}}\, \frac{\AA{q3}^2}{\AA{\hat{K}q}}
\label{pre_amp_qb_3}
\eeq
for which we find the value of $z$ and of shifted quantities to be
\bea
(\hat{p}_3+p_q)^2 &=& 0 \Leftrightarrow \AA{q3}\SSS{3q} - z\, \AA{3q}\SSS{q1} = 0 \Leftrightarrow z= \frac{\SSS{q3}}{\SSS{q1}} \, ,
\nn \\
\AR{\hat{1}} &=& \AR{1} + \frac{\SSS{q3}}{\SSS{q1}} \AR{3} \, , \quad \SR{\hat{3}} = \SR{3} + \frac{\SSS{3q}}{\SSS{q1}} \SR{1} = \frac{\SSS{31}}{\SSS{q1}} \SR{q} \, ,
\nn \\ 
\AR{\hat{K}}\SL{\hat{K}} &=& \AR{q}\SL{q} + \AR{3}\SL{\hat{3}} = \left( \AR{q} + \frac{\SSS{31}}{\SSS{q1}}\, \AR{3} \right) \SL{q} \, , 
\nn \\ 
\AA{\qb\hat{K}} &=& \AA{\qb q} + \frac{\SSS{31}}{\SSS{q1}} \AA{\qb3} = \frac{\AL{\qb}\slashp_3+\slashp_q\SR{1}}{\SSS{q1}} \, ,
\nn \\
\AA{\hat{1}2} &=& \AA{12} + \frac{\SSS{q3}}{\SSS{q1}}\AA{32} = - \frac{\AL{2} \slashk_{\qb} \SR{q}}{\SSS{q1}} \, ,
\nn \\
\AA{\hat{K}1} &=& - \frac{(p_2+k_{\qb})^2}{\SSS{q1}} \, , \quad \AA{q\hat{K}} = \frac{\AA{q3}\SSS{31}}{\SSS{q1}} \, ,
\eea
which turn (\ref{pre_amp_qb_3}) into
\beq
\frac{1}{(p_2+k_{\qb})^2}\, \frac{\SSS{1q}\AL{\qb} \slashp_3+\slashp_q \SR{1}^2}{\AA{2\qb}\SSS{q3}\SSS{31}\AL{2}\slashk_{\qb}\SR{q}} \, .
\eeq
So the final amplitude is given by
\beq
\Amp(g_1^+,g_2^+,\bar{q}^*,q^-,g_3^-) \stackrel{e^\mu=\frac{\AL{3}\g^\mu\SR{1}}{2}}{=}
\frac{1}{\AL{2}\slashk_{\qb}\SR{q}} \left(
\frac{1}{\kappa_{\qb}}\,  \frac{\AL{3}\slashp_1+\slashk_{\qb}\SR{\qb}^2\AL{3}\slashk_{\qb}\SR{q}}{(p_q+k_{\qb})^2\AA{12}\AA{13}\SSS{q\qb}} +
\frac{\SSS{1q}\AL{\qb} \slashp_3+\slashp_q \SR{1}^2}{(p_2+k_{\qb})^2\AA{2\qb}\SSS{q3}\SSS{31}}
\right) \, .
\label{3ways1}
\eeq
\begin{figure}[h]
\begin{center}
\includegraphics[scale=0.7]{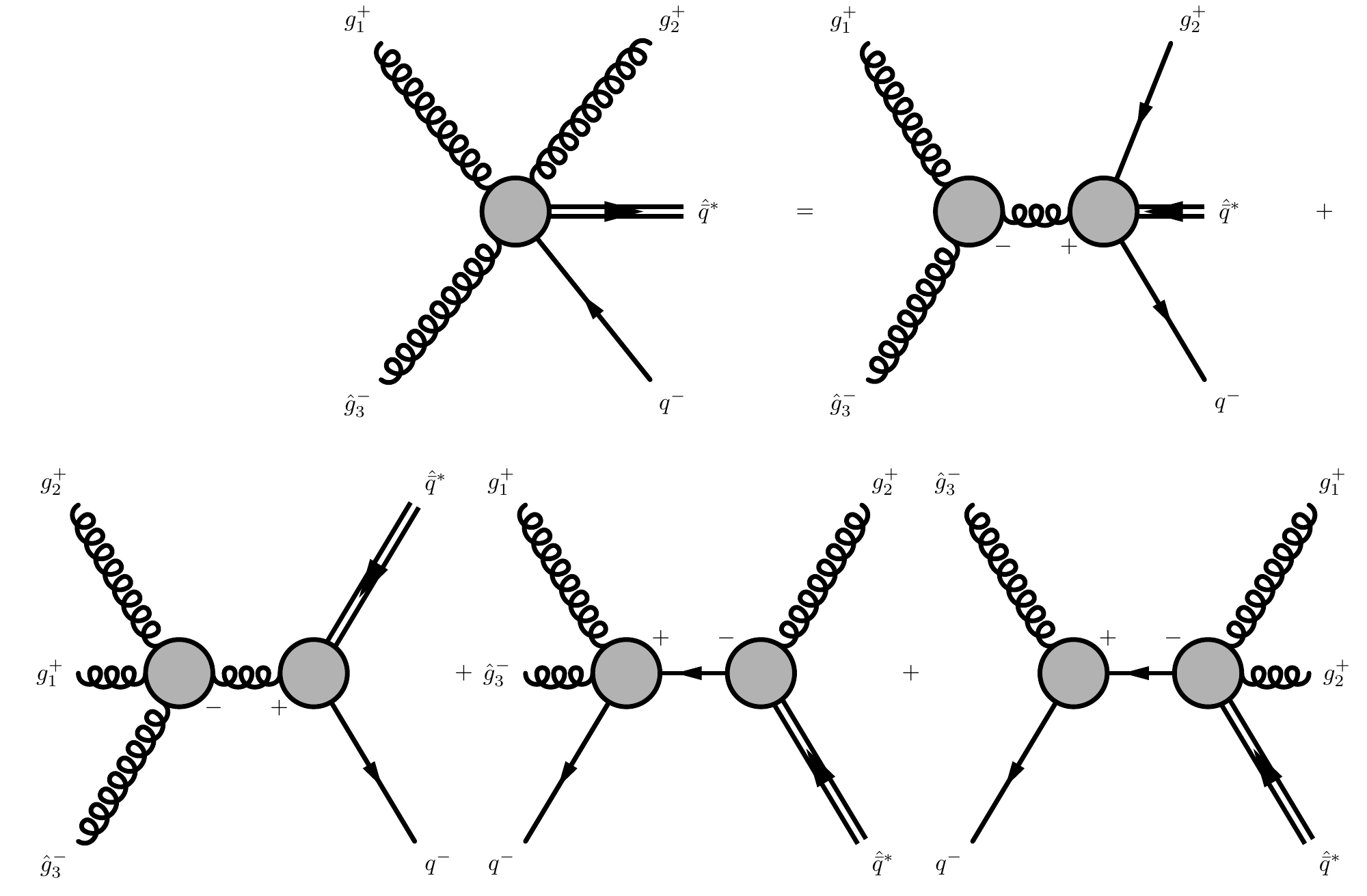}
\caption{The recursion for $\Amp(g_1^+,g_2^+,\bar{q}^*,q^-,g_3^-)$ with shift vector 
$e^\mu = \frac{1}{2}\AL{3}\g^\mu\SR{\qb}$, which {\em is legitimate in the on-shell case}.}
\label{5_point_recursion_2_fig}
\end{center}
\end{figure}
Now we will derive $\Amp(g_1^+,g_2^+,\bar{q}^*,q^-,g_3^-)$ in the second way, with $e^\mu = \frac{\AL{3}\g^\mu\SR{\qb}}{2}$,
which can be used for the on-shell recursion. The four residues are shown in fig. \ref{5_point_recursion_2_fig}.

The first one is
\beq
\frac{\AA{\hat{K}3}^3}{\AA{31}\AA{1\hat{K}}}\, \frac{1}{\AA{13}\SSS{31}}\, \frac{\AA{\qb q}^3}{\AA{\qb q}\AA{q\hat{K}}\AA{\hat{K}2}\AA{2\qb}} \, .
\eeq
with pole location and shifted quantities
\bea
z &=& \frac{\SSS{31}}{\SSS{1\qb}} \, ,
\nn \\
\SR{\hat{3}} &=& \SR{3} + \frac{\SSS{31}}{\SSS{1\qb}} \SR{\qb} = \frac{\SSS{3\qb}}{\SSS{1\qb}} \SR{1} \, ,
\nn \\
\AR{\hat{K}} \SL{\hat{K}} &=& \left( \AR{1} + \frac{\SSS{3\qb}}{\SSS{1\qb}}\AR{3} \right) \SL{1}  \, ,
\nn \\
\AA{\hat{K}3} &=& \AA{13} \, , 
\quad 
\AA{1\hat{K}} = \frac{\AA{13}\SSS{3\qb}}{\SSS{1\qb}}\, , 
\quad 
\AA{q\hat{K}} = \frac{\AL{q} \slashp_1+\slashp_3 \SR{\qb}}{\SSS{1\qb}} \, ,
\quad
\AA{\hat{K}2} = - \frac{\AL{2} \slashp_1+\slashp_3 \SR{\qb}}{\SSS{1\qb}} \, ,
\eea
which lead to the form
\beq
\frac{\AA{\qb q}^2\SSS{1\qb}^3}{\SSS{13}\SSS{\qb 3}\AA{2\qb}\AL{q}\slashp_1+\slashp_3\SR{\qb}\AL{2}\slashp_1+\slashp_3\SR{\qb}}
\label{pre_amp_3rd_1}
\eeq
The second term is 
\beq
\frac{\AA{3\hat{K}}^4}{\AA{\hat{K}3}\AA{31}\AA{12}\AA{2\hat{K}}}\, \frac{1}{(k_{\qb}+p_q)^2} \, 
\frac{1}{\kappa_{\qb}} \, \frac{\SSS{\qb\hat{K}}^3\SSS{q\hat{K}}}{\SSS{\qb\hat{K}}\SSS{\hat{K}q}\SSS{q\qb}} = 
\frac{1}{\kappa_{\qb}(k_{\qb}+p_q)^2} \, \frac{\AL{3}\slashK\SR{\qb}^3}{\AA{31}\AA{12}\SSS{\qb q}\AL{2}\hat{\slashK}\SR{\qb}} \, ,
\eeq
for which
\bea
z &=& \frac{(k_{\qb}+p_q)^2}{\AL{3} \slashk_{\qb}+\slashp_q \SR{\qb}} \, ,
\nn \\
\hat{\slashK} &=& \slashk_{\qb} + \slashp_q - \frac{(k_{\qb}+p_q)^2}{\AL{3}\slashk_{\qb}+\slashp_q\SR{\qb}} \left( \AR{3}\SL{\qb} + \SR{\qb}\AL{3} \right) \, ,
\nn \\
\AL{3} \hat{\slashK} \SR{\qb} &=& \AL{3} \slashk_{\qb} + \slashp_q \SR{\qb} \, ,
\nn \\
\AL{2} \hat{\slashK} \SR{\qb} &=& \AL{2} \slashk_{\qb} + \slashp_q \SR{\qb} \, ,
\eea
bringing the second term into the form
\beq
\frac{1}{\kappa_{\qb}(k_{\qb}+p_q)^2} \, \frac{\AL{3} \slashk_{\qb} + \slashp_q \SR{\qb}^3}{\SSS{q\qb}\AA{12}\AA{13}\AL{2} \slashk_{\qb}+\slashp_q \SR{\qb}} \, .
\label{pre_amp_3rd_2}
\eeq
The third terms is written as
\beq
\frac{\AA{3q}^3\AA{3\hat{K}}}{\AA{q3}\AA{31}\AA{1\hat{K}}\AA{\hat{K}q}} \, \frac{1}{(p_2+k_{\qb})^2} \, \frac{1}{\kappa_{\qb}} \,
\frac{\SSS{2\qb}^3\SSS{2\hat{K}}}{\SSS{\hat{K}\qb}\SSS{\qb2}\SSS{2\hat{K}}} =
\frac{1}{\kappa_{\qb}(p_2+k_{\qb})^2} \, \frac{\AA{3q}^2\SSS{2\qb}^2\AL{3}\hat{\slashK}\SR{2}}{\AA{31}\AL{1}\hat{\slashK}\SR{2}\AL{q}\hat{\slashK}\SR{\qb}} \, .
\eeq
The shifted spinors and vectors are
\bea
z &=& \frac{(p_2+k_{\qb})^2}{\AL{3} \slashp_2+\slashk_{\qb} \SR{\qb}} \, , 
\nn \\
\hat{\slashK} &=& \slashk_{\qb} + \slashp_2 - \frac{(p_2+k_{\qb})^2}{\AL{3} \slashp_2+\slashk_{\qb} \SR{\qb}} \left( \AR{3}\SL{\qb} + \SR{\qb}\AL{3} \right) \, ,
\nn \\
\AL{3}\hat{\slashK}\SR{2} &=& \AL{3} \slashk_{\qb}\SR{2} \, , \quad \AL{q}\hat{\slashK}\SR{\qb} = \AL{q} \slashk_{\qb} + \slashp_2  \SR{\qb} \, ,
\nn \\
\AL{1}\hat{\slashK}\SR{2} &=&  \frac{ \AL{1} \slashk_{\qb} \SR{2}\AL{3} \slashk_{\qb}+\slashp_2 \SR{\qb}+(k_{\qb}+p_2)^2\AA{13}\SSS{2\qb} }{\AL{3}\slashk_{\qb}+\slashp_2\SR{\qb}} \, .
\eea
with which we write our third residue as
\beq
\frac{1}{\kappa_{\qb}(p_2+k_{\qb})^2} \, 
\frac{\AA{3q}^2\SSS{2\qb}^2\AL{3}\slashk_{\qb}\SR{2}}{\AA{13}\AL{q}\slashp_2+\slashk_{\qb}\SR{\qb}} \, 
\frac{\AL{3} \slashk_{\qb}+\slashp_2 \SR{\qb}}{\AL{1} \slashk_{\qb} \SR{2}\AL{3} \slashk_{\qb}+\slashp_2 \SR{\qb}+(k_{\qb}+p_2)^2\AA{13}\SSS{2\qb} } \, .
\label{pre_amp_3rd_3}
\eeq
We come to the fourth and last term, 
\beq
\frac{\AA{3q}^3\AA{3\hat{K}}}{\AA{3\hat{K}}\AA{\hat{K}q}\AA{q3}} \, 
\frac{1}{\AA{q3}\SSS{3q}} \, 
\frac{\AA{\qb\hat{K}}^3}{\AA{\hat{K}1}\AA{12}\AA{2\qb}\AA{\qb\hat{K}}} = 
\frac{\AA{3q}\AA{\qb\hat{K}}^2}{\AA{12}\AA{2\qb}\SSS{3q}\AA{3\hat{K}}\AA{q\hat{K}}} \, ,
\eeq
which, through
\bea
&&
z = -\frac{\SSS{q3}}{\SSS{q\qb}} \, ,  
\nn \\
&&
\SR{\hat{3}} = \SR{3} + \frac{\SSS{3q}}{\SSS{q\qb}} \SR{\qb} = \frac{\SSS{3\qb}}{\SSS{q\qb}} \SR{q} \, , 
\nn \\
&&
\AR{\hat{K}}\SL{\hat{K}} =  - \left(  \AR{q} + \frac{\SSS{3\qb}}{\SSS{q\qb}}\AR{3} \right) \SL{q} \, ,
\nn \\
&&
\AA{3\hat{K}} = \AA{q3} \, , 
\quad 
\AA{\qb\hat{K}} = \frac{\AL{\qb} \slashp_3+\slashp_q \SR{\qb}}{\SSS{\qb q}} \, , 
\nn \\
&&
\AA{\hat{K}1} = \frac{\AL{1} \slashp_3+\slashp_q \SR{\qb}}{\SSS{q\qb}} \, ,
\quad 
\AA{q\hat{K}} = \frac{\AA{q3}\SSS{3\qb}}{\SSS{\qb q}} \, ,
\eea
becomes
\beq
\frac{\AL{\qb}\slashp_3+\slashp_q\SR{\qb}^2}{\AA{12}\AA{2\qb}\SSS{3q}\SSS{\qb3}\AL{1} \slashp_3+\slashp_q \SR{\qb}} \, .
\label{pre_amp_3rd_4}
\eeq

Thus, adding (\ref{pre_amp_3rd_1}), (\ref{pre_amp_3rd_2}), (\ref{pre_amp_3rd_3}) and (\ref{pre_amp_3rd_4}) we get our final result,
\bea
\Amp(g_1^+,g_2^+,\bar{q}^*,q^-,g_3^-)
&\stackrel{e^\mu=\frac{\AL{3}\g^\mu\SR{\qb}}{2}}{=}&
\frac{\AA{\qb q}^2\SSS{1\qb}^3}{\SSS{13}\SSS{\qb 3}\AA{2\qb}\AL{q}\slashp_1+\slashp_3\SR{\qb}\AL{2}\slashp_1+\slashp_3\SR{\qb}} 
\nn \\
&&\hspace{-25mm}
+\frac{1}{\kappa_{\qb}(k_{\qb}+p_q)^2} \, \frac{\AL{3} \slashk_{\qb} + \slashp_q \SR{\qb}^3}{\SSS{q\qb}\AA{12}\AA{13}\AL{2} \slashk_{\qb}+\slashp_q \SR{\qb}}
\nn \\
&& \hspace{-25mm}
+ \frac{1}{\kappa_{\qb}(p_2+k_{\qb})^2} \, 
\frac{\AA{3q}^2\SSS{2\qb}^2\AL{3}\slashk_{\qb}\SR{2}}{\AA{13}\AL{q}\slashp_2+\slashk_{\qb}\SR{\qb}} \, 
\frac{\AL{3} \slashk_{\qb}+\slashp_2 \SR{\qb}}{\AL{1} \slashk_{\qb} \SR{2}\AL{3} \slashk_{\qb}+\slashp_2 \SR{\qb}+(k_{\qb}+p_2)^2\AA{13}\SSS{2\qb} } 
\nn \\
&&\hspace{-25mm}
+ \frac{\AL{\qb}\slashp_3+\slashp_q\SR{\qb}^2}{\AA{12}\AA{2\qb}\SSS{3q}\SSS{\qb3}\AL{1} \slashp_3+\slashp_q \SR{\qb}} \, .
\label{3ways2}
\eea
Using the same shift vector, one can check that the on-shell recursion works as well: the first and fourth contributions contain subleading sub-amplitudes,
so that only the second and the third contributions do not vanish and their sum is exactly the MHV amplitude which is expected. 
This can also be seen very quickly by taking our derivation and stripping off
the factors $1/\kappa_{\qb}$ from the final forms for the second and third term and taking $k_{\qb}$ to be on-shell. \\
\begin{figure}[h]
\begin{center}
\includegraphics[scale=0.7]{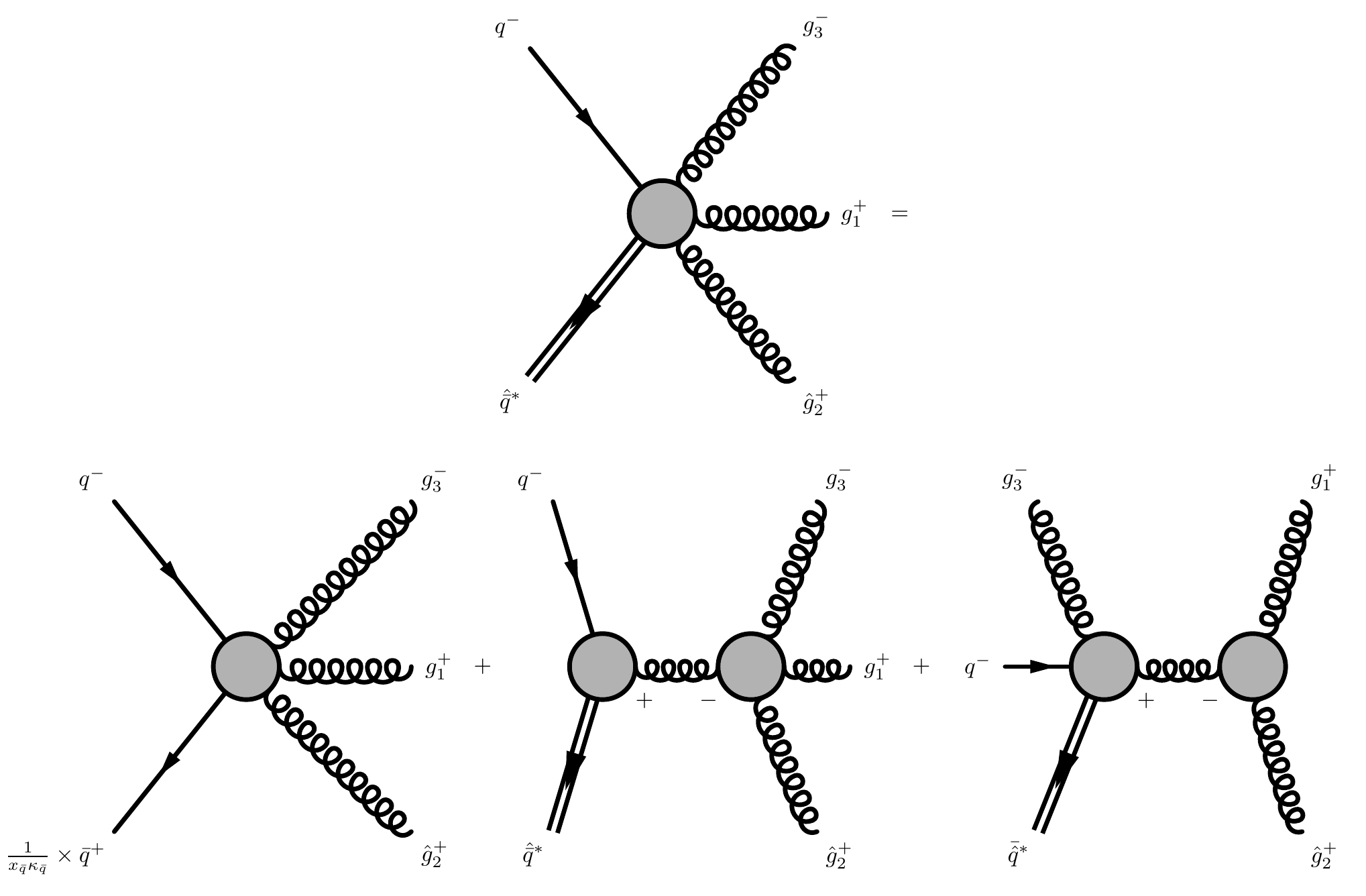}
\caption{The recursion for $\Amp(g_1^+,g_2^+,\bar{q}^*,q^-,g_3^-)$ with shift vector $e^\mu = \frac{1}{2}\AL{\qb}\g^\mu\SR{2}$, 
which {\em is not legitimate in the on-shell case}.}
\label{5_point_recursion_3_fig}
\end{center}
\end{figure}
And now let us go a bit further and use the shift vector $e^\mu=\frac{\AL{\qb}\g^\mu\SR{2}}{2}$, which,
remembering the discussion of section \ref{fancy_shift}, is not legitimate in the on-shell case but
should work when the shifted antifermion is off-shell.
The three contributions are those in fig. \ref{5_point_recursion_3_fig}. \\
The first one, the $C^f$-residue for the antifermion, is
\beq
\frac{1}{\sqrt{x_{\qb}}\kappa_{\qb}}\, \frac{\AA{3q}^3\AA{3\hat{\qb}}}{\AA{1\hat{2}}\AA{\hat{2}\hat{\qb}}\AA{\hat{\qb}q}\AA{q3}\AA{31}} \, .
\eeq
Looking at Table \ref{CDf}, we immediately see that the shifted quantities are
\beq
\AR{\hat{2}} = \frac{(\slashp_2+\slashk_{\qb})\SR{\qb}}{\SSS{2\qb}} \, , \quad \AR{\hat{\qb}} = \frac{\slashk_{\qb}\SR{2}}{\sqrt{x_{\qb}}\SSS{\qb2}} \, ,
\eeq
from which it is immediate to conclude that the first term is
\beq
\frac{1}{\kappa_{\qb}} \, 
\frac{\AA{3q}^2\SSS{2\qb}^3 \AL{3}\slashk_{\qb}\SR{2}}{\AA{13}\AL{1}\slashp_2+\slashk_{\qb}\SR{\qb} \AL{q}\slashk_{\qb}\SR{2}\SL{2}\slashk_{\qb}(\slashk_{\qb}+\slashp_2)\SR{\qb}}
\label{amp_ill_1}
\eeq
The second term, with the fermion pair on the left, reads
\beq
\frac{1}{\hat{\kappa}_{\qb}}\, \frac{\SSS{\hat{K}\qb}^3\SSS{\hat{K} q}}{\SSS{\hat{K}q}\SSS{q\qb}\SSS{\qb\hat{K}}}\, 
\frac{1}{(k_{\qb}+p_q)^2}\, \frac{\AA{3\hat{K}}^4}{\AA{\hat{K}3}\AA{31}\AA{1\hat{2}}\AA{\hat{2}\hat{K}}} =
\frac{1}{\hat{\kappa}_{\qb}}\, \frac{1}{(k_{\qb}+p_q)^2} \frac{\AL{3}\hat{\slashK}\SR{\qb}^3}{\SSS{q\qb}\AA{31}\AA{1\hat{2}}\AL{\hat{2}}\hat{\slashK} \SR{\qb}} \, ,
\label{amp_ill_2}
\eeq
where we have to replace
\bea
(\hat{k}_{\qb} + p_q)^2 
&=& 
0 \Leftrightarrow (k_{\qb}+p_q)^2 + z\,\AL{\qb}\slashp_q+\slashk_{\qb}\SR{2} = 0 \Leftrightarrow z = - \frac{(k_{\qb}+p_q)^2}{\AL{\qb}\slashk_{\qb}+\slashp_q\SR{2}}
\nn \\
\AR{\hat{2}} 
&=& 
\AR{2} + \frac{(k_{\qb}+p_q)^2}{\AL{\qb}\slashk_{\qb}+\slashp_q\SR{2}} \AR{\qb} \, ,
\nn \\
\hat{\slashK} &=& - \left(  \slashk_{\qb} + \slashp_q \right) + \frac{(k_{\qb}+p_q)^2}{\AL{\qb}\slashk_{\qb}+\slashp_q\SR{2}} \, \left(\AR{\qb}\SL{2}+\SR{2}\AL{\qb} \right) \, ,
\nn \\
\hat{\kappa}_{\qb} 
&=&
\frac{\AL{\qb} (\slashk_{\qb}+\slashp_q)\slashp_2\slashk_{\qb} + ( k_{\qb}+p_q)^2 \slashp_2 \SR{\qb}}{\AL{\qb}(\slashk_{\qb}+\slashp_q)\slashp_2\AR{\qb}} \, ,
\nn \\
\AA{1\hat{2}} 
&=&
\frac{(k_{\qb}+p_q)^2\AA{1\qb}-\AL{\qb}(\slashk_{\qb}+\slashp_q)\slashp_2\SR{1}}{\AL{\qb}\slashk_{\qb}+\slashp_q\SR{2}}\, ,
\nn \\
\AL{3}\hat{\slashK}\SR{\qb}
&=&
- \frac{\AL{3}\slashp_q+\slashk_{\qb}\SR{\qb}\AL{\qb}\slashp_q+\slashk_{\qb}\SR{2}+(k_{\qb}+p_q)^2 \AA{\qb3}\SSS{2\qb}}{\AL{\qb}\slashp_q+\slashk_{\qb}\SR{2}} \, ,
\nn \\
\AL{\hat{2}} \hat{\slashK} \SR{\qb}
&=&
- \frac{\AL{\qb} (\slashk_{\qb}+\slashp_q)\slashp_2 (\slashk_{\qb}+\slashp_q) + (k_{\qb}+p_q)^2(\slashk_{\qb}+\slashp_2+\slashp_q) \SR{\qb}}{\AL{\qb}\slashk_{\qb}+\slashp_q\SR{2}} \, ,
\eea
obtaining
\bea
&&
\frac{1}{(k_{\qb}+p_q)^2\SSS{q \qb }\AA{13}}\, 
\frac{\AL{\qb}(\slashk_{\qb}+\slashp_q)\slashp_2\AR{\qb}}{\AL{\qb} (\slashk_{\qb}+\slashp_q)\slashp_2 \slashk_{\qb} +(k_{\qb}+p_q)^2\slashp_2\SR{\qb}}
\\
&& \hspace{-5mm}
\times \frac{\left\{\AL{3}\slashp_q+\slashk_{\qb}\SR{\qb}\AL{\qb}\slashp_q+\slashk_{\qb}\SR{2}+(k_{\qb}+p_q)^2 \AA{\qb3}\SSS{2\qb} \right\}^3}
{\AL{\qb}\slashk_{\qb}+\slashp_q\SR{2}\left\{ (k_{\qb}+p_q)^2\AA{1\qb}-\AL{\qb}(\slashk_{\qb}+\slashp_q)\slashp_2\SR{1} \right\}
\AL{\qb} (\slashk_{\qb}+\slashp_q)\slashp_2 (\slashk_{\qb}+\slashp_q) + (k_{\qb}+p_q)^2(\slashk_{\qb}+\slashp_2+\slashp_q) \SR{\qb}} \, .
\nn
\eea
Finally, we can write the third term
\beq
\frac{1}{\hat{\kappa}_{\qb}} \, \frac{\SSS{\qb\hat{K}}^3\SSS{q\hat{K}}}{\SSS{\hat{K}3}\SSS{3q}\SSS{q \qb}\SSS{\qb\hat{K}}} \, 
\frac{1}{\AA{12}\SSS{21}} \, \frac{\SSS{21}^3}{\SSS{1\hat{K}}\SSS{\hat{K}2}} = 
\frac{1}{\hat{\kappa}_{\qb}} \, \frac{\SSS{12}^2 \SSS{\qb \hat{K}}^2\SSS{q\hat{K}}}{\AA{12}\SSS{\hat{K}3}\SSS{3q}\SSS{q\qb}\SSS{1\hat{K}}\SSS{2\hat{K}}} \, ,
\eeq
obtaining the shifted variables
\bea
z &=& \frac{\AA{12}}{\AA{1\qb}} \, , 
\nn \\
\AR{\hat{2}} &=& \AR{2} - \frac{\AA{21}}{\AA{\qb1}} \AR{\qb} = \frac{\AA{\qb2}}{\AA{\qb1}} \AR{1} \, ,
\nn \\
\AR{\hat{K}}\SL{\hat{K}} &=& \AR{1}\, \left(  \SL{1} + \frac{\AA{\qb2}}{\AA{\qb1}} \SL{2} \right) \, ,
\nn \\
\SSS{1\hat{K}} &=& \frac{\SSS{12}\AA{\qb2}}{\AA{\qb1}} \, , \quad \SSS{2\hat{K}} = \SSS{21} \, , \quad \SSS{\hat{K}3} = - \frac{\AL{\qb}\slashp_1+\slashp_2\SR{3}}{\AA{1\qb}}
\nn \\
\SSS{\qb\hat{K}} &=& \frac{\AL{\qb} \slashp_1+\slashp_2 \SR{\qb}}{\AA{1\qb}} \, , 
\quad 
\SSS{q\hat{K}} = \frac{\AL{\qb} \slashp_1+\slashp_2 \SR{\qb}}{\AA{1\qb}} \, ,
\quad
\hat{\kappa}_{\qb} = \frac{\AL{1} \slashk_{\qb}+\slashp_2 \SR{\qb}}{\AA{1\qb}} \, ,
\eea
from which we get the final form of the third term
\beq
\frac{\AL{\qb} \slashp_1+\slashp_2 \SR{\qb}^2 \AL{\qb} \slashp_1+\slashp_2 \SR{q}}
{\AA{12}\AA{2\qb}\SSS{q3}\SSS{\qb q}\AL{1} \slashk_{\qb}+\slashp_2 \SR{\qb}\AL{\qb} \slashp_1+\slashp_2 \SR{3} } \, .
\label{amp_ill_3}
\eeq
Putting together (\ref{amp_ill_1}), (\ref{amp_ill_2}) and (\ref{amp_ill_3}) we get our result
\bea
\Amp(g_1^+,g_2^+,\bar{q}^*,q^-,g_3^-) &\stackrel{e^\mu=\frac{\AL{\qb}\g^\mu\SR{2}}{2}}{=}&
\frac{1}{(k_{\qb}+p_q)^2\SSS{\qb q}\AA{31}}\, 
\frac{\AL{\qb}(\slashk_{\qb}+\slashp_q)\slashp_2\AR{\qb}}{\AL{\qb} (\slashk_{\qb}+\slashp_q)\slashp_2 \slashk_{\qb} +(k_{\qb}+p_q)^2\slashp_2\SR{\qb}}
\nn \\
&&\hspace{-5.5cm}
\times
\frac{\left\{\AL{3}\slashp_q+\slashk_{\qb}\SR{\qb}\AL{\qb}\slashp_q+\slashk_{\qb}\SR{2}+(k_{\qb}+p_q)^2 \AA{\qb3}\SSS{2\qb} \right\}^3}
{\AL{\qb}\slashk_{\qb}+\slashp_q\SR{2}\left\{ (k_{\qb}+p_q)^2\AA{1\qb}-\AL{\qb}(\slashk_{\qb}+\slashp_q)\slashp_2\SR{1} \right\}
\AL{\qb} (\slashk_{\qb}+\slashp_q)\slashp_2 (\slashk_{\qb}+\slashp_q) + (k_{\qb}+p_q)^2(\slashk_{\qb}+\slashp_2+\slashp_q) \SR{\qb}} \, .
\nn \\
&&\hspace{-5.5cm}
+ \frac{1}{\kappa_{\qb}} \, 
\frac{\AA{3q}^2\SSS{2\qb}^3 \AL{3}\slashk_{\qb}\SR{2}}{\AA{31}\AL{1}\slashp_2+\slashk_{\qb}\SR{\qb} \AL{q}\slashk_{\qb}\SR{2}\SL{qb}(\slashp_2+\slashk_{\qb})\slashk_{\qb}\SR{2}} +
\frac{\AL{\qb} \slashp_1+\slashp_2 \SR{\qb}^2 \AL{\qb} \slashp_1+\slashp_2 \SR{q}}
{\AA{12}\AA{2\qb}\SSS{q3}\SSS{\qb q}\AL{1} \slashk_{\qb}+\slashp_2 \SR{\qb}\AL{\qb} \slashp_1+\slashp_2 \SR{3} }
\label{3ways3}
\eea

Despite this, if one tries to use the same shift vector in the on-shell case, then the MHV amplitude
which is expected is not obtained. 
We must stress that we have checked numerically that (\ref{3ways1}), (\ref{3ways2}) and (\ref{3ways3}) 
have exactly the same values at multiple phase space points, despite their appearance is completely different.
Actually, the difference in the structure of non-MHV amplitudes is a phenomenon already known from the on-shell case. \\

For the other two amplitudes we just give the results, specifying the shifts used in order to obtain them.
The procedure is similar to the way of getting the first form of $\Amp(g_1^+,g_2^+,\bar{q}^*,q^-,g_3^-)$, 
i.e. by shifting the two gluons as indicated by the shift vectors:
\bea
\Amp(g_1^+,g_2^-,\bar{q}^*,q^-,g_3^+)
&\stackrel{e^\mu=\AL{1}\g^\mu\SR{3}/2}{=}&
\frac{1}{\kappa_{\qb}}\, \frac{\AL{q} \slashp_2+\slashk_{\qb} \SR{\qb}^3}{\AA{13}\AA{q3}\SSS{2\qb}\AL{q}\slashk_{\qb}\SR{2}\AL{1}\slashp_2+\slashk_{\qb}\SR{\qb}}
\nn \\
&& \hspace{-15mm}
+\, \frac{1}{(k_{\qb}+p_q)^2}\frac{\AA{\qb q}^2\SSS{13}^3}{\SSS{21}\AL{\qb}\slashp_q+\slashk_{\qb}\SR{3}\AL{q}\slashk_{\qb}\SR{2}}
\nn \\
&& \hspace{-15mm}
+\, \frac{\AA{2\qb}^3\SSS{3\qb }^3}{\AA{12}\SSS{q\qb}\AL{\qb}\slashp_1+\slashp_2\SR{3}\AL{1}\slashk_{\qb}+\slashp_2\SR{\qb}\AL{\qb}\slashp_1+\slashp_2\SR{\qb}} \, ,
\eea
\bea
\Amp(g_1^-,g_2^+,\bar{q}^*,q^-,g_3^+)
&\stackrel{e^\mu=\AL{1}\g^\mu\SR{3}/2}{=}&
\frac{1}{\kappa_{\qb}}\,\frac{1}{(p_2+k_{\qb})^2}\, \frac{\AA{q1}^3\SSS{2\qb}^2\AL{1}\slashk_{\qb}\SR{2}}{\AA{13}\AA{q3}\AL{q} \slashk_{\qb} \SR{2}\AL{1} \slashp_2+\slashk_{\qb} \SR{\qb}}
\nn \\
&& \hspace{-15mm}
+\, \frac{\SSS{23}^4\AA{q\qb}^2}{(k_{\qb}+p_q)^2 \SSS{13}\SSS{21}\AL{q}\slashk_{\qb}\SR{2}\AL{\qb} \slashk_{\qb}+\slashp_q\SR{\qb} }
\nn \\
&& \hspace{-15mm}
+\, \frac{\AA{1\qb}^4\SSS{3\qb}^3}{\AA{12}\AA{2\qb}\SSS{q\qb}\AL{1} \slashk_{\qb}+\slashp_2 \SR{\qb}\AL{\qb}\slashp_1+\slashp_2\SR{3}\AL{\qb}\slashp_1+\slashp_2\SR{\qb}} \, .
\eea
Now that we have all the necessary amplitudes in the case of the off-shell antifermion, the ones with the off-shell fermion
are simply got by $ \qb \leftrightarrow q $, as already mentioned.

\subsection{Off-shell gluon}

When the gluon is off-shell, on one hand there are no subleading contributions, so that the independent amplitudes are only four;
on the other hand, the position of the quark pair relative to the gluons matters, and there are two possible positions which are
not equivalent. We assume the fermion pair helicities $(h_{\bar{q}},h_q) = (+,-)$, obtaining the inequivalent amplitudes
\begin{eqnarray}
\Amp(g^*,\bar{q}^+,q^-,g_1,g_2)\quad 
&&
\left\{ \begin{array}{c}
\Amp(g^*,\bar{q}^+,q^-,g_1^-,g_2^-) \quad \textrm{MHV}
\\
\Amp(g^*,\bar{q}^+,q^-,g_1^+,g_2^+)  \quad \textrm{MHV}
\\
\Amp(g^*,\bar{q}^+,q^-,g_1^+,g_2^-)  \quad \textrm{non-MHV}
\\
\Amp(g^*,\bar{q}^+,q^-,g_1^-,g_2^+) \quad \textrm{non-MHV}
\end{array} \right.
\nn \\
\Amp(g^*,g_1,\bar{q}^+,q^-,g_2)\quad 
&&
\left\{ \begin{array}{c}
\Amp(g^*,g_1^-,\bar{q}^+,q^-,g_2^-)  \quad \textrm{MHV}
\\
\Amp(g^*,g_1^+,\bar{q}^+,q^-,g_2^+)  \quad \textrm{MHV}
\\
\Amp(g^*,g_1^+,\bar{q}^+,q^-,g_2^-)  \quad \textrm{non-MHV}
\\
\Amp(g^*,g_1^-,\bar{q}^+,q^-,g_2^+)  \quad \textrm{non-MHV}
\end{array} \right.
\end{eqnarray}
Also here, we will show in detail how to perform the calculation only in one case, i.e. $\Amp(g^*,\bar{q}^+,q^-,g_1^-,g_2^-)$,
the others being similarly worked out.
\begin{figure}[h]
\begin{center}
\includegraphics[scale=0.7]{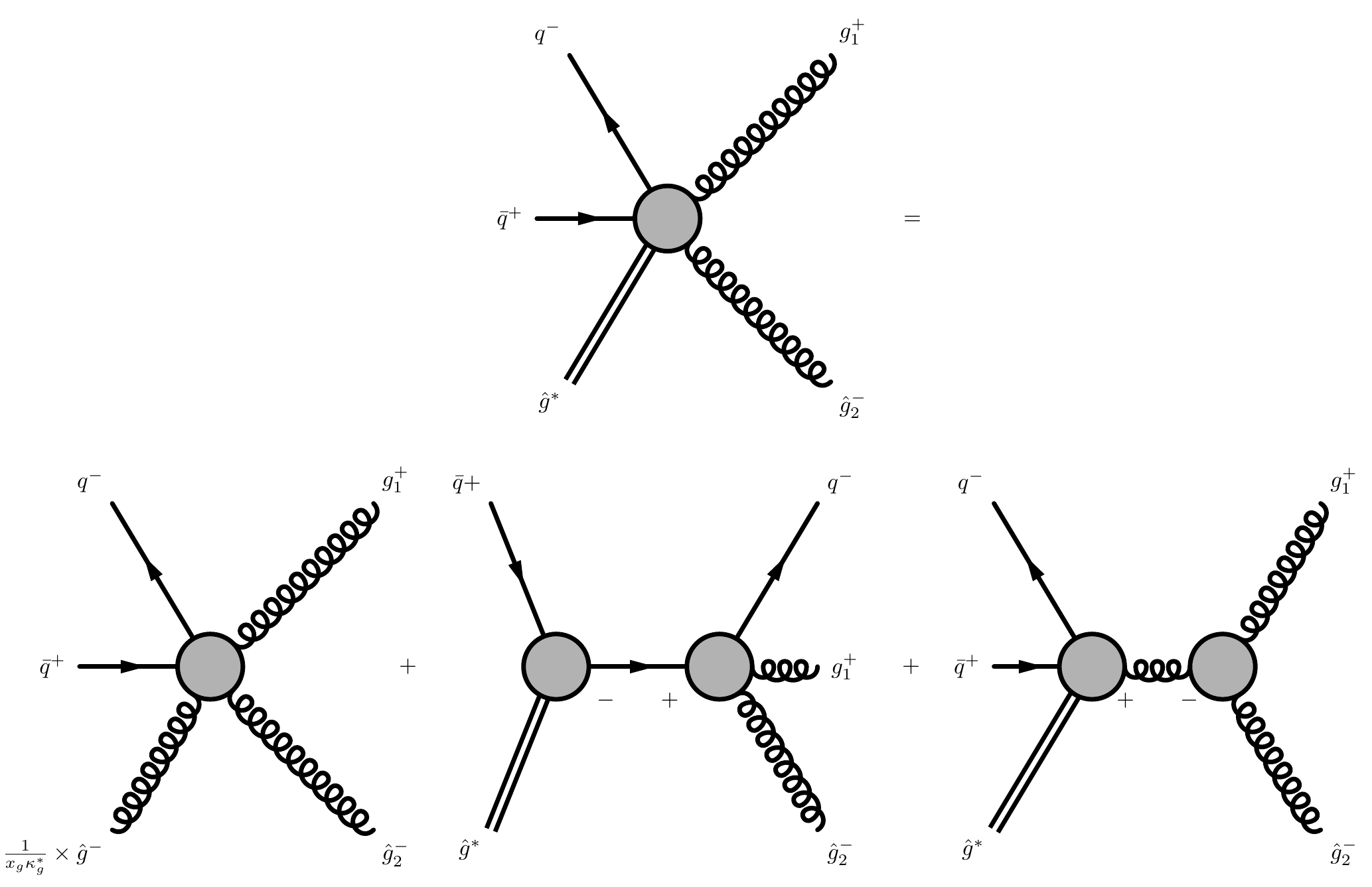}
\caption{The recursion for $\Amp(g^*,\bar{q}^+,q^-,g_1^-,g_2^-)$.}
\label{5_point_recursion_4_fig}
\end{center}
\end{figure}
It can be seen in fig. \ref{5_point_recursion_4_fig} that there are three contributions. \\
The first is the $C^g$-pole and is given by
\beq
\frac{1}{x_g\,\kstr_g}\, \frac{\SSS{\qb 1}^3 \SSS{q1}}{\SSS{1q}\SSS{q\qb}\SSS{\qb\hat{k}_g}\SSS{\hat{k_g}\hat{2}} \SSS{\hat{2}1}}\, ,
\eeq
in which we have to insert the shifted quantities in Table \ref{CD}, thereby obtaining
\beq
\frac{1}{\kstr_g}\, 
\frac{\SSS{\qb 1}^3 \AA{2g}^4 }{\SSS{\qb q} \AL{g} \slashp_2 +\slashk_g  \SR{1} \AL{2}  \slashk_g\,\left( \slashk_g + \slashp_2 \right)  \SR{g} \AL{2}  \slashk_g \SR{\qb} }\, .
\eeq
The second contribution is
\beq
\frac{1}{\kappa_g}\, \frac{\SSS{g\qb}^2}{\SSS{\qb \hat{K}}} \, \frac{1}{\left( k_g+p_{\qb} \right)^2} \, \frac{\AA{2 q}^3\AA{2\hat{K}}}{\AA{\hat{K}q}\AA{q1}\AA{12}\AA{2\hat{K}}} = 
\frac{1}{\kappa_g}\,\frac{1}{\left( k_g+p_{\qb} \right)^2} \,   \frac{\AA{2 q}^3 \SSS{g\qb}^2 }{\SSS{\qb\hat{K}}\AA{\hat{K}q}\AA{q1}\AA{12} } \, .
\label{pre_amp5_offs_g}
\eeq
The location of the pole $z$ and, thus, the values of shifted quantities are determined by
\bea
(\hat{k}_g+p_{\qb})^2 &=& 0 
\Leftrightarrow 
( k_g+p_{\qb})^2 + z \, \AL{2} \slashk_g + \slashp_{\qb} \SR{g} = 0 
\Leftrightarrow
z = - \frac{(k_g + p_{\qb})^2}{\AL{2} \slashk_g+\slashp_{\qb}  \SR{g}} \, ,
\nn \\
\hat{k}_g^\mu &=& k_g^\mu - \frac{(k_g + p_{\qb})^2}{\AL{2} \slashk_g+\slashp_{\qb}  \SR{g}}\, \frac{\AL{2}\g^\mu\SR{g}}{2}
\Rightarrow
\hat{\slashk_g} - \frac{(k_g + p_{\qb})^2}{\AL{2} \slashk_g+\slashp_{\qb}  \SR{g}} \, \left( \AR{2}\SL{g} + \SR{g}\AL{2} \right) \, ,
\nn \\
\SSS{\qb \hat{K}}\AA{\hat{K} q} &=& \frac{ (k_g+p_{\qb})^2 \SSS{\qb g}\AA{2q} - \AL{2} \slashk_g+\slashp_{\qb} \SR{g}\AL{q}\slashk_g\SR{\qb}  }{\AL{2} \slashk_g+\slashp_{\qb} \SR{g}} \, ,
\eea
which, after replacing into (\ref{pre_amp5_offs_g}), give
\beq
\frac{1}{\kappa_g}\,\frac{1}{\left( k_g+p_{\qb} \right)^2} \, 
\frac{\SSS{g\qb}^2\AA{2q}^3\AL{2} \slashk_g+\slashp_{\qb} \SR{g} }
{\AA{1q}\AA{12}\left\{ (k_g+p_{\qb})^2 \SSS{\qb g}\AA{2q} - \AL{2} \slashk_g+\slashp_{\qb} \SR{g}\AL{q}\slashk_g\SR{\qb} \right\}} \, .
\eeq
Finally, the third term is 
\beq
\frac{1}{\hat{\kappa}^*_g} \frac{\AA{gq}^3 \AA{g\qb} }{\AA{g\qb}\AA{\qb q}\AA{q\hat{K}}\AA{\hat{K}g}}\, \frac{1}{\AA{12}\SSS{21}}\,
\frac{\AA{2\hat{K}}^3}{\AA{\hat{K}1}\AA{12}} =
\frac{1}{\hat{\kappa}^*_g} \frac{\AA{gq}^3 \AA{2\hat{K}}^3 }{\AA{12}^2\SSS{21}\AA{\qb q}\AA{\hat{K}1}\AA{q\hat{K}}\AA{\hat{K}g}} \, .
\eeq
Pole and shifted variables are
\bea
z &=& \frac{\SSS{12}}{\SSS{1g}} \Rightarrow 
\SR{\hat{2}} = \frac{\SSS{2g}}{\SSS{1g}} \SR{1} 
\nn \\
\AR{\hat{K}}\SL{\hat{K}} &=& \left( \AR{1} + \frac{\SSS{2g}}{\SSS{1g}} \AR{2} \right) \SL{1} \, , \quad
\hat{\kappa}^*_g = \frac{\AL{g} \slashk_g+\slashp_2 \SR{1} }{\SSS{g1}}
\nn \\
\AA{q\hat{K}} &=& \frac{\AL{q} \slashp_1 + \slashp_2 \SR{g} }{\SSS{1g}} \, , 
\hspace{.95cm}
\AA{\hat{K}g} = - \frac{ \AL{g} \slashp_1+\slashp_2 \SR{g} }{\SSS{1g}} \, ,
\nn \\
\AA{\hat{K}1} &=& \frac{\AA{21}\SSS{2g}}{\SSS{1g}} \, , 
\hspace{1.7cm}
 \AA{2\hat{K}} = \AA{21} \, ,
\eea
leading to the expression
\beq
\frac{\AA{gq}^3\SSS{g1}^4}{\AA{\qb q}\SSS{12}\SSS{g2}\AL{q} \slashp_1+\slashp_2 \SR{g}\AL{g} \slashp_1+\slashp_2 \SR{g}\AL{g} \slashk_g+\slashp_2 \SR{1}} \, .
\eeq
We report the full expression of the amplitude here computed
\bea
\Amp(g^*,\qb^+,q^-,g_1^+,g_2^-) 
&\stackrel{e^\mu=\AL{2}\g^\mu\SR{g}/2}{=}&
\frac{1}{\kstr_g}\, 
\frac{\SSS{\qb 1}^3 \AA{2g}^4 }{\SSS{\qb q} \AL{g} \slashp_2 +\slashk_g  \SR{1} \AL{2}  \slashk_g\,\left( \slashk_g + \slashp_2 \right)  \SR{g} \AL{2}  \slashk_g \SR{\qb} } 
\nn \\
&& \hspace{-20mm}
+\, \frac{1}{\kappa_g}\,\frac{1}{\left( k_g+p_{\qb} \right)^2}\,
\frac{\SSS{g\qb}^2\AA{2q}^3\AL{2} \slashk_g+\slashp_{\qb} \SR{g} }
{\AA{1q}\AA{12}\left\{ (k_g+p_{\qb})^2 \SSS{\qb g}\AA{2q} - \AL{2} \slashk_g+\slashp_{\qb} \SR{g}\AL{q}\slashk_g\SR{\qb} \right\}} 
\nn \\
&& \hspace{-20mm}
+\, \frac{\AA{gq}^3\SSS{g1}^4}{\AA{\qb q}\SSS{12}\SSS{g2}\AL{q} \slashp_1+\slashp_2 \SR{g}\AL{g} \slashp_1+\slashp_2 \SR{g}\AL{g} \slashk_g+\slashp_2 \SR{1}} \, . 
\eea
The second non-MHV amplitude with the first ordering is
\bea
\Amp(g^*,\qb^+,q^-,g_1^-,g_2^+) 
&\stackrel{e^\mu = \AL{g}\g^\mu\SR{2}/2}{=}&
\frac{1}{\kappa_g}\, \frac{\AA{q1}^2\AA{\qb1}\SSS{2g}^4}{\AA{\qb q} \AL{1} \slashp_2 + \slashk_g \SR{g} \SL{g} \left(\slashk_g+\slashp_2 \right)\slashk_g\SR{2} \AL{\qb}\slashk_g\SR{2} }
\nn \\
&&\hspace{-20mm}
+\, \frac{1}{\kstr_g}\,\frac{1}{(k_g+p_{\qb})^2}\, \frac{\SSS{2q} \AL{g} \slashk_g+\slashp_{\qb} \SR{2}^3 }
{\SSS{1q}\SSS{12}\left\{ (k_g+p_{\qb})^2 \AA{\qb g}\SSS{2q} -  \AL{\qb}\slashk_g\SR{q}\AL{g}\slashk_g+\slashp_{\qb} \SR{2}\right\}}
\nn \\
&&\hspace{-20mm}
+\, \frac{ \AA{1g}^4\SSS{g\qb}^2 \SSS{gq}}{\SSS{\qb q}\AA{12}\AA{2g} \AL{g} \slashp_2+\slashk_g\SR{g}\AL{g} \slashp_1+\slashp_2\SR{q}\AL{g} \slashp_1+\slashp_2\SR{g} }
\eea
Concerning the second ordering, the two non-MHV amplitudes are found to be
\bea
\Amp(g^*,g_1^+,\qb^+,q^-,g_2^-) 
&\stackrel{e^\mu=\AL{2}\g^\mu\SR{g}/2}{=}&
\frac{1}{\kstr_g}\, \frac{\SSS{\qb1}^2\SSS{q1}\AA{2g}^4}{\SSS{\qb q} \AL{g} \slashp_2 + \slashk_g \SR{q} \AL{g}\left( \slashk_g+\slashp_2 \right) \slashk_g \AR{2} \AL{2}\slashk_g\SR{1} }
\nn \\
&&
+\, \frac{1}{\kappa_g}\,\frac{1}{(k_g+p_1)^2}\, \frac{\SSS{g1}^3 \AA{2q}^2 \AA{2\qb}}{\AA{\qb q} \AL{2} \slashk_g \SR{1} \AL{\qb} \slashp_1 + \slashk_g  \SR{g}} \, ,
\\
&&
+\, \frac{ \AL{g} \slashp_2+\slashp_q \SR{g}^2\AA{g\qb}\SSS{gq} }{\AA{1\qb}\AA{g1}\SSS{q2} \SSS{2g} \AL{g} \slashp_2+\slashk_g \SR{q} \AL{\qb} \slashp_2+\slashp_q \SR{g} }
\nn \\
\Amp(g^*,g_1^-,\qb^+,q^-,g_2^+) 
&\stackrel{e^\mu=\AL{g}\g^\mu\SR{2}/2}{=}&
\frac{1}{\kappa_g}\, \frac{ \AA{1q}^3 \SSS{2g}^4 }{ \AA{\qb q}\AL{1}\slashk_g\SR{2}\AL{q}\slashp_2+\slashk_g\SR{g}\SL{g}(\slashp_2+\slashk_g)\slashk_g\SR{2} } 
\nn \\
&&
+\, \frac{1}{\kstr_g} \, \frac{1}{(k_g+p_1)^2} \, \frac{\AA{1g}^3\SSS{2\qb }^3}{\SSS{q\qb}\AL{1}\slashk_g\SR{2}\AL{g}\slashk_g+\slashp_1\SR{\qb}}
\nn \\
&&
+\, \frac{\SSS{g\qb}^3\AA{gq}^3}{\SSS{1\qb }\SSS{g1}\AA{q2}\AA{g2}\AL{q} \slashk_g+\slashp_2 \SR{g}\AL{g}\slashp_2+\slashp_q\SR{\qb}} \, .
\eea
%

\section{Summary}\label{summary}

We have discussed in detail the BCFW recursion relation for processes with a fermion pair plus any number of gluons.
One of the partons is always off the mass shell, as required in the framework of hybrid HEF~\cite{Deak:2009xt,Kutak:2012rf}.
To apply this description to the case of full HEF, one can extend the calculations presented here to get 
the same amplitudes with one more parton off-shell.
We have seen that, in the case when one of the fermions is off-shell and shifted, 
one can choose between two different shift vectors, but only one of these choices stays legitimate in the on-shell limit.

In the case of MHV amplitudes, we have provided results holding for any number of particles.
They completely determine the 3- and 4-point scattering amplitudes, whereas in all the other cases non-MHV contributions show up.

The case of the 5-point amplitude was completely worked out for any of the partons off the mass shell for all the non equivalent 
orderings of the scattering particles by performing the calculation of all the non-MHV contributions. \\

{\bf Note added in proof}

A Mathematica notebook containing the results we have discussed and 
allowing prompt numerical evaluation is included with the submission of this paper on the arXiv.

\acknowledgments


The authors would like to thank K. Kutak for useful discussions and comments.
This work was partially supported by NCN grant DEC-2013/10/E/ST2/00656.
M. S. also gratefully acknowledges support from the "Angelo Della Riccia" foundation.
All figures were drawn with feynMF.

\appendix

\section{The construction of 3-point amplitudes}\label{App3Point}%

In order to apply the BCFW construction, it is necessary that
the fundamental building blocks, namely the lowest-order non-vanishing color-stripped amplitudes, are available.
They can be computed in a straightforward way by applying Feynman rules and
we show how to derive two of them this way, in order to illustrate how the rules listed in section \ref{definitions} have to be applied;
we derive one of these amplitude via the BCFW recursion as well, in order to justify the claim of section \ref{definitions},
that the BCFW recursion program is completely independent from the diagrammatic approach also in the off-shell case.

We provide the complete set of results.

\subsection{On-shell amplitudes}

The on-shell 3-point amplitudes for gluons are well known from the literature \cite{Schwartz:2013pla}.
They vanish for real momenta, but not when some of these become complex variables.
The same is true for on-shell 3-point amplitudes with one fermion pair and one gluon, which can be
seen from the general MHV representation in \cite{Luo:2005rx}.
We list them here:
\begin{eqnarray}
\Amp(g_1^+,g_2^-,g_3^-) = \frac{\AA{23}^3}{\AA{12}\AA{31}} \, 
\quad
&&
\Amp(g_1^-,g_2^+,g_3^+) = \frac{\SSS{32}^3}{\SSS{21}\SSS{13}} \nn
\\
\Amp(g^-,\qb^+,q^-) = \frac{\AA{g q}^3\AA{g \qb}}{\AA{g \qb}\AA{\qb q}\AA{q g}}\, 
\quad
&&
\Amp(g^+,\qb^+,q^-) = \frac{\SSS{g \qb}^3\SSS{g q}}{\SSS{g q}\SSS{q \qb}\SSS{\qb g}} \nn
\\
\Amp(g^-,\qb^-,q^+) = \frac{\AA{g \qb}^3\AA{g q}}{\AA{g \qb}\AA{\qb q}\AA{q g}}
\quad
&&
\Amp(g^+,\qb^-,q^+) = \frac{\SSS{g q}^3\SSS{g \qb}}{\SSS{g q}\SSS{q\qb}\SSS{\qb g}} \nn
\end{eqnarray}
%

\subsection{$0\rightarrow g^* g g$}

Amplitudes with three gluons, one of which is off-shell, were computed in \cite{vanHameren:2014iua},
so we simply report them here.
First, equal helicity amplitudes vanish,
\begin{equation}
\Amp(g_1^*,g_2^+,g_3^+) = \Amp(g_1^*,g_2^-,g_3^-) = 0 \, . 
\end{equation}
Amplitudes for which the two on-shell gluons have opposite helicity are given by a simple generalization of the on-shell formula,
\begin{equation}
\Amp(g_1^*,g_2^+,g_3^-) = \frac{1}{\kstr_1}\frac{\AA{31}^3}{\AA{12}\AA{23}} = \frac{1}{\kappa_1}\frac{\SSS{21}^3}{\SSS{13}\SSS{32}}   \, .
\end{equation}
We will see in a moment that amplitudes with one off-shell fermion are a simple generalisation of the on-shell formula as well.

\subsection{$0\rightarrow g^* \qb q $}\label{App3PointSpecial}%

\begin{figure}[h]
\begin{center}
\includegraphics[scale=0.15]{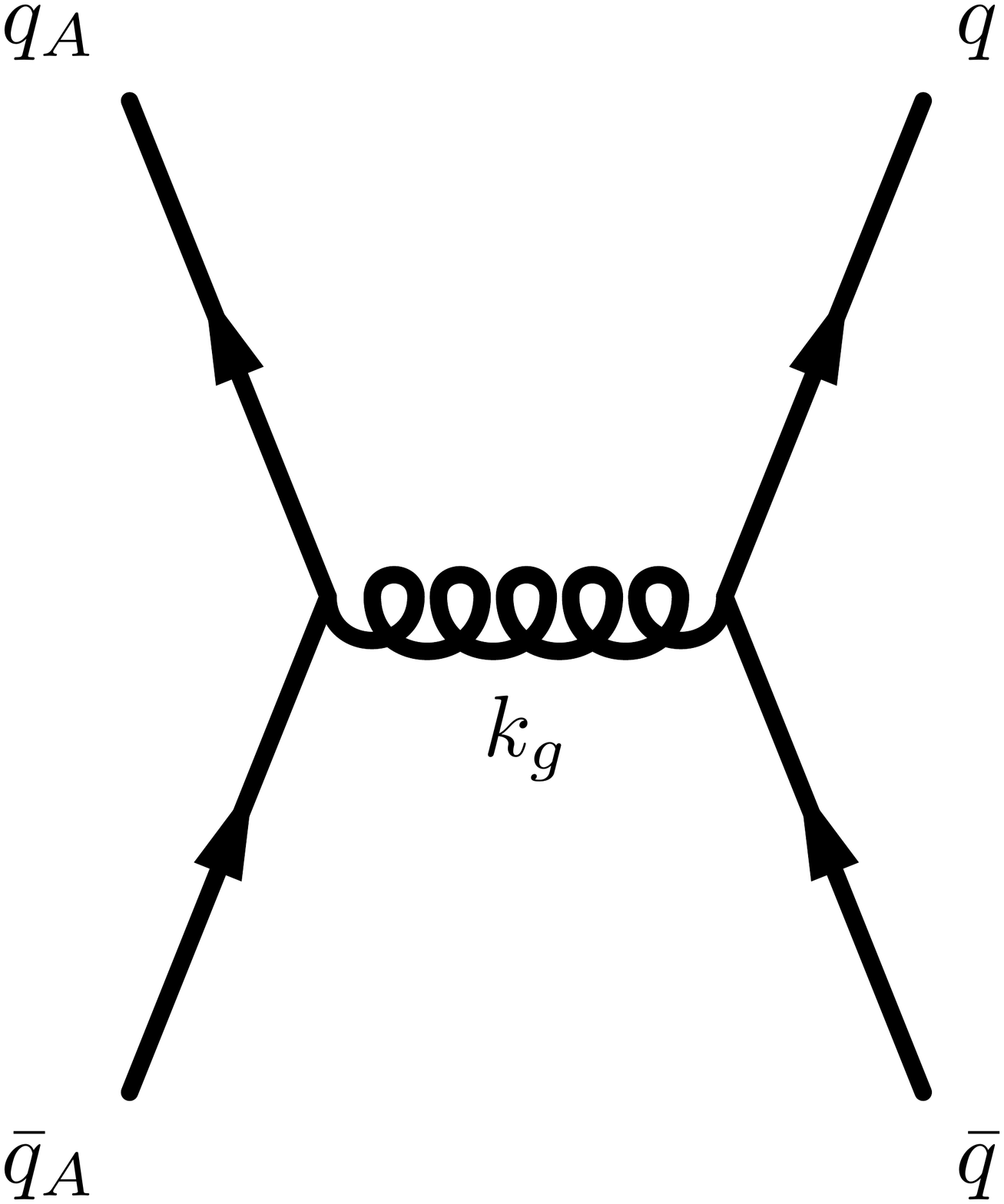}
\caption{The only diagram contributing to the gauge-invariant scattering of an off-shell gluon and two on-shell fermions}
\label{gsqbq_amp}
\end{center}
\end{figure}

Now we present the diagrammatic computation of the 3-point amplitudes with one quark-antiquark pair and 
an off-shell gluon. 
According to the Feynman rules listed in section \ref{definitions}, 
only one diagram contributes to this kind of process, as depicted in fig. \ref{gsqbq_amp}.
One can see the auxiliary quark-antiquark pair going into the off-shell gluon which is
coupled in turn to the on-shell fermion pair.
We will do the explicit computation for one single helicity configuration, 
namely $\Amp(g^*,\qb^+,q^-)$, the other one being completely analogous.
\begin{eqnarray}
\Amp(g^*,\qb^+,q^-) 
&=& 
\AL{q} \gamma^\mu  \SR{\qb} \, \frac{\eta_{\mu\nu}}{k_g^2} \, \frac{\AL{g}\gamma^\nu\SR{g}}{2}\,   = 
\AL{q} \gamma^\mu  \SR{\qb} \, \frac{1}{(p_q + p_{\qb})^2} \, \frac{\AL{g}\gamma_\mu\SR{g}}{2} \,
\nn \\
&=&
\frac{\SSS{g \qb} \AA{qg} }{\AA{q \qb} \SSS{q \qb} } \, \frac{\AL{q} {\slashk}_g \SR{g} }{\AL{q} {\slashk}_g \SR{g}}  = 
- \frac{1}{\kappa_g} \, \frac{\SSS{g \qb}}{ \AA{q \qb}\SSS{\qb q}} \, \AL{q} {\slashp}_{\qb} \SR{g} =
\frac{1}{\kappa_g}\, \frac{\SSS{g \qb}^3 \SSS{g q} }{\SSS{g q} \SSS{q \bar{q} } \SSS{\bar{q}  g} } \, ,
\end{eqnarray}
where, passing from the first to the second line, we have purportedly multiplied ad divided by the same quantity so as to isolate a $1/ \kappa_g$ prefactor.
Notice that it would have also been possible to isolate $1/ \kappa^*_g$, thereby getting the \emph{two equivalent representations}
\begin{equation}
\Amp(g^*,\qb^+,q^-)  = 
\frac{1}{\kappa_g}\, \frac{\SSS{g \qb}^3 \SSS{g q} }{\SSS{g q} \SSS{q \bar{q} } \SSS{\bar{q}  g} }  = 
\frac{1}{\kappa^*_g}\, \frac{\AA{g q}^3 \AA{g \qb} }{\AA{g \bar{q}  } \AA{ \bar{q} q } \AA{q g}   } 
\end{equation}
Similarly, for the other helicity configuration we find the two representations
\begin{equation}
\Amp(g^*,\qb^-,q^+)  = 
\frac{1}{\kappa_g}\, \frac{\SSS{g q}^3 \SSS{g \qb} }{\SSS{g q} \SSS{q \bar{q} } \SSS{\bar{q}  g} }  = 
\frac{1}{\kappa^*_g}\, \frac{\AA{g \qb}^3 \AA{g q} }{\AA{g \bar{q}  } \AA{ \bar{q} q } \AA{q g}   }   \, .
\end{equation}

And now we show how this result can be recovered by BCFW recursion.
The following derivation can be carried out also for all the other $3$-point functions.

The gluon being off shell, the only pole which appears in this amplitude is the $C^g$-type one due to the 
off-shell leg squared momentum vanishing, so that
\beq
\Amp(g^*,\qb^+,q^-)  = \frac{1}{x_g\kappa_g}\,\Amp(\hat{g}^+,\hat{\qb}^+,q^-)  = 
\frac{1}{x_g\kappa_g}\, \frac{\SSS{\hat{g}\hat{\qb}}^3\SSS{\hat{g}q}}{\SSS{\hat{g} q} \SSS{q \hat{\qb}}\SSS{\hat{\qb} \hat{g}}} \, .
\label{rec3pt}
\eeq
A look at Table \ref{CD} shows that (\ref{rec3pt}) is just
\beq
\Amp(g^*,\qb^+,q^-)  = \frac{1}{\kappa_g}\, \frac{\SSS{g \qb}^3 \SSS{g q} }{\SSS{g q} \SSS{q \bar{q} } \SSS{\bar{q}  g} } \, ,
\eeq
as claimed.

\subsection{$0\rightarrow g \bar{q}^{*} q $}

\begin{figure}[h]
\begin{center}
\includegraphics[scale=0.5]{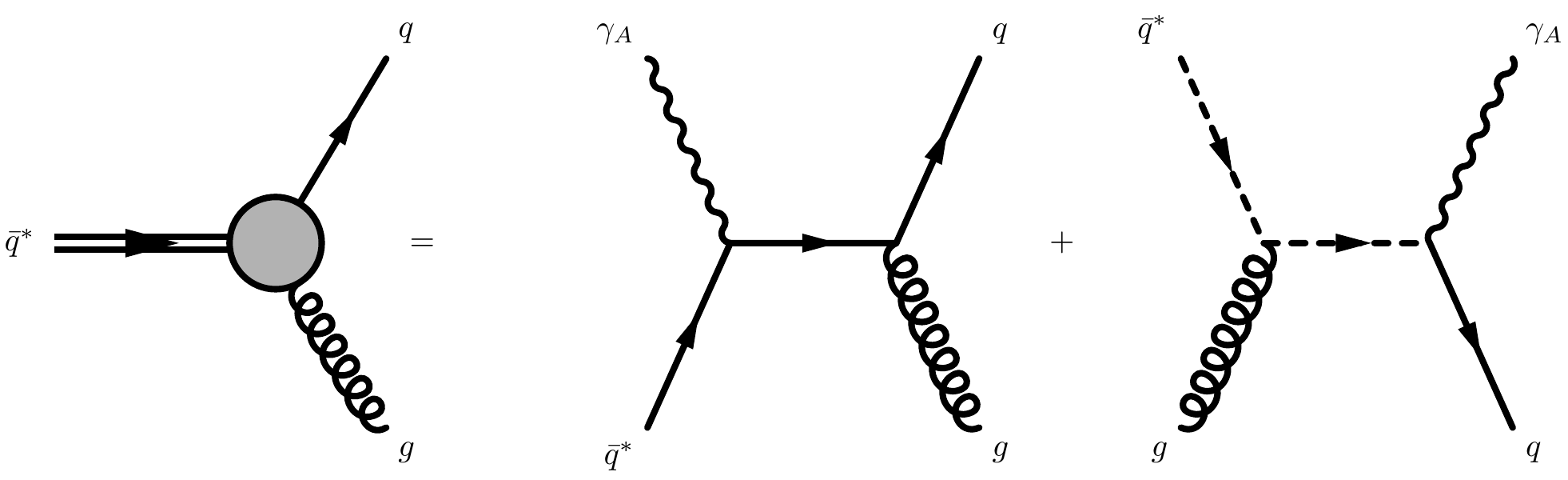}
\caption{The diagrams describing the gauge-invariant scattering of a gluon, a fermion and an off-shell antifermion}
\label{qbsqg_amp}
\end{center}
\end{figure}

Now we move to the case of a 3-point amplitude with an off-shell antifermion.
This time the amplitude, for a general helicity configuration, is given by the sum of two contributions, as illustrated in fig. \ref{qbsqg_amp}.
Using again the Feynman rules and the definitions of the polarization vectors 
for massless gauge bosons given in section \ref{definitions}, we find, for instance,
\begin{eqnarray}
\Amp(g^+,\qb^*,q^-) &=& 
\sum_{h=\pm} \left\{
\AL{q} \frac{{\slasheps}_g^+}{\sqrt{2}}\, \frac{- {\slashk}_{\qb}}{k_{\qb}^2} \, \frac{{\slasheps}^h_\gamma}{\sqrt{2}} \SR{\qb} + 
\AL{q} \frac{{\slasheps}_\gamma^h}{\sqrt{2}}\, \frac{{\slashp_{\qb}}}{2 \lid{p_{\qb}}{p_q}} \, \frac{{\slasheps}^+_g}{\sqrt{2}}\SR{\qb}
\right\}
\end{eqnarray}
The sum is over the two possible helicities of the zero momentum auxiliary photon, but we will see that only one actually contributes.
In fact, choosing $p_q$ to be the reference momentum for the gluon polarization vector, it turns out, recalling (\ref{polarization}), that
\begin{equation}
\slasheps^+_g = \frac{\sqrt{2}}{\AA{qg}}\, \left(  \SR{g}\AL{q} + \AR{q}\SL{g} \right)\, ,
\end{equation}
so that 
\begin{equation}
\Amp(g^+,\qb^*,q^-) =
0 + \frac{ \AL{q} {\slasheps}^-_\gamma \SR{\qb} \AA{\qb q} \SSS{g \qb}}{\sqrt{2}\,\AA{\qb q}\SSS{q \qb}\,\AA{q g}   }\, ,
\label{preqbsqg}
\end{equation}
where it is apparent that the helicity of the auxiliary photon cannot be but negative, because, according to the prescription detailed in 
\cite{vanHameren:2013csa}, the polarization vectors of the photon used to introduce an antifermion in a gauge-invariant way are
\begin{eqnarray}
\slasheps^+_\gamma
&=&
\frac{\sqrt{2}}{\AA{q_{\qb} \qb }}\, \left(  \SR{\qb}\AL{q_{\qb}} + \AR{q_{\qb}}\SL{\qb} \right)
\nn \\ 
\slasheps^-_\gamma
&=&
\frac{\sqrt{2}}{\SSS{\qb q_{\qb}}}\, \left(  \AR{\qb}\SL{q_{\qb}} + \SR{q_{\qb}}\AL{\qb} \right) \, ,
\label{gamma_polarization}
\end{eqnarray}
where $q_{\qb}$ stands for the auxiliary vector for the photon, whereas its massless momentum is 
identified with the direction of $\qb$. \\
Inserting (\ref{gamma_polarization}) into (\ref{preqbsqg}), we find, after some simplifications,
\begin{equation}
\Amp(g^+,\qb^*,q^-) = \frac{\AA{\qb q} \SSS{g \qb} }{\SSS{q \qb} \AA{q g} }\, \frac{\AL{q} \slashk_{\qb} \SR{\qb}}{\AL{q} \slashk_{\qb} \SR{\qb} } =
\frac{1}{\kappa_{\qb}}\, \frac{ \SSS{g \qb}^3 \SSS{g q} }{\SSS{g q} \SSS{q \qb} \SSS{\qb g} }
\end{equation}
Along similar lines, we find also the other amplitudes with an off-shell quark, the final results being
\begin{eqnarray}
\Amp(g^+,\qb^*,q^-) &=&  \frac{1}{\kappa_{\qb}}\, \frac{ \SSS{g \qb}^3 \SSS{g q} }{ \SSS{g q} \SSS{q \qb} \SSS{\qb g} }  \, ,
\nn \\
\Amp(g^-,\qb^*,q^+) &=&  \frac{1}{\kstr_{\qb}}\, \frac{ \AA{g \qb}^3 \AA{g q} }{\AA{g \qb} \AA{\qb q} \AA{q g} } \, ,
\nn \\
\Amp(g^\pm,\qb^*,q^\pm) &=& 0 \, .
\label{3ptFermions}
\end{eqnarray}
%

\subsection{$0\rightarrow g \bar{q} q^*$}%

At this point, it is easy to perform the computation of the 3-point amplitudes with one off-shell fermion,
as it closely mimics that of the previous sections, so that we just list the results:
\begin{eqnarray}
\Amp(g^+,\qb^-,q^*) &=&  \frac{1}{\kappa_{q}}\, \frac{ \SSS{g q}^3 \SSS{g \qb} }{ \SSS{g q} \SSS{q \qb} \SSS{\qb g} }  \, ,
\nn \\
\Amp(g^-,\qb^+,q^*) &=&  \frac{1}{\kstr_{q}}\, \frac{ \AA{g q}^3 \AA{g \qb} }{\AA{g \qb} \AA{\qb q} \AA{q g} } \, ,
\nn \\
\Amp(g^\pm,\qb^\pm,q^*) &=& 0 \, .
\end{eqnarray}
%

\section{Spinors}\label{AppSpinors}%

\subsection{Conventions}

Following the conventions in \cite{vanHameren:2014iua},
the left- and right-handed spinors used in this paper can be defined in the following way:
\begin{eqnarray}
\SR{p}
&=& 
\begin{pmatrix} L(p) \\ \mathbf{0} \end{pmatrix}
\quad \quad
L(p) = \frac{1}{\sqrt{|p^0+p^3|} }\, \begin{pmatrix} -p^1+ \imag\, p^2 \\ p^0+p^3 \end{pmatrix} 
\nn \\
\AR{p}
&=& 
\begin{pmatrix} \mathbf{0} \\ R(p) \end{pmatrix}
\quad \quad
R(p) = \frac{\sqrt{|p^0 + p^3|} }{ p^0 + p^3 } \, \begin{pmatrix} p^0 + p^3 \\ p^1+ \imag \, p^2 \end{pmatrix}  
\end{eqnarray}
The "dual" spinors are defined as
\begin{equation}
\SL{p} = \big(\, (\mathcal{E}L(p))^T \, ,\,\mathbf{0} \,\big)
\quad \quad
\AL{p} = \big(\, \mathbf{0} \, \, ( \mathcal{E}^T R(p) )^T \,\big)
\quad \quad  \textrm{where} \quad
\mathcal{E} = \begin{pmatrix} 0 & 1 \\ -1 & 0 \end{pmatrix}
\end{equation}
The definition of the ``dual'' spinors does not involve complex conjugation and all spinors are well defined for complex momenta. \\
Defining the Pauli vector as
\begin{equation}
\sigma^\mu = \left( \mathbf{1}_{2\times 2}, \stackrel{\rightarrow}{\sigma} \right) \equiv \left( \sigma^0, \sigma^1,\sigma^2,\sigma^3 \right) \, ,
\end{equation}
our conventions for the Dirac matrices are
\begin{equation}
\gamma^0 = 
\left(
\begin{array}{cc}
            0 & \sigma^0 \\
\sigma^0 & 0
\end{array}
\right)
\, , \quad
\gamma^i = 
\left(
\begin{array}{cc}
            0 & \sigma^i \\
- \sigma^i & 0
\end{array}
\right) \, , 
\quad
i = 1,2,3 \, .
\end{equation}
so that $\gamma^5$ is given by
\begin{equation}
\gamma^5 \equiv \imag\, \gamma^0 \gamma^1\gamma^2\gamma^3 = 
\left(
\begin{array}{cc}
-\mathbf{1} & 0 \\
0 & \mathbf{1}
\end{array}
\right)
\, .
\end{equation}
The dyadic product satisfies the identity
\begin{equation}
\AR{p}\SL{p} + \SR{p}\AL{p} = \slashp = \gamma_\mu p^\mu
\end{equation}
The following spinor identities hold and are extensively used in the paper, with the convention that $p^\mu$, $q^\mu$ and $r^\mu$ are always light like,
whereas $k^\mu$ does not have to be light-like
\begin{eqnarray}
\AL{p}\SR{q} &=& \SL{p}\AR{q} = 0 \nn
\\
\AL{p}\AR{p} &=& \SL{p}\SR{p} = 0 \nn
\\
\slashp\AR{p} &=& \slashp\SR{p} = \AL{p}\slashp = \SL{p}\slashp = 0 \nn
\\
\AL{p}\AR{q} &\equiv& \AA{pq} = - \AA{qp} \nn
\\
\SL{p}\SR{q} &\equiv& \SSS{pq} = -\SSS{qp}  \nn
\\
\AA{pq}\SSS{qp} &=& 2\lid{p}{q} \nn
\\
\AS{p|\slashr|q} &=& \AA{pr}\SSS{rq} \nn
\\
\gamma_\mu \AL{p} \gamma^\mu \SR{q}  &=& 2\, \left( \SR{q}\AL{p} + \AR{p}\SL{q}   \right)
\nn
\\
\AL{r}\gamma_\mu \SR{s} \AL{p} \gamma^\mu \SR{q}  &=& 2\, \AA{r p} \SSS{q s}
\nn
\\
\AL{p}\gamma^{\mu_1}\dots \gamma^{\mu_n}\SR{q} &=& \SL{q}\gamma^{\mu_n}\dots \gamma^{\mu_1}\AR{p}
\nn\\
\AL{p} \gamma^{\mu_1}\dots \gamma^{\mu_n} \AR{q} &=& -  \AL{q} \gamma^{\mu_n} \dots \gamma^{\mu_1} \AR{p}
\nn\\
\SL{p} \gamma^{\mu_1}\dots \gamma^{\mu_n} \SR{q} &=& -  \SL{q} \gamma^{\mu_n} \dots \gamma^{\mu_1} \SR{p}
\end{eqnarray}
The last two relations trivially reduce to $0=0$ for odd $n$, and the one before reduces to $0=0$ for even $n$.

\subsection{Schouten identity}\label{AppSchouten}

A very insightful way to express the Schouten identity is to write it as a completeness relation.
For any $p^\mu,r^\mu$ with $p^2=r^2=0$ and $\lid{p}{r}\neq0$ we have
\begin{equation}
\frac{\AR{r}\AL{p}}{\AA{pr}} + \frac{\AR{p}\AL{r}}{\AA{rp}} + \frac{\SR{r}\SL{p}}{\SSS{pr}} + \frac{\SR{p}\SL{r}}{\SSS{rp}} = 1 \, .
\label{Schouten}
\end{equation}
A simple application is the proof that $\kappa$ and $\kstr$ are independent of the auxiliary momentum.
Inserting the identity operator strategically, we find for $\kappa$ 
\begin{equation}
\kappa = \frac{\AS{q|\slashk|p}}{\AA{qp}} 
= \frac{\AS{q|1\slashk|p}}{\AA{qp}}
= \frac{\AA{qr}\AS{p|\slashk|p}}{\AA{pr}\AA{qp}} + \frac{\AA{qp}\AS{r|\slashk|p}}{\AA{rp}\AA{qp}}
= \frac{\AS{r|\slashk|p}}{\AA{rp}} \, ,
\end{equation}
where we used the fact that $\AS{p|\slashk|p}=2\lid{p}{k}=0$.
The same can be shown for $\kstr$.




\begin{thebibliography}{99}


\bibitem{Kleiss:1985yh}
R.~Kleiss and W.~J. Stirling, {\it {Spinor Techniques for Calculating p anti-p $\to$ W+- / Z0 + Jets}},  
{\em Nucl.Phys.} {\bf B262} (1985) 235--262.

\bibitem{Gunion:1985vca}
J.~Gunion and Z.~Kunszt, {\it {Improved Analytic Techniques for Tree Graph Calculations and the G g q anti-q Lepton anti-Lepton Subprocess}},  
{\em Phys.Lett.} {\bf B161} (1985) 333.

\bibitem{Xu:1986xb}
Z.~Xu, D.-H. Zhang, and L.~Chang, {\it {Helicity Amplitudes for Multiple Bremsstrahlung in Massless Nonabelian Gauge Theories}},  
{\em Nucl.Phys.} {\bf B291} (1987) 392.

\bibitem{Berends:1987me}
F.~A. Berends and W.~Giele, {\it {Recursive Calculations for Processes with n Gluons}},  
{\em Nucl.Phys.} {\bf B306} (1988) 759.

\bibitem{Mangano:1987xk}
M.~L. Mangano, S.~J. Parke, and Z.~Xu, {\it {Duality and Multi - Gluon Scattering}},  
{\em Nucl.Phys.} {\bf B298} (1988) 653.

\bibitem{Mangano:1990by}
M.~L. Mangano and S.~J. Parke, {\it {Multiparton amplitudes in gauge theories}},  
{\em Phys.Rept.} {\bf 200} (1991) 301--367, [{\tt arXiv:hep-th/0509223}].

\bibitem{Berends:1989hf}
F.~A. Berends, W.~Giele, and H.~Kuijf, {\it {Exact and Approximate Expressions for Multi - Gluon Scattering}}, 
 {\em Nucl.Phys.} {\bf B333} (1990) 120.

\bibitem{Kosower:1989xy}
D.~A. Kosower, {\it {Light Cone Recurrence Relations for QCD Amplitudes}},
{\em Nucl.Phys.} {\bf B335} (1990) 23.

\bibitem{Britto:2004ap}
R.~Britto, F.~Cachazo, and B.~Feng, {\it {New recursion relations for tree amplitudes of gluons}},  
{\em Nucl.Phys.} {\bf B715} (2005) 499--522, [{\tt arXiv:hep-th/0412308}].

\bibitem{Britto:2005fq}
R.~Britto, F.~Cachazo, B.~Feng, and E.~Witten, {\it {Direct proof of tree-level recursion relation in Yang-Mills theory}},  
{\em Phys.Rev.Lett.} {\bf 94} (2005) 181602, [{\tt arXiv:hep-th/0501052}].

\bibitem{Luo:2005rx}
M.-x. Luo and C.-k. Wen, {\it {Recursion relations for tree amplitudes in super gauge theories}},  
{\em JHEP} {\bf 0503} (2005) 004, [{\tt arXiv:hep-th/0501121}].

\bibitem{vanHameren:2014iua}
A.~van Hameren, {\it {BCFW recursion for off-shell gluons}},  
{\em JHEP} {\bf 1407} (2014) 138, [{\tt arXiv:1404.7818}].

\bibitem{Catani:1990eg}
S.~Catani, M.~Ciafaloni, and F.~Hautmann, {\it {High-energy factorization and
  small x heavy flavor production}},  {\em Nucl.Phys.} {\bf B366} (1991) 135--188.

\bibitem{Collins:1991ty}
J.~C. Collins and R.~K. Ellis, {\it {Heavy quark production in very high-energy hadron collisions}},  
{\em Nucl.Phys.} {\bf B360} (1991) 3--30.

\bibitem{Fadin:1993wh}
V.~S. Fadin and L.~Lipatov, {\it {Radiative corrections to QCD scattering amplitudes in a multi - Regge kinematics}},  
{\em Nucl.Phys.} {\bf B406} (1993) 259--292.

\bibitem{Fadin:1996nw}
V.~S. Fadin and L.~Lipatov, {\it {Next-to-leading corrections to the BFKL equation from the gluon and quark production}}, 
{\em Nucl.Phys.} {\bf B477} (1996) 767--808, [{\tt arXiv:hep-ph/9602287}].

\bibitem{Lipatov:1995pn}
L.~Lipatov, {\it {Gauge invariant effective action for high-energy processes in QCD}},  
{\em Nucl.Phys.} {\bf B452} (1995) 369--400, [{\tt arXiv:hep-ph/9502308}].

\bibitem{Antonov:2004hh}
E.~Antonov, L.~Lipatov, E.~Kuraev, and I.~Cherednikov, {\it {Feynman rules for effective Regge action}},  
{\em Nucl.Phys.} {\bf B721} (2005) 111--135, [\href{http://xxx.lanl.gov/abs/hep-ph/0411185}{{\tt hep-ph/0411185}}].

\bibitem{vanHameren:2012if}
A.~van Hameren, P.~Kotko, and K.~Kutak, {\it {Helicity amplitudes for high-energy scattering}},  
{\em JHEP} {\bf 1301} (2013) 078, [{\tt arXiv:1211.0961}].

\bibitem{Kotko:2014aba}
P.~Kotko, {\it {Wilson lines and gauge invariant off-shell amplitudes}}, 
{\em JHEP} {\bf 1407} (2014) 128, [{\tt arXiv:1403.4824}].

\bibitem{Lipatov:2000se}
L.~Lipatov and M.~Vyazovsky, {\it {QuasimultiRegge processes with a quark exchange in the t channel}},  
{\em Nucl.Phys.} {\bf B597} (2001) 399--409, [{\tt arXiv:hep-ph/0009340}].

\bibitem{vanHameren:2013csa}
A.~van Hameren, K.~Kutak, and T.~Salwa, {\it {Scattering amplitudes with off-shell quarks}}, 
{\em Phys.Lett.} {\bf B727} (2013) 226--233, [{\tt arXiv:1308.2861}].

\bibitem{Nefedov:2013ywa}
M.~Nefedov, V.~Saleev, and A.~V. Shipilova, {\it {Dijet azimuthal decorrelations at the LHC in the parton Reggeization approach}},
{\em Phys.Rev.} {\bf D87} (2013) 094030, [{\tt arXiv:1304.3549}].

\bibitem{Kniehl:2014qva}
B.~Kniehl, M.~Nefedov, and V.~Saleev, {\it {Prompt-photon plus jet associated photoproduction at HERA in the parton Reggeization approach}},  
{\em Phys.Rev.} {\bf D89} (2014), no.~11 114016, [{\tt arXiv:1404.3513}].

\bibitem{Nefedov:2014qea}
A.~Karpishkov, M.~Nefedov, V.~Saleev, and A.~Shipilova, {\it {Open charm production in the parton Reggeization approach: Tevatron and the LHC}},  
{\em Phys.Rev.} {\bf D91} (2015), no.~5 054009, [{\tt arXiv:1410.7139}].

\bibitem{Karpishkov:2014epa}
A.~Karpishkov, V.~Saleev, M.~Nefedov, and A.~Shipilova, {\it {B-meson production in the Parton Reggeization Approach at Tevatron and the LHC}},
{\em Int.J.Mod.Phys.} {\bf A30} (2015) 0023, [{\tt arXiv:1411.7672}].

\bibitem{Cachazo:2004kj}
F.~Cachazo, P.~Svrcek, and E.~Witten, {\it {MHV vertices and tree amplitudes in gauge theory}},  
{\em JHEP} {\bf 0409} (2004) 006, [{\tt arXiv:hep-th/0403047}].

\bibitem{Parke:1986gb}
S.~J. Parke and T.~Taylor, {\it {An Amplitude for $n$ Gluon Scattering}},
{\em Phys.Rev.Lett.} {\bf 56} (1986) 2459.

\bibitem{Bury:2015dla}
M.~Bury and A.~van Hameren, {\it {Numerical evaluation of multi-gluon amplitudes for High Energy Factorization}},
 [{\tt arXiv:1503.08612}].

\bibitem{Deak:2009xt}
M.~Deak, F.~Hautmann, H.~Jung, and K.~Kutak, {\it {Forward Jet Production at the Large Hadron Collider}},  
{\em JHEP} {\bf 0909} (2009) 121, [{\tt arXiv:0908.0538}].

\bibitem{Kutak:2012rf}
K.~Kutak and S.~Sapeta, {\it {Gluon saturation in dijet production in p-Pb collisions at Large Hadron Collider}},  
{\em Phys.Rev.} {\bf D86} (2012) 094043, [{\tt arXiv:1205.5035}].

\bibitem{Schwartz:2013pla}
M.~D. Schwartz, {\em {Quantum Field Theory and the Standard Model}}.
\newblock 2013.








\end{thebibliography}
\end{document}